\documentclass[epj,nopacs]{svjour}
\usepackage{epsfig,array}
\usepackage{multirow,color}
\usepackage{epsf,graphicx}
\usepackage{times,latexsym,euscript,amstext,amssymb,amsbsy,amsmath,amsfonts}
\usepackage{mdwlist}\usepackage{epsfig}
\usepackage{slashbox,multirow}
\newcommand{\be}{\begin{eqnarray}}
\newcommand{\ee}{\end{eqnarray}}
\newcommand{\bc}{\begin{center}}
\newcommand{\ec}{\end{center}}
\newcommand{\bea}{\begin{eqnarray}}
\newcommand{\eea}{\end{eqnarray}}

\newcommand{\beq}{\begin{equation}}
\newcommand{\eeq}{\end{equation}}

\newcommand{\nn}{\nonumber \\ }
\newcommand{\nnn}{\nonumber  }

\def\fun#1#2{\lower3.6pt\vbox{\baselineskip0pt\lineskip.9pt
\ialign{$\mathsurround=0pt#1\hfil##\hfil$\crcr#2\crcr\sim\crcr}}}

\begin{document}

\title{P-wave excited baryons from pion- and photo-induced hyperon production
}
\titlerunning{P-wave excited baryons }
\author{
 A.V.~Anisovich$\,^{1,2}$, E.~Klempt$\,^1$, V.A.~Nikonov$\,^{1,2}$,
 A.V.~Sarantsev$\,^{1,2}$ and U.~Thoma$\,^{1}$}
\authorrunning{A.V.~Anisovich \it et al.}
\institute{$^1\,$Helmholtz-Institut f\"ur Strahlen- und Kernphysik,
Universit\"at Bonn, Germany\\
$^2\,$Petersburg Nuclear Physics Institute, Gatchina, Russia}

\date{Received: \today / Revised version:}

\abstract{We report evidence for $N(1710)P_{11}$, $N(1875)P_{11}$,
$N(1900)P_{13}$, $\Delta(1600)P_{33}$, $\Delta(1910)P_{31}$, and
$\Delta(1920)P_{33}$, and find indications that $N(1900)P_{13}$
might have a companion state at 1970\,MeV. The controversial
$\Delta(1750)P_{31}$ is not seen. The evidence is derived from a
study of data on pion- and photo-induced hyperon production, but
other data are included as well. Most of the resonances reported
here were found in the Karlsruhe-Helsinki (KH84)
\cite{Hohler:1979yr} and the Carnegie-Mellon (CM)
\cite{Cutkosky:1980rh} analyses but were challenged recently by the
Data Analysis Center at GWU \cite{Arndt:2006bf}. Our analysis is
constrained by the energy independent $\pi N$ scattering amplitudes
from either KH84 or GWU. The two $\pi N$ amplitudes from KH84 or
GWU, respectively, lead to slightly different $\pi N$ branching
ratios of contributing resonances but the debated resonances are
required in both series of fits.
 \vspace{1mm}   }
\mail{klempt@hiskp.uni-bonn.de\\
The $\pi N$ induced amplitudes, photoproduction observables and
multipoles for both solutions (BG2010-01 and 02) can be downloaded
from our web site as figures or in the numerical form
(http://pwa.hiskp.uni-bonn.de). }

\maketitle

\section{Introduction}

The existence of radially excited resonances and - if affirmed -
their mass pattern represents one of the most controversial issues
in baryon spectroscopy (see, e.g., \cite{Klempt:2009pi} for a recent
review). Well known is the problem of the lowest mass nucleon
excitation $N(1440)P_{11}$, the so-called Roper resonance
\cite{Arndt:1985vj}. It is studied in many reactions, and its
existence is beyond doubt. But its - compared to quark model
calculations - low mass and its broad width has invited speculations
that it could be dynamically generated and unrelated to $(qqq)$
spectroscopy. In the $P_{11}$-wave, discrepancies between different
analyses show up above the Roper resonance. H\"ohler and
collaborators \cite{Hohler:1979yr} and Cutkosky and collaborators
\cite{Cutkosky:1980rh} identified two further states in the $P_{11}$
partial wave, $N(1710)P_{11}$ and $N(2100)P_{11}$. Manley and
Saleski \cite{Manley:1992yb} found two resonances as well,
$N(1710)P_{11}$ and a state at 1885\,MeV (listed under the
$N(2100)P_{11}$ PDG entry \cite{Amsler:2008zzb}) which is confirmed
in a coupled channel analysis of photoproduction data
\cite{Anisovich:2009zy} but not listed in the PDG listing as
separate state. In \cite{Vrana:1999nt}, observation of $N(1710)$
$P_{11}$ and $N(2100)P_{11}$ was reported. Hence there seem to be
two or possibly even three $P_{11}$ states with spin-parity
$J^P=1/2^+$ above the Roper resonance. However, in a recent analysis
of a large body of $\pi N$ elastic and charge exchange scattering
data, only the Roper resonance was confirmed~\cite{Arndt:2006bf}
putting into doubt the existence of the three candidate resonances
above it. The authors in \cite{Suzuki:2009nj} suggest that the two
lowest $P_{11}$ nucleon resonances, the Roper $N(1440)P_{11}$ and
$N(1710)P_{11}$, originate from a single bare state. In the $P_{13}$
wave, the existence of $N(1720)P_{13}$ is beyond doubt (even though
not seen in \cite{Suzuki:2009nj}) but the next state,
$N(1900)P_{13}$, was observed in none of the $\pi N$ amplitude
analyses \cite{Hohler:1979yr,Cutkosky:1980rh,Arndt:2006bf} but only
in coupled channel analyses
\cite{Manley:1992yb,Nikonov:2007br,Penner:2002md} which included
some inelastic channels.

A similar situation is found for $\Delta$ excitations.
In~\cite{Arndt:2006bf}, the lowest-mass resonance above
$\Delta(1232)P_{33}$, the $\Delta(1600)P_{33}$ resonance, is found
with a much larger width than reported in analyses of elastic
scattering data \cite{Hohler:1979yr,Cutkosky:1980rh,Manley:1992yb}
and of inelastic reactions
\cite{Manley:1992yb,Penner:2002md,Horn:2007pp}. The
$\Delta(1920)P_{33}$ resonance is absent in~\cite{Arndt:2006bf}. To
complete the low-spin positive-parity states, we mention
$\Delta(1750)P_{31}$ which is seen in some inelastic reactions
\cite{Manley:1992yb,Vrana:1999nt,Penner:2002md} and not seen in the
analyses of \cite{Hohler:1979yr,Cutkosky:1980rh}. In
\cite{Arndt:2006bf}, one pole is found in the $P_{31}$ wave at $(M,
\Gamma)=(1771, 479)$\,MeV which is consistent with
$\Delta(1750)P_{31}$. A Breit-Wigner resonance yields
$M=2067.9\pm1.7$\,MeV and $\Gamma=543\pm10$\,MeV. The four-star
$\Delta(1910)P_{31}$ is argued to be highly questionable. In
\cite{Suzuki:2009nj}, the 3-star $\Delta(1600)P_{33}$ and the two
4-star resonances $\Delta(1910)P_{31}$ and $\Delta(1920)P_{33}$ are
all missing.

The situation is thus very unsatisfactory. The Particle Data Group
bases the evidence for the existence of states nearly entirely on
four analyses
\cite{Hohler:1979yr,Cutkosky:1980rh,Arndt:2006bf,Manley:1992yb}. The
latest analysis by Arndt {\it et al.}~\cite{Arndt:2006bf} includes
high precision data from the meson factories at {\sc LAMPF}, {\sc
PSI}, and {\sc TRIUMF} which constrain the low-energy region very
precisely, and important measurements of spin rotation parameters
(see \cite{Arndt:2006bf} for a list of data). Hence one should
expect that the analysis \cite{Arndt:2006bf} should also be the most
reliable one. But this analysis challenges the existence of many
resonances. A strong argument in favor of the Arndt analysis is the
correct prediction of spin rotation parameters
\cite{Alekseev:1996gs,Alekseev:2000nk,Alekseev:2005zr} and of the
backward asymmetry \cite{Alekseev:2008cw} in the elastic pion-proton
scattering from ITEP/PNPI  while the predictions from
\cite{Hohler:1979yr,Cutkosky:1980rh} show clear discrepancies with
the data. On the other hand, it is difficult to believe that the
consistency between the older analyses is just fortuitous.

A difficulty in the partial wave analysis of $\pi N$ elastic
scattering data lies in the fact that the amplitudes cannot be
constructed from the data on $\pi N$ elastic (and charge exchange)
scattering  without theoretical input. For most energies and most
angles, only the differential cross section $d\sigma/d\Omega$ and
the target asymmetry (describing the distribution of the angle
between scattering plane and target polarization) are known. From
this data, the absolute values of the spin-flip and spin non-flip
amplitudes $|H|$ and $|G|$ can be determined but not their phases.
This continuum ambiguity can only be resolved by enforcing
dispersion relations. Likely, the different use of dispersion
relations is responsible for the differences between the old
analyses of \cite{Hohler:1979yr,Cutkosky:1980rh} and the analysis
presented in \cite{Arndt:2006bf}.

In this paper we study the consistency of the $\pi N\to \pi N$
scattering amplitudes for low-spin positive-parity resonances from
the analyses \cite{Hohler:1979yr,Arndt:2006bf} with the $\pi N$ and
$\gamma N$ transition amplitudes into kaon-hyperon and $\eta N$
final states:
\begin{subequations}
\bea
\pi^- p\quad\to\quad &\Lambda\ K^0 & \qquad
\label{pip_KL}\\
\pi^+ p\quad\to\quad &\Sigma^+ K^+ & \qquad
\label{pip_KS++}\\
\pi^- p\quad\to\quad &\Sigma^0 K^0 & \qquad
\label{pip_KS00} \\
\pi^- p\quad\to\quad &n\,\eta & \qquad
\label{pip_neta}
\eea\vspace{-10mm}
\end{subequations}
\begin{subequations}
\bea
\hspace{-20mm}{\rm and}\qquad\gamma p\quad\to\quad &\Lambda\ K^+ &
\qquad
\label{gp_KL}\\
\gamma p\quad\to\quad &\Sigma^0 K^+ & \qquad
\label{gp_KS++}\\
\gamma p\quad\to\quad &\Sigma^+ K^0 & \qquad
\label{gp_KS00} \\
\gamma p\quad\to\quad &p\,\pi^0 & \qquad
\label{gp_npi0}\\
\gamma p\quad\to\quad &p\,\eta & \qquad
\label{gp_neta}
\eea
\end{subequations}
The study of strangeness production has a few distinctive
advantages. The differential cross sections for these reactions are
known with reasonable accuracy and, due to the self-analyzing power
of the final-state hyperons, the recoil polarization can be
determined from the hyperon decay. Reactions (\ref{pip_KL},
\ref{pip_neta}, \ref{gp_KL}, \ref{gp_neta}) are restricted to
nucleon resonances, reaction (\ref{pip_KS++}) to $\Delta$
resonances, reactions (\ref{pip_KS00}, \ref{gp_KS++}, \ref{gp_KS00})
receive contributions from both. Hence the isospin decomposition of
the transition amplitude is defined by the data.

Here we use, alternatively, the energy independent elastic $\pi N$
amplitudes from \cite{Hohler:1979yr} or \cite{Arndt:2006bf},
respectively, to find energy dependent amplitudes satisfying the
data on elastic scattering and on reactions
(\ref{pip_KL})-(\ref{gp_neta}). In addition, the amplitudes are
constrained by a large number of photoproduction data. The Cutkosky
amplitudes \cite{Cutkosky:1980rh} are mostly in between the Arndt
and the H\"ohler amplitudes. We did not use these systematically
since we did not expect additional insight.

The use of a large number of final states constrains the
inelasticity of the $\pi N$ amplitude which otherwise is a free fit
parameter to be determined for every bin. This is a major advantage
of this analysis compared to those of
\cite{Hohler:1979yr,Cutkosky:1980rh,Arndt:2006bf}. A major result of
this analysis is that the existence and the properties of all
resonances used here are hardly affected by the choice of elastic
$\pi N$ amplitudes \cite{Hohler:1979yr,Arndt:2006bf}. Only the $\pi
N$ coupling constants of resonances change, and even this change is
moderate. We account for the difference by increasing the error in
the $\pi N$ branching ratios.

Finally, we comment on the naming scheme we adopt. We use the
conventional names of the Particle Data Group \cite{Amsler:2008zzb}:
$N({\rm mass})L_{2I,2J}$ and $\Delta({\rm mass})L_{2I,2J}$ where $I$
and $J$ are isospin and total spin of the resonance and $L$ the
orbital angular momentum in the decay of the resonance into nucleon
and pion. For resonances not included in \cite{Amsler:2008zzb}, we
use $N_{J^P}({\rm mass})$ and  $\Delta_{J^P}({\rm mass})$ which
gives the spin-parity of the resonance. The latter scheme is adopted
from the meson naming scheme and easily understood also outside of
the baryon community. In this way, the reader can easily see which
resonances are introduced here and which resonances found here are
compatible with PDG values.

\begin{table}[pb]
\caption{\label{piN_data_table}Pion induced reactions fitted in the
coupled-channel analysis and $\chi^2$ contributions for the solution
BG2010-02.}
\bc\begin{tabular}{ccccc}
\hline\hline\\[-2ex]
$\pi N \rightarrow \pi N$& Wave & $N_{\rm data}$ &$w_i$ &$\chi^2_i/N_{\rm data}$\\[1ex]\hline\\[-2ex]
\cite{Arndt:2006bf}& $S_{11}$ & 104 & 30 & 1.95 \\
& $S_{31}$ & 112 & 20 & 2.07 \\
& $P_{11}$ & 112 & 50 & 2.12 \\
& $P_{31}$ & 104 & 20 & 3.86\\
& $P_{13}$ & 112 & 25 & 1.22 \\
& $P_{33}$ & 120 & 15 & 2.87\\
& $D_{13}$ & 96 & 10 & 2.87\\
& $D_{33}$ & 108 & 12 &2.68 \\
& $D_{15}$ & 96 & 20 & 3.67\\
& $F_{35}$ & 62 & 20 & 1.48\\
& $F_{37}$ & 72 & 10 & 2.76\\[0.5ex]\hline \hline\\[-2ex]
$\pi^- p \rightarrow \eta n$ & Observ. & $N_{\rm data}$&$w_i$ & $\chi^2_i/N_{\rm data}$\\[1ex]\hline\\[-2.3ex]
\cite{Richards:1970cy}&  $d\sigma/d\Omega$ & 68 & 20 &1.68 \\
\cite{Prakhov:2005qb}&  $d\sigma/d\Omega$ & 84 & 30 & 2.50 \\[0.5ex]\hline\\[-2.3ex]
\hline\\
[-2.0ex] $\pi^- p \rightarrow K^0\Lambda$ & Observ. & $N_{\rm data}$&$w_i$ & $\chi^2_i/N_{\rm data}$\\[1ex]\hline\\[-2.3ex]
\cite{Knasel:1975rr}&  $d\sigma/d\Omega$ & 298 & 30 &2.31 \\
\cite{Baker:1978qm,Saxon:1979xu}&  $d\sigma/d\Omega$ & 299 & 30 & 0.90 \\
\cite{Baker:1978qm,Saxon:1979xu}&  $P$ & 354 & 30 & 1.98 \\
\cite{Bell:1983dm}&  $\beta$ & 72 & 100 & 2.45 \\
[0.5ex]\hline\\[-2.3ex]
\hline \\[-2.0ex]
$\pi^+ p \rightarrow K^+\Sigma^+$ & Observ. & $N_{\rm data}$&$w_i$ & $\chi^2_i/N_{\rm data}$\\[1ex]\hline\\[-2.3ex]
\cite{Candlin:1982yv}&  $d\sigma/d\Omega$ & 609 & 35 &1.27 \\
\cite{Candlin:1982yv}&  $P$ & 304 & 30 & 1.58 \\
\cite{Candlin:1988pn}&  $\beta$ & 7 &1000 & 1.97 \\
[0.5ex]\hline\\[-2.3ex]
\hline\\[-2.0ex]
$\pi^- p \rightarrow K^0\Sigma^0$ & Observ. & $N_{\rm data}$&$w_i$ & $\chi^2_i/N_{\rm data}$\\[1ex]\hline\\[-2.3ex]
\cite{Hart:1979jx}&  $d\sigma/d\Omega$ & 259 & 30 &0.77 \\
\cite{Hart:1979jx}&  $P$ & 90 & 30 & 1.36 \\
[0.5ex]\hline\\[-2.3ex]
\hline
\end{tabular}\ec
\end{table}
\section{Data, PWA method, and fits}
\subsection{Data}
Tables~\ref{piN_data_table}-\ref{chisquare1} summarize the data used
here. Given are the reaction, the observables and references to the
data, the number of data points, the weight with which the data are
used in the fits, and the $\chi^2$ per data point of our final
solution BG2010-02.

\begin{table}[pt] \caption{\label{3BodyReactions}Reactions
leading to 3-body final states are included in event-based
likelihood fits for the solution BG2010-02. CB stands for CB-ELSA;
CBT for CBELSA/TAPS.}
\bc\begin{tabular}{lcccc}
\hline\hline\\[-1ex]
\multicolumn{2}{c}{$d\sigma/d\Omega(\pi^-p \rightarrow \pi^0\pi^0
n)$} &\hspace{-3mm} $N_{\rm data}$ &$w_i$& $-\ln L$\\[1ex]\hline\\[-2ex]
T=373 MeV   &&\hspace{-3mm} 5248 & 10& -939\\
\multicolumn{2}{l}{T=472 MeV \hspace{16mm} Crystal }  &\hspace{-3mm} 10641& 5& -2605\\
\multicolumn{2}{l}{T=551 MeV \hspace{15mm}  Ball \cite{Prakhov:2004zv}} &\hspace{-3mm} 41172 & 2.5& -7245\\
\multicolumn{2}{l}{T=655 MeV \hspace{16mm}  (BNL)}  &\hspace{-3mm} 63514 & 2& -14926\\
T=691 MeV & &\hspace{-3mm} 30030 & 3.5& -8055\\
T=748 MeV   &&\hspace{-3mm} 30379 & 4& -6952\\[0.4ex]\hline\\[-2.1ex]
\hspace{-2mm}$d\sigma/d\Omega(\gamma p \rightarrow \pi^0\pi^0 p)$   &\hspace{-5mm}CB\hspace{2mm}\cite{Thoma:2007bm,Sarantsev:2007bk}&\hspace{-3mm} 110601 & 4 & -26795\\
\hspace{-2mm}$d\sigma/d\Omega(\gamma p \rightarrow \pi^0\eta p)$
&\hspace{-4mm}CB \cite{Horn:2007pp,Weinheimer:2003ng,Horn:2008qv}&
\hspace{-3mm}17468 & 8 &
-5652\\
\hline\hline\\[-2ex]
\multicolumn{2}{c}{} &\hspace{-3mm} $N_{\rm data}$ &$w_i$& $\chi^2/N_{\rm data}$\\[1ex]\hline\\[-2ex]
$\Sigma(\gamma p \rightarrow \pi^0\pi^0 p)$
&\hspace{-5mm}GRAAL\hspace{2mm}\cite{Assafiri_03}
&\hspace{-3mm} 128 & 35 & 0.96\\
$\Sigma(\gamma p \rightarrow \pi^0\eta p)$
&\hspace{-5mm}CBT\hspace{2mm}\cite{Gutz:2008zz}&\hspace{-3mm} 180
& 15 & 2.41\\
 $E(\gamma p \rightarrow \pi^0\pi^0 p)$
&\hspace{-5mm}GDH/A2\hspace{2mm}\cite{Ahrens_07} &\hspace{-3mm} 16 & 35 & 1.31
\\\hline\hline
\end{tabular}\ec
\caption{\label{chisquare}Observables from $\pi$ photoproduction
fitted in the coupled-channel analysis and $\chi^2$ contributions
for the solution BG2010-02.\vspace{-2mm}}
\bc\begin{tabular}{lcccc}
\hline\hline\\[-1ex]
$\gamma p \rightarrow \pi^0 p$ &\hspace{-3mm} Observ. & $N_{\rm data}$&$w_i$ & $\chi^2_i/N_{\rm data}$\\[1ex]\hline\\[-2ex]
\cite{Fuchs:1996ja} (TAPS@MAMI)&\hspace{-3mm} $d\sigma/d\Omega$  & 1691 & 0.8&1.81 \\
\cite{Ahrens:2002gu,Ahrens:2004pf} (GDH A2)&\hspace{-3mm} $d\sigma/d\Omega$  & 164 & 7&1.17 \\
\cite{Bartalini:2005wx} (GRAAL)&\hspace{-3mm} $d\sigma/d\Omega$  & 861 & 2&1.59 \\
\cite{Bartholomy:2004uz,vanPee:2007tw} (CB)&\hspace{-3mm} $d\sigma/d\Omega$  & 1106 & 3.5&1.65 \\
\cite{Dugger:2007bt} (CLAS)&\hspace{-3mm} $d\sigma/d\Omega$ & 592 & 6 &1.81 \\
\cite{Bartalini:2005wx,Barbiellini:1970qu,Gorbenko:1974sz,Gorbenko:1978re,Belyaev:1983xf,%
Blanpied:1992nn,Beck:1997ew,Adamian:2000yi,Blanpied:2001ae}(GRAAL a.o.)&  $\Sigma$ &1492& 3 &2.82\\
\cite{Sparks:2010vb} (CBT) &  $\Sigma$& 374 & 30 & 1.03\\
\cite{Gorbenko:1974sz,Gorbenko:1978re,Belyaev:1983xf,Booth:1976es,Feller:1976ta,%
Gorbenko:1977rd,Herr:1977vx,Fukushima:1977xj,Bussey:1979wt,Agababian:1989kd,%
Asaturian:1986bj,Bock:1998rk,Maloy:1961qy}&  $T$&389& 8 &3.16\\
\cite{Gorbenko:1974sz,%
Gorbenko:1978re,Belyaev:1983xf,Maloy:1961qy,Gorbenko:1975pz,Kato:1979br,Bratashevsky:1980dk,%
Bratashevsky:1986xz}&  $P$&607& 3 &3.10\\
\cite{Bussey:1979wr,Ahrens:2005zq} &  $G$&75& 5 &1.20\\
\cite{Bussey:1979wr} &  $H$&71& 5 &1.21\\
\cite{Ahrens:2002gu,Ahrens:2004pf} &  $E$&140& 7 &1.30\\
\cite{Bratashevsky:1980dk,Avakyan:1991pj}&  $O_x$&7& 10 &0.95\\
\cite{Bratashevsky:1980dk,Avakyan:1991pj}&  $O_z$&7& 10 &0.42\\\hline\\[-2.3ex]
\hline\\[-2ex]
$\gamma p \rightarrow \pi^+ n$ & Observ. & $N_{\rm data}$&$w_i$ & $\chi^2_i/N_{\rm data}$\\[1ex]\hline\\[-2ex]
\cite{Ecklund:1967zz,Betourne:1968bd,Bouquet:1971cv,Fujii:1971qe,%
Ekstrand:1972rt,Fujii:1976jg,Arai:1977kb,Durwen:1980mq,Althoff:1983te,%
Heise:1988ag,Buechler:1994jg,Dannhausen:2001yz}&  $d\sigma/d\Omega$ & 1583 & 2 &1.75  \\
\cite{Ahrens:2004pf,Ahrens:2006gp} (GDH A2)&  $d\sigma/d\Omega$ & 408 & 14 &0.55  \\
\cite{Dugger:2009pn} (CLAS)&  $d\sigma/d\Omega$ & 484 & 4 &1.54  \\
\cite{Blanpied:2001ae,Taylor:1960dn,Smith:1963zza,Alspector:1972pw,Knies:1974zx,%
Ganenko:1976rf,Bussey:1979ju,Getman:1981qt,Hampe:1980jb,Beck:1999ge,%
Ajaka:2000rj,Bocquet:2001ny}&  $\Sigma$ &899 & 3 &3.07\\
\cite{Bussey:1979ju,Getman:1981qt,Althoff:1973kb,Arai:1973xs,Feller:1974qf,Althoff:1975kt,Genzel:1975tx,%
Althoff:1976gq,Althoff:1977ef,Fukushima:1977xh,Getman:1980pw,%
Fujii:1981kx,Dutz:1996uc}&  $T$&661 & 3 &2.66\\
\cite{Bussey:1979ju,Getman:1981qt,Egawa:1981uj}&  $P$&252 & 3 &2.11\\
\cite{Ahrens:2005zq,Bussey:1980fb,Belyaev:1985sp} &  $G$&86 & 8 &4.97\\
\cite{Bussey:1980fb,Belyaev:1985sp,Belyaev:1986va} &  $H$&128& 3& 4.59\\
\cite{Ahrens:2004pf,Ahrens:2006gp} &  $E$&231& 14 & 1.58\\\hline\\[-2.3ex]
\hline
\end{tabular}
\ec
\caption{\label{chisquare-eta}Observables from $\eta$
photoproduction fitted in the coupled-channel analysis and $\chi^2$
contributions for the solution BG2010-02.}
\bc\begin{tabular}{lcccc}$\gamma p \rightarrow \eta p$ & Observ. & $N_{\rm data}$&$w_i$ & $\chi^2_i/N_{\rm data}$\\[1ex]\hline\\[-2ex]
\cite{Krusche:nv} TAPS & $d\sigma/d\Omega$ &100 & 7 &2.45 \\
\cite{Crede:2009zzb} CBT& $d\sigma/d\Omega$ &680 & 40 &1.29 \\
\cite{Ajaka:1998zi} GRAAL& $\Sigma$ &51 & 10 &1.91 \\
\cite{Bartalini:2007fg} GRAAL&  $\Sigma$ &100& 15 &2.88\\
\cite{Bock:1998rk} PHOENICS&  $T$ &50& 70 &1.29\\
\hline\\[-2ex]
\hline\\[-2ex]
\end{tabular}\vspace{-2mm}\ec
\end{table}
In the list of $\pi N$ elastic scattering waves
(Table~\ref{piN_data_table}), we included the waves $F_{35}$ and
$F_{37}$ but not $F_{15}$ and $F_{17}$. The low-mass part of the
$F_{15}$ $\pi N$ elastic scattering amplitude is easily described by
$N(1680)F_{15}$ but the contribution of higher mass states to the
pion-induced reactions (\ref{pip_KL}), (\ref{pip_KS00}), and
(\ref{pip_neta}) is small and ambiguous. The Arndt solution
\cite{Arndt:2006bf} shows more structure than the H\"ohler solution
\cite{Hohler:1979yr} while in most other cases, the Arndt solution
is smoother. At the moment, we have no reason to prefer one over the
other one. The other partial waves are, however, not affected by
this uncertainty. Higher-mass $F$-wave nucleon resonances do
contribute to photoproduction and are included as multichannel
relativistic Breit-Wigner amplitudes in the partial wave analysis of
all reactions (1) and (2).

\begin{table}[pt]
\caption{\label{chisquare1}Hyperon photoproduction observables
fitted in the coupled-channel analysis and $\chi^2$ contributions
for the solution BG2010-02.}
\bc\begin{tabular}{lcccc}
\hline\hline\\[-2ex]
$\gamma p \rightarrow K^+ \Lambda$ & Observ. & $N_{\rm data}$&$w_i$ & $\chi^2_i/N_{\rm data}$\\[1ex]\hline\\[-2ex]
\cite{McCracken:2009ra} CLAS&  $d\sigma/d\Omega$&1320 &14 &0.81 \\
\cite{Zegers:2003ux} LEPS&  $\Sigma$ &45& 10 & 3.32\\
\cite{Lleres:2007tx} GRAAL&  $\Sigma$ &66& 8 & 1.68\\
\cite{McCracken:2009ra} CLAS&  $P$&1270& 8   &1.90\\
\cite{Lleres:2007tx} GRAAL&  $P$&66 &10 &0.70\\
\cite{Lleres:2008em} GRAAL&  $T$&66 & 15 &1.33\\
\cite{Bradford:2006ba} CLAS&  $C_x$&160 &15 &1.74\\
\cite{Bradford:2006ba} CLAS&  $C_z$&159 & 15 &1.45\\
\cite{Lleres:2008em} GRAAL&  $O_{x'}$&66 & 12 &1.48\\
\cite{Lleres:2008em} GRAAL&  $O_{z'}$&66 & 15 &1.40\\\hline\\[-2.3ex]
\hline\\[-2ex]
$\gamma p \rightarrow K^+ \Sigma$ & Observ. & $N_{\rm data}$&$w_i$ & $\chi^2_i/N_{\rm data}$\\[1ex]\hline\\[-2ex]
\cite{Bradford:2005pt} CLAS& $d\sigma/d\Omega$ & 1280& 3.5 &1.97 \\
\cite{Zegers:2003ux} LEPS&  $\Sigma$ &45& 10 &1.63\\
\cite{Lleres:2007tx} GRAAL&  $\Sigma$ &42& 10 &1.60\\
\cite{McNabb:2003nf} CLAS&  $P$&95 & 10 &1.71\\
\cite{Bradford:2006ba} CLAS&  $C_x$&94 &15 &2.89\\
\cite{Bradford:2006ba} CLAS&  $C_z$&94 &15 &1.86\\\hline\\[-2.3ex]
\hline\\[-2ex]
$\gamma p \rightarrow K^0 \Sigma^+$ & Obsv. & $N_{\rm data}$&$w_i$ & $\chi^2_i/N_{\rm data}$\\[1ex]\hline\\[-2ex]
\cite{McNabb:2003nf} CLAS&  $d\sigma/d\Omega$ & 48 &3   &3.25 \\
\cite{Lawall:2005np} SAPHIR&  $d\sigma/d\Omega$ & 156 &5 &1.34 \\
\cite{Castelijns:2007qt} CBT&  $d\sigma/d\Omega$ & 72 &10&0.77 \\
\cite{Castelijns:2007qt} CBT&  $P$&72 & 15 &0.95\\
\hline\hline
\end{tabular}\ec
\end{table}
The fit minimizes the total log likelihood defined by
\be -\ln {\cal L}_{\rm tot}= ( \frac 12\sum w_i\chi^2_i-\sum w_i\ln{\cal L}_i ) \ \frac{\sum
N_i}{\sum w_i N_i}
\ee
where the summation over binned data contributes to the $\chi^2$
while unbinned data contribute to the likelihoods ${\cal L}_i$. For
convenience of the reader, we quote differences in fit quality as
$\chi^2$ difference. $\Delta\chi^2=-2\Delta{\cal L}_{\rm tot}$.  For
new data, the weight is increased from $w_i=1$ until a visually
acceptable fit is reached. Without weights, low-statistics data e.g.
on polarization variables may be reproduced unsatisfactorily without
significant deterioration of the total ${\cal L}_{\rm tot}$. The
likelihood function is normalized to avoid an artificial increase in
statistics.

Tables~\ref{piN_data_table}-\ref{chisquare1} display the large
number of pion- and photo-induced reactions used in the coupled
channel analysis presented here. The data comprise nearly all
important reactions including multiparticle final states. Resonances
with sizable coupling constants to $\pi N$ and $\gamma N$ are thus
unlikely to escape the fits even though further single and double
polarization experiments are certainly needed to unambiguously
constrain the contributing amplitudes.

\subsection{Recent partial wave analyses}

At the time when new photo-production experiments came into sight
promising precise data on differential cross section and on
asymmetries due to photon-, target- and/or recoil-polarization,
several groups enforced their efforts to re-fit older data on pion-
and photo-induced reactions, and new groups were formed. The aim was
to be prepared for an analysis of the forthcoming data, to fit the
data, to extract properties of baryon resonances and to give
interpretations of the spectrum of resonances  and of the reaction
dynamics. This is not a review, hence we just quote from major
groups a few recent papers or a review where references to earlier
work can be found.

From the three ``classical" analysis groups
\cite{Hohler:1979yr,Cutkosky:1980rh,Arndt:2006bf}, only the GWU
group is still active in methodology \cite{Workman:2008iv} and data
analysis \cite{Arndt:2008zz}. Some of their recent
results can be found in \cite{Workman:1999wt,Arndt:2002xv,%
Azimov:2003bb,Arndt:2003if,Arndt:2005wk}. We mention here the work
of Bennhold and Haberzettl and collaborators
\cite{Mart:1999ed,Haberzettl:2001kr,Nakayama:2009he}, also working
at GWU. The physics at MAMI (among other data) has been closely
followed by the MAID group of Drechsel and Tiator
\cite{Kamalov:2001yi,Kamalov:2002wk,Chen:2002mn,Chiang:2002vq,%
Yang:2003jx,Chen:2007cy,Drechsel:2007if,Tiator:2010rp}. Smaller
groups at Zagreb
\cite{Batinic:1995kr,Ceci:2006ra,Ceci:2006jj,Ceci:2009zz},
Gent \cite{Janssen:2001pe,Corthals:2006nz,Corthals:2007kc,%
Vancraeyveld:2009qt} and KVI made significant contributions
to the field \cite{Korchin:1998ff,%
Usov:2005wy,Usov:2006wg,Shyam:2008fr,Scholten:2008zz}. A few further
papers are to be mentioned \cite{Williams:1992tp,David:1995pi,%
Kaiser:1996js,delaPuente:2008bw,He:2009zzi}.

Strong groups have been formed at Jlab and Bonn/J\"ulich. At Jlab,
the {\sc\small Excited Baryon Analysis Center} \cite{Lee:2009zzo}
was formed with the ambitious goal to extract and interpret
properties of nucleon resonances from the world data of meson
production reactions induced by pions, photons and electrons. In a
dynamical coupled-channel model, $\pi N$ elastic scattering
\cite{JuliaDiaz:2007kz}, photoproduction of pions
\cite{JuliaDiaz:2007fa}, the reaction $\pi^-p\to n\eta$
\cite{Durand:2008es}, and pion and photo-induced production of
hyperons \cite{JuliaDiaz:2006is} are analyzed and properties of
contributing resonances are determined. The Bonn/J\"ulich group has
over the years developed a unitary coupled channel exchange model
that obeys the strictures from analyticity
\cite{Krehl:1999km,Doring:2009yv,Doring:2009bi,Doring:2010ap}. For
kaon photo- and electroproduction off protons, a gauge invariant
chiral unitarity framework was developed in \cite{Borasoy:2007ku}.
Also considered are Regge models
\cite{Sibirtsev:2007wk,Sibirtsev:2009bj,Huang:2009pv,Sibirtsev:2010yj}
and heavy
mesons \cite{Guo:2008zg}. The Gie\ss en group \cite{Feuster:1996ww,%
Feuster:1997pq,Feuster:1998cj,Penner:2002ma,Penner:2002md,%
Shklyar:2004ba,Shklyar:2004dy,Shklyar:2005xg,Shklyar:2006xw,Shyam:2009za}
pioneered coupled-channel analyses of large data sets.

\subsection{The PWA method}

The approach used for the construction of amplitudes for pion and
photo-induced reactions is described in
\cite{Anisovich:2004zz,Anisovich:2006bc}. A shorter survey can be
found in \cite{Anisovich:2009zy}. Here, we give explicit formulas
for differential cross section and recoil polarization for
pion-induced production of a spin-1/2 baryon and a pseudoscalar
meson.

For $\pi N$ transition into the channels $\pi N$, $\eta N$,
$K\Lambda$ and $K\Sigma$, the amplitude can be written as
\be
&&A_{\pi N}=\omega^*\left [G(s,t)+H(s,t)i(\vec \sigma \vec n) \right
]\omega' \;,
\nonumber \\[0.3ex]
&&G(s,t)=\sum\limits_L \big [(L\!+\!1)F_L^+(s)+ L F_L^-(s)\big ]
P_L(z) \;, \nonumber \\[-1ex]
&&H(s,t)=\sum\limits_L \big [F_L^+(s)- F_L^-(s)\big ] P'_L(z) \;.
\label{piN_others}
\ee
with $z=\cos \Theta$, $\Theta$ the scattering angle of the outgoing
meson in the center-of-mass system (cms), and the decay plane normal
\be
\vec n_j=\varepsilon_{\mu\nu j} \frac{q_\mu k_\nu}{|\vec k||\vec
q|}\,.
\ee
The amplitudes $F_L^\pm(s)$ are related to the scattering amplitude
$T_L^\pm(s)$ known from $S$-matrix theory by
\be
T_L^\pm(s) = \frac{2q}{\sqrt s}\,F_L^\pm(s)
\ee
Here, $\vec q$ is the initial cms momentum, $q$ its modulus, $\vec
k$ is the final cms momentum.  $\omega$ and $\omega'$ are the
spinors of the nucleons in the initial and the final state, and
$\varepsilon_{\mu\nu j}$ is the antisymmetric tensor. The functions
$F^\pm_L$ depend only on the invariant mass squared $s$, the '+'
functions describe the $1/2^-\!$, $3/2^+\!$, $5/2^-\!$, $\ldots$
states and '-' functions describe $1/2^+\!$, $3/2^-\!$, $5/2^+\!$,
$\ldots$ states. $P_L$ are Legendre polynomials in $z$ and $P_L'$
are their derivatives.

At fixed energy the unpolarized cross section is proportional to the
amplitude squared
\be
|A|^2=\frac 12 {\rm Tr}\left[A^*_{\pi N}A_{\pi N}\right
]=|G(s,t)|^2\!+\!|H(s,t)|^2(1\!-\!z^2)~~
\ee
and the recoil asymmetry can be calculated as:
\be
P=\frac{{\rm  Tr}\left[A^*_{\pi N}\sigma_2A_{\pi N}\right
]}{2|A|^2\cos \phi}=\sin\Theta\frac{2Im\left (H^*(s,t)G(s,t)\right
)}{|A|^2}\,.
\ee

Near threshold, only contributions from $S$ and $P$-waves are
expected. For the $S_{2I,2J}$ and $P_{2I,2J}$ amplitudes we have
\begin{subequations}\label{seven}
\bea
\label{sevena}\hspace{-5mm}\underline{S_{2I,1}};\hspace{-3mm}&  G\!=F^+_0; \quad\ \ H\!=0;       &\qquad |A|^2\!=|F^+_0|^2\\
\label{sevenb}\hspace{-5mm}\underline{P_{2I,1}};\hspace{-3mm}&\quad\ \ G\!=F^-_1z; \ \ H\!=-F^-_1; &\qquad |A|^2\!=|F^-_1|^2\\
\label{sevenc}\hspace{-5mm}\underline{P_{2I,3}};\hspace{-3mm}&\quad\  G\!=2F^+_1z;
\ \ H\!=F^+_1; \ &|A|^2\!=|F^+_1|^2(3z^2\!\!+\!1)
\eea
\end{subequations}
where the indices $(2I,2J)$ remind of the isospin $I$ and the spin
$J$ of the partial waves.

The recoil asymmetry vanishes unless different amplitudes interfere.
Thus,
\begin{subequations}\label{eight}
\bea
\label{eighta}\underline{S_{2I,1}\!+\!P_{2I,1}}:&\quad P\frac{|A|^2}{\sin\Theta}=
-2Im(F^+_0F^{-*}_1) \\
|A|^2=&|F^+_0|^2+|F^-_1|^2+2zRe(F^{+*}_0F^-_1)\nn
\label{eightb}\underline{S_{2I,1}\!+\!P_{2I,3}}:&\quad P\frac{|A|^2}{\sin\Theta}=
2Im(F^+_0F^{+*}_1) \\
|A|^2=&|F^+_0|^2+|F^+_1|^2(3z^2\!+\!1)+4zRe(F^{+*}_0F^+_1)\nn
\label{eightc}\underline{P_{2I,1}\!+\!P_{2I,3}}:&\quad P\frac{|A|^2}{\sin\Theta}=
6zIm(F^{+*}_1F^-_1)\\ |A|^2=&|F^+_1-F^-_1|^2+z^2\left
(3|F^+_1|^2-2Re(F^{+*}_1F^-_1)\right ).\nnn
\eea
\end{subequations}
where $|A|^2$ represents the angular distribution and
$P\,|A|^2/\sin\Theta$ an observable proportional to the recoil
polarization parameter $P$.

The interference of $1/2^-$ and $1/2^+$ waves leads to a dependence
of the differential cross section linear in $z$ while the recoil
asymmetry multiplied by the differential cross section and divided
by $\sin\Theta$ should be flat. The interference of $1/2^-$ and
$3/2^+$ also produces a flat distribution for $P|A_n|^2/\sin\Theta$
while the differential cross section has a $z^2$ term. The
interference between the $1/2^+$ and $3/2^+$ waves provides a
symmetric differential cross section and a $P|A_n|^2/\sin\Theta$
distribution proportional to~$z$.

The $F_{37}$ wave is one of the dominant waves in the $\pi^+p\to
K^+\Sigma^+$ reaction. Its amplitude can be cast into the form
\be
\underline{F_{2I,7}}\qquad G=2(5z^3\!-\!z)F^+_3 \qquad
H=\frac{F^+_3}{2}(15z^2\!-\!3)
\ee
and the amplitude squared:
\be\label{ten}
|A|^2\!= |F^+_3|^2\frac
14(175z^6\!\!-\!165z^4\!\!+\!45z^2\!\!+\!9)\,.
\ee
Its interference with one of the $S$- or $P$-waves, written in the
form $P\frac{|A_n|^2}{\sin\Theta}$, is calculated to
\begin{subequations}\label{eleven}
\bea
\label{elevena}\hspace{-3mm}\underline{S_{2I1}\!+\!F_{2I7}}:\quad\
P\frac{|A|^2}{\sin\Theta}=
3(5z^2\!-\!1)Im(F^+_0F^{+*}_3) \\
\label{elevenb}\hspace{-3mm}\underline{P_{2I,1}\!+\!F_{2I,7}:}\quad
P\frac{|A|^2}{\sin\Theta}= 5(7z^3\!-\!3z)Im(F^-_1F^{+*}_3)\\
\label{elevenc}\hspace{-3mm}\underline{P_{2I,3}\!+\!F_{2I,7}:}\quad
P\frac{|A|^2}{\sin\Theta}= 2(5z^3\!+\!3z)Im(F^{+}_1F^{+*}_3)
\eea
\end{subequations}

\begin{figure}[pb]
\bc
\includegraphics[width=0.48\textwidth,height=0.55\textheight]{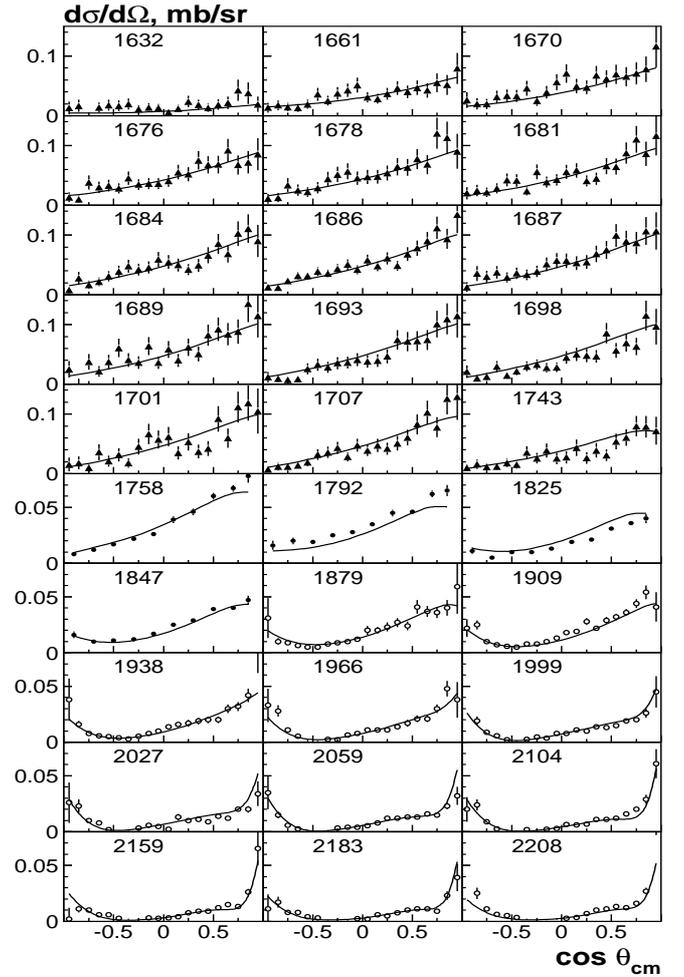}\\
\ec
\caption{\label{plam_dcs}The differential cross section for the
reaction $\pi^- p\to K^0\Lambda$. The triangles denote the data from
\cite{Knasel:1975rr} and open circles the data from
\cite{Baker:1978qm,Saxon:1979xu}, the curves represent our fit
BG2010-02.}
\end{figure}
\subsection{Features of the data in the threshold region}
\boldmath\subsubsection{\label{KL}The reaction $\pi^- p\to
K^0\Lambda$}\unboldmath

In a first attempt we try to identify evolving features of the data
without or with minimal use of the partial wave analysis. The
differential cross section for the reaction $\pi^- p\to K^0\Lambda$
is shown in Fig.~\ref{plam_dcs}. We mention that resonances which
\begin{figure}[pt]
\bc
\epsfig{file=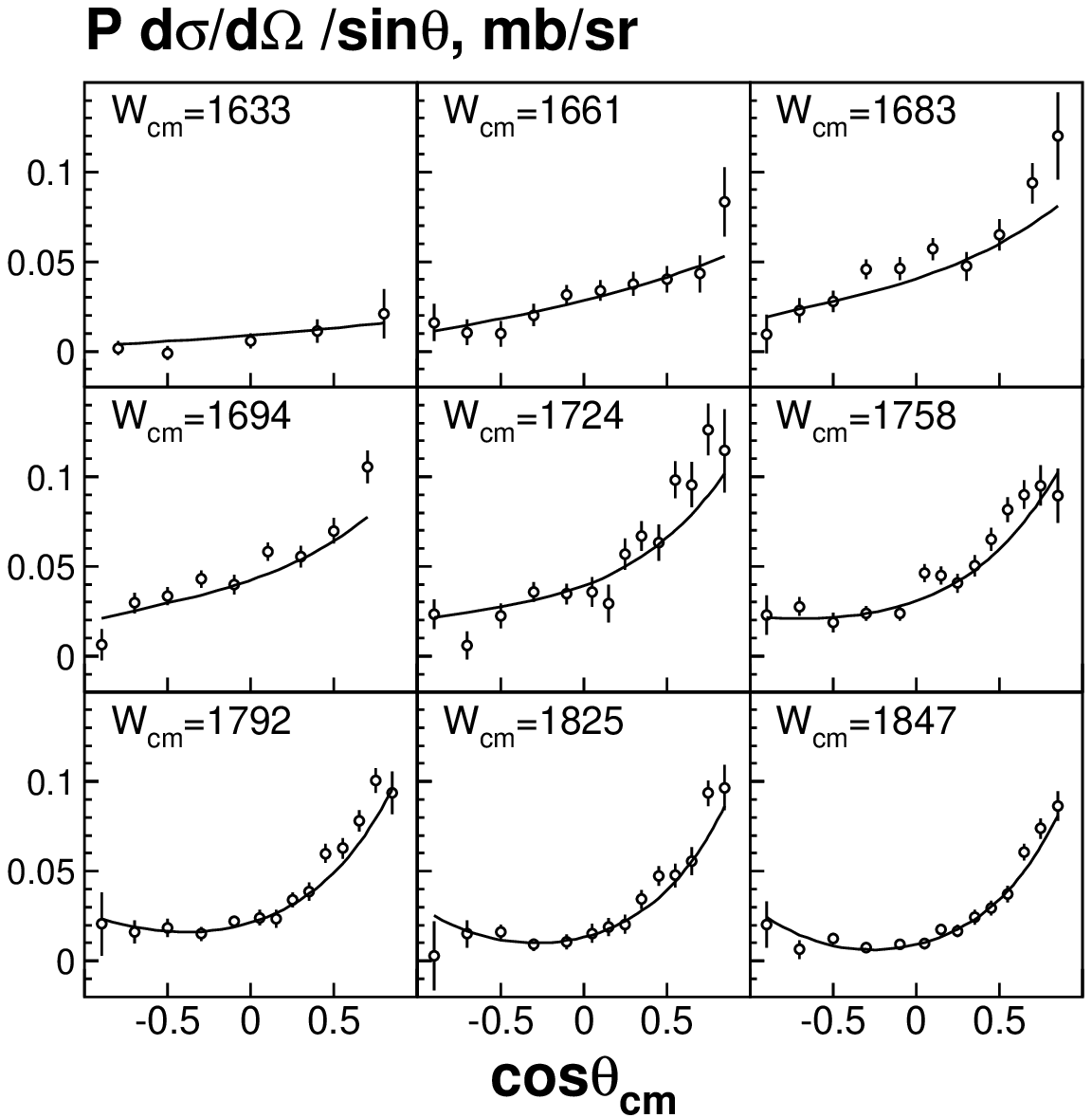,width=0.48\textwidth,height=0.30\textheight,clip=on}
\ec
\caption{\label{plam_pa}The $Pd\sigma/(d\Omega\sin\Theta)$
observable for the reaction $\pi^- p\,\to\,K^0\Lambda $. The recoil
asymmetry data are from \cite{Baker:1978qm,Saxon:1979xu}, the curves
represent our fit BG2010-02.}
\bc
\epsfig{file=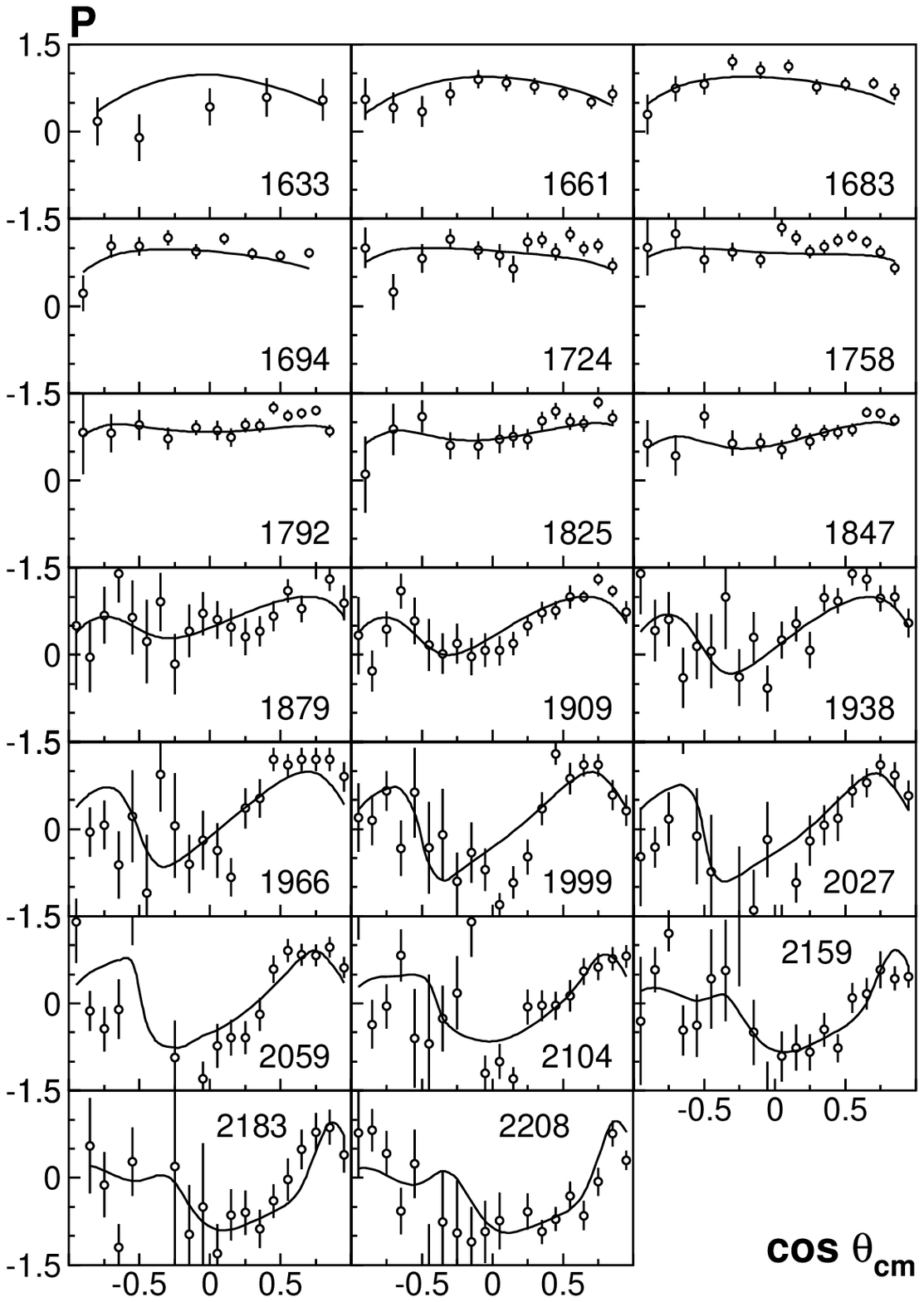,width=0.48\textwidth,height=0.40\textheight,clip=on}
\ec
\caption{\label{plam}The recoil asymmetry for $\pi^- p\,\to\,K^0
\Lambda$. The open circles denote the data from
\cite{Baker:1978qm,Saxon:1979xu}, the curves represent our fit
BG2010-02.}
\end{figure}
may be observed in this reaction belong to the nucleon excitation
series; $\Delta$ resonances do not couple to $\Lambda K$. The
observable $Pd\sigma/(d\Omega\sin\Theta)$ in the region near
threshold is shown in Fig.~\ref{plam_pa}. To construct this
observable the differential cross section was taken from the result
of the fit shown as curves in Fig.~\ref{plam_dcs}. We use the
differential cross sections from the fit to avoid additional
statistical fluctuations. Of course, a numerical summation of the
differential cross sections would result in very similar
distributions. Hence the distributions in Fig.~\ref{plam_pa} can be
considered as purely experimental ones. Several observations can be
made:
\begin{enumerate}
\item From the threshold region up to $\approx$1750\,MeV, the angular
distribution ($d\sigma/d\Omega$) rises nearly linearly in
$\cos\Theta$ due to the interference of $S_{11}$ and $P_{11}$ waves
(see eq.~\ref{eighta}).
\item There is a small quadratic part, indicating contributions
from the $P_{13}$ wave (eq.~\ref{sevenc}) or from the interference
of $S_{11}$ and/or $P_{11}$ with the $P_{13}$ wave
(eqs.~\ref{eightb}, \ref{eightc}).
\item Above 1850\,MeV the angular distribution has a strong
$1+ 3\cos^2\Theta$ contribution signaling a large $P_{13}$ wave
(eq.~\ref{sevenc}), and a linear part due to interference with the
$S_{11}$ wave (eq.~\ref{eightb}).
\item Near threshold, the $Pd\sigma/(d\Omega\sin\Theta)$ observable
is non-zero, almost flat, and shows a small linear rise. The flat
part indicates interference between $S$ and $P$-waves
(eqs.~\ref{eighta}, \ref{eightb}), the linear rise interference
between the $P_{11}$ and $P_{13}$ waves (eq.~\ref{eightc}). The
linear part becomes more significant in the 1650-1725\,MeV region.
\item In the region from 1800 to 2000\,MeV, the
$Pd\sigma/(d\Omega\sin\Theta)$ polarization has a more complicated
angular dependence which indicates the presence of higher waves.
\item The form of the angular distribution and of
$Pd\sigma/(d\Omega\sin\Theta)$ change rapidly with energy suggesting
that resonances play an important role in the dynamics.
\end{enumerate}

\begin{figure}[pt]
\bc
\epsfig{file=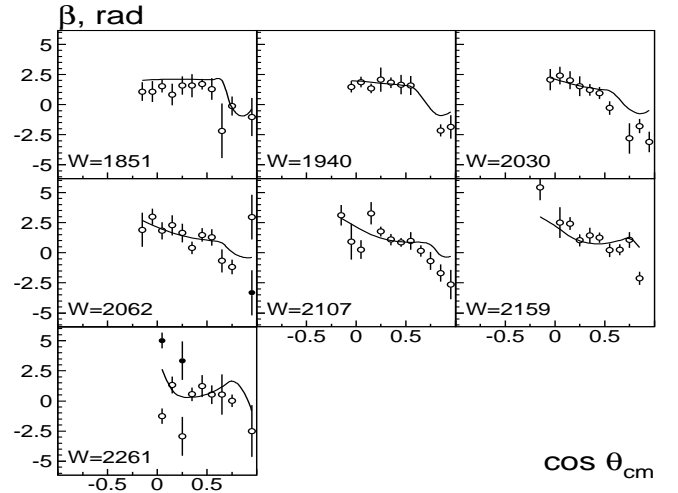,width=0.48\textwidth,height=0.28\textheight,clip=on}
\ec
\caption{\label{beta_klam}The rotation parameter for $\pi^-
p\,\to\,K^0 \Lambda$. The open circles denote the data from
\cite{Bell:1983dm}, the full circles show the mirror points (due to
the $2\pi$ ambiguity of $\beta$), and the curves represent our fit
BG2010-02.}
\end{figure}

These qualitative results are confirmed in the quantitative partial
wave analysis which will be discussed below. Here we just emphasize,
in Figs. \ref{plam_dcs}-\ref{plam}, that the data are well
reproduced by our fits. The description of the rotation parameter
$\beta$  measured in the experiment \cite{Bell:1983dm} is shown in
Fig.~\ref{beta_klam}. There are some systematic deviations in the
very forward region. However, there might be a problem with these
data points or/and with the given errors. The observable $A$ can be
written as
\be
A=(1-P^2)^{\frac
12}\cos\beta=\frac{|G|^2-|H|^2(1-z^2)}{|G|^2+|H|^2(1-z^2)}.
\ee
At $z=\pm 1$, $A$ must be equal to $+1$  while for the very forward
measured points at $\cos\Theta=0.95$ this observable are close to
-1. We did not find such an extreme oscillatory behavior of this
observable.

\boldmath\subsubsection{\label{KS}The reaction $\pi^+ p\to
K^+\Sigma^+$}\unboldmath In a next step, we discuss the reaction
$\pi^+ p\to K^+\Sigma^+$. In this reaction, all contributing
s-channel resonances must belong to the $\Delta$ series. Again, we
can draw some qualitative conclusions which do not depend on the
partial wave analysis.

\begin{figure}[pt]
\bc
\epsfig{file=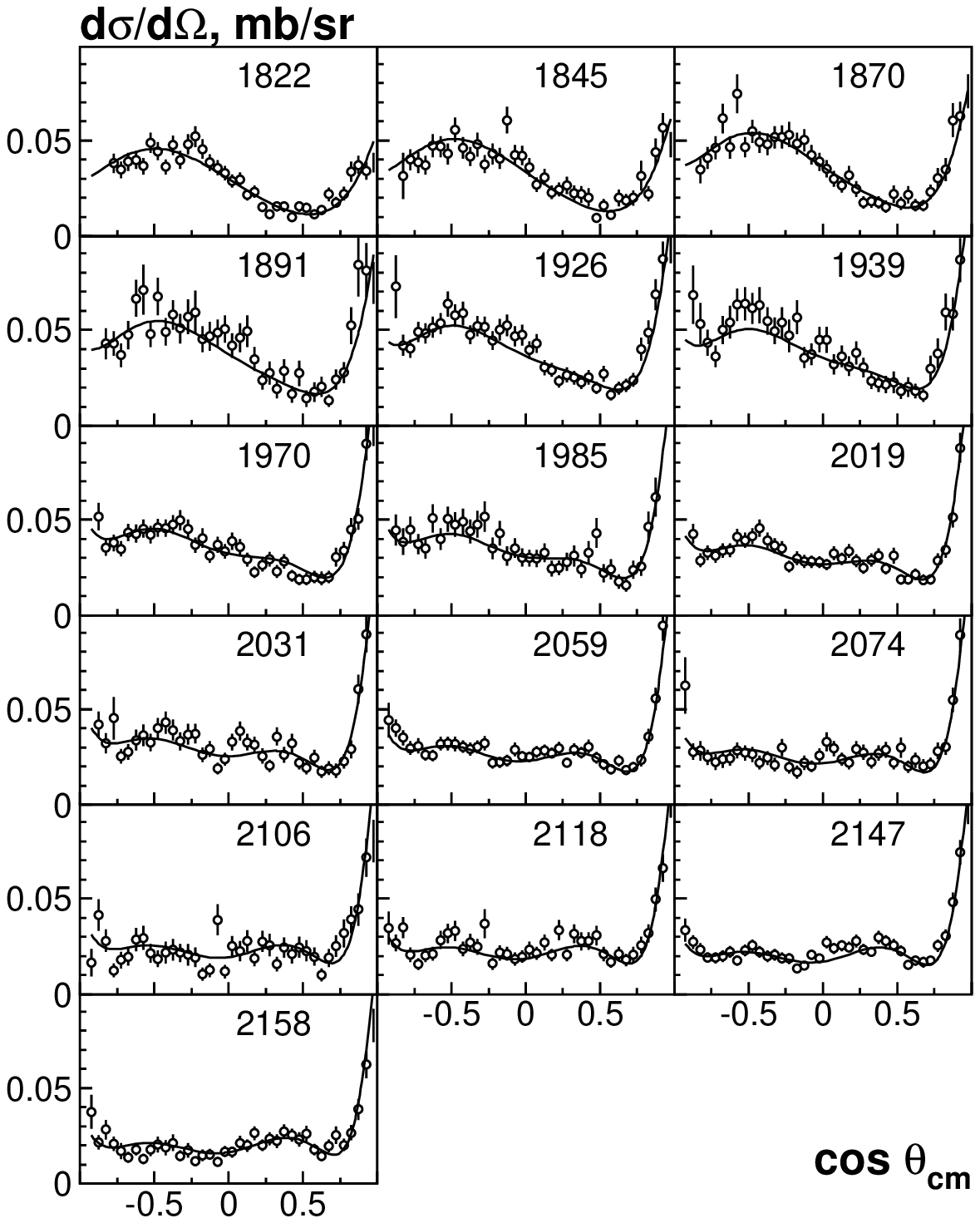,width=0.48\textwidth,height=0.40\textheight,clip=on}
\ec
\caption{\label{ksig_dcs}The differential cross section for the
reaction $\pi^+ p\,\to\,K^+\Sigma^+$. The data are taken from
\cite{Candlin:1982yv}. The curves represent our fit BG2010-02.}
\bc
\epsfig{file=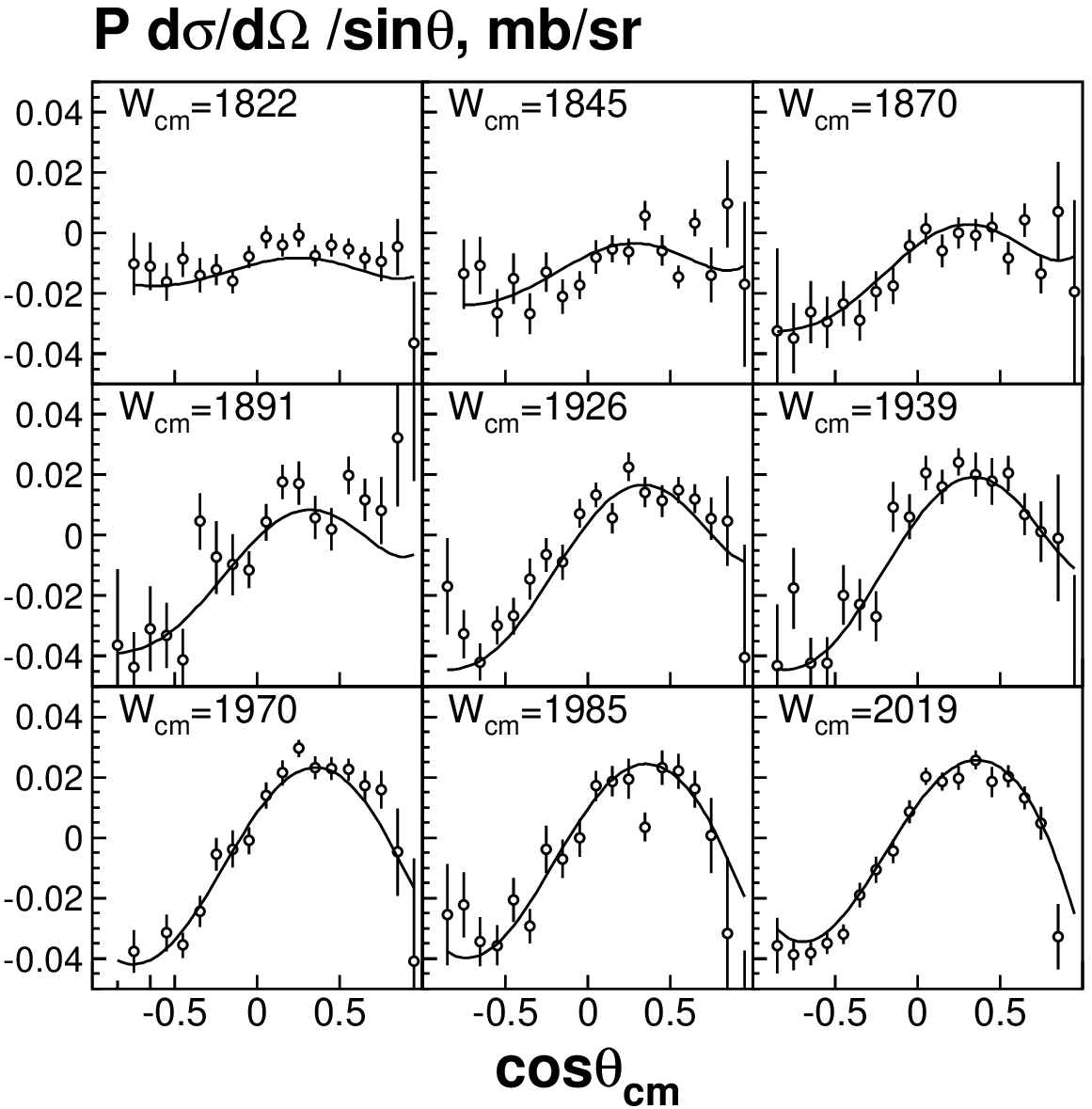,width=0.45\textwidth,height=0.30\textheight,clip=on}
\ec
\caption{\label{psig_pa} The $Pd\sigma/(d\Omega\sin\Theta)$
observable for the reaction $\pi^+ p\,\to\,\Sigma^+K^+$. The data
are from \cite{Candlin:1982yv}, the curves represent our fit
BG2010-02.}
\end{figure}

The differential cross section for $\pi^+ p\to K^+\Sigma^+$ is shown
in Fig.~\ref{ksig_dcs} and the observable
$Pd\sigma/(d\Omega\sin\Theta)$ in the first nine bins is shown in
Fig~\ref{psig_pa}. The best evidence for the existence of
$\Delta(1920)P_{33}$ is derived from this reaction.

\begin{figure}[pt]
\bc
\epsfig{file=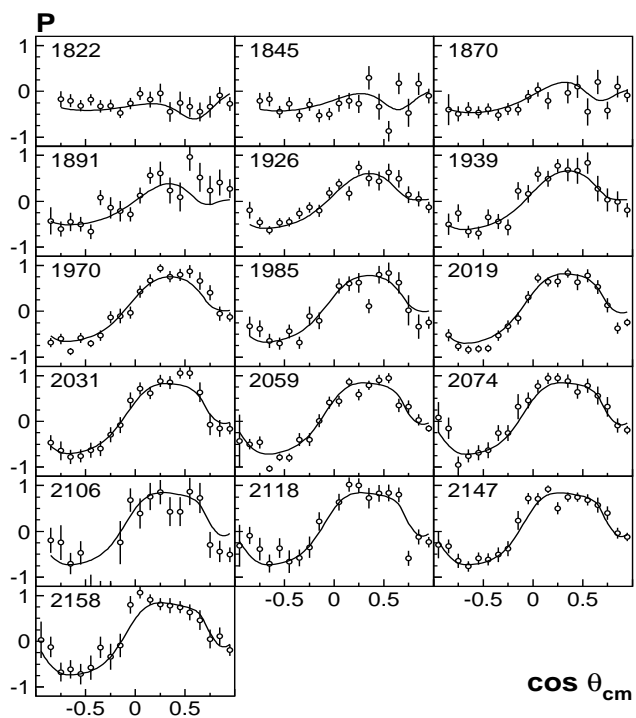,width=0.48\textwidth,height=0.40\textheight,clip=on}
\ec
\caption{\label{ksig_p} The recoil asymmetry $P$ for the reaction
$\pi^+ p\,\to\,K^+\Sigma^+$. The data are from
\cite{Candlin:1982yv}, the curves represent our fit BG2010-02.}
\end{figure}

\begin{enumerate}
\item In the threshold region, the angular distribution exhibits two
extreme values indicating the presence of a $z^3$ term, likely due
to interference of $P_{31}$ or $P_{33}$ with the $F_{37}$ wave
(eqs.~\ref{elevenb}, \ref{elevenc}).
\item The recoil polarization below 1900 MeV is comparatively small which
points at the dominance of one particular wave.
\item The angular distribution prefers $P_{33}$ (eq.~\ref{sevenc})
compared to $P_{31}$ (eq.~\ref{sevenb}) as dominant wave.
\item Above 1940 MeV the recoil polarization is large and even approaches the
extreme values $\pm1$. Hence there should be two dominant waves;
other partial wave distributions are likely small.
\item One of the two waves must be the $F_{37}$ wave.
In the mass range from 2000 to 2150\,MeV, the angular distribution
requires $J^P=7/2^+$ and interference with other waves. The partial
wave analysis identifies the second wave as $5/2^+$ wave.
\item The angular distributions and the recoil asymmetry undergo
rapid changes. Hence the partial waves must be resonating.
\end{enumerate}

\begin{figure}[pt]
\bc
\epsfig{file=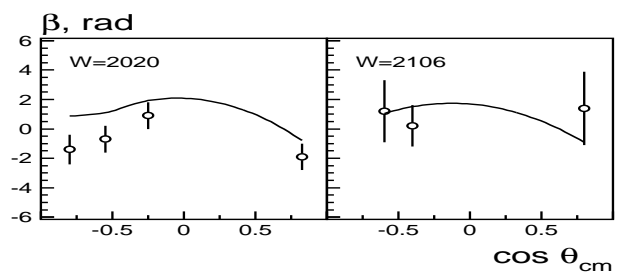,width=0.45\textwidth,height=0.15\textheight,clip=on}~~~~~
\ec
\caption{\label{beta_ksig}The rotation parameter for $\pi^+
p\,\to\,K^+ \Sigma^+$. The open circles denote the data from
\cite{Bell:1983dm} and the curves represent our fit BG2010-02.}
\end{figure}

The experimental cross sections, recoil polarization $P$ and
rotation parameter $\beta$ of the $\Sigma^+$ hyperon and the curves
resulting from our fits are shown in
Fig.~\ref{ksig_dcs}-\ref{beta_ksig}, for the mass region used in the
fits.

\boldmath\subsubsection{The reaction $\pi^- p\to
K^0\Sigma^0$}\unboldmath

\begin{figure}[pt]
\bc
\epsfig{file=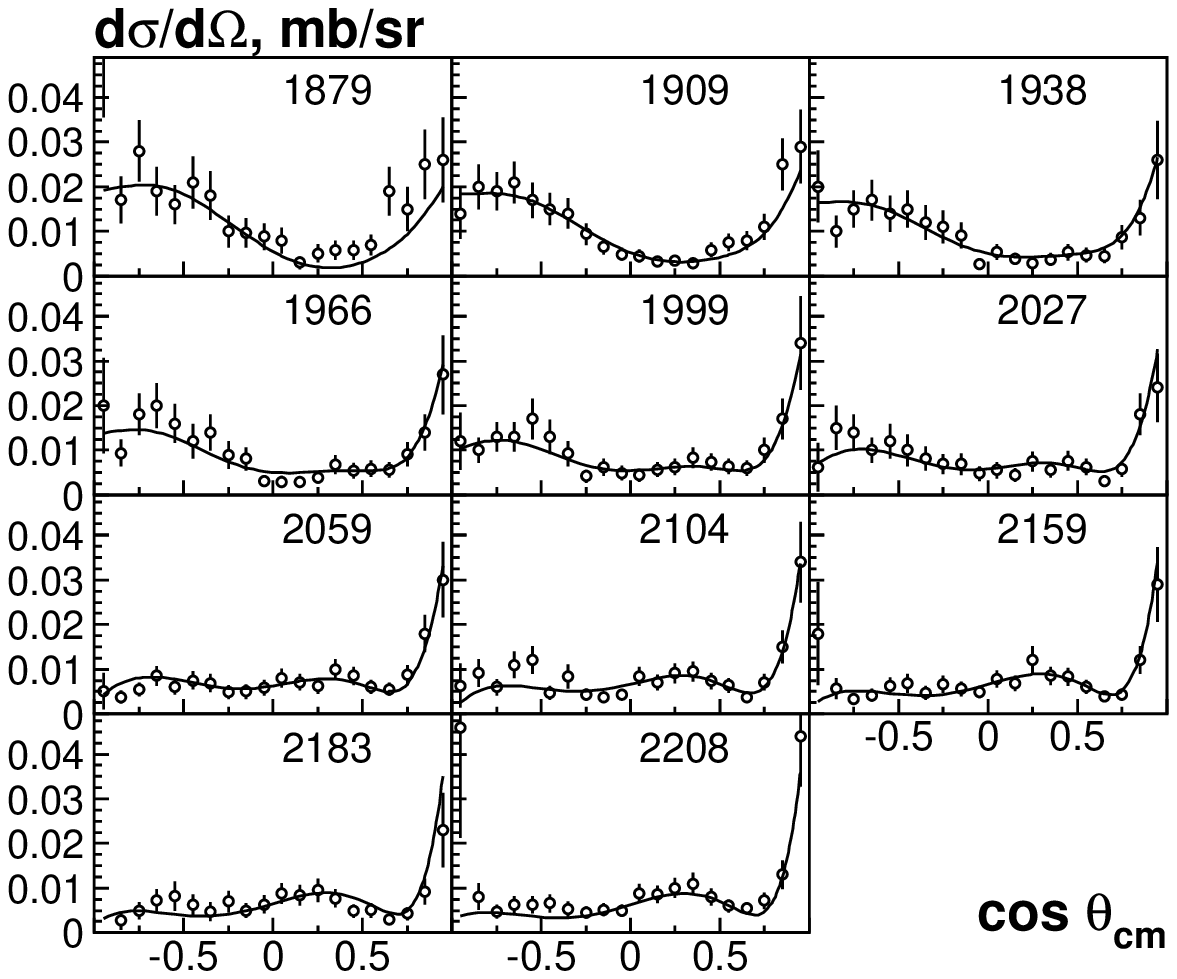,width=0.48\textwidth,clip=on}
\ec
\caption{\label{ksig_dcs_00}The differential cross section for the
reaction $\pi^- p\,\to\,K^0\Sigma^0$. The open circles denote the
data from \cite{Hart:1979jx}; the curves represent our fit
BG2010-02.}
\bc
\epsfig{file=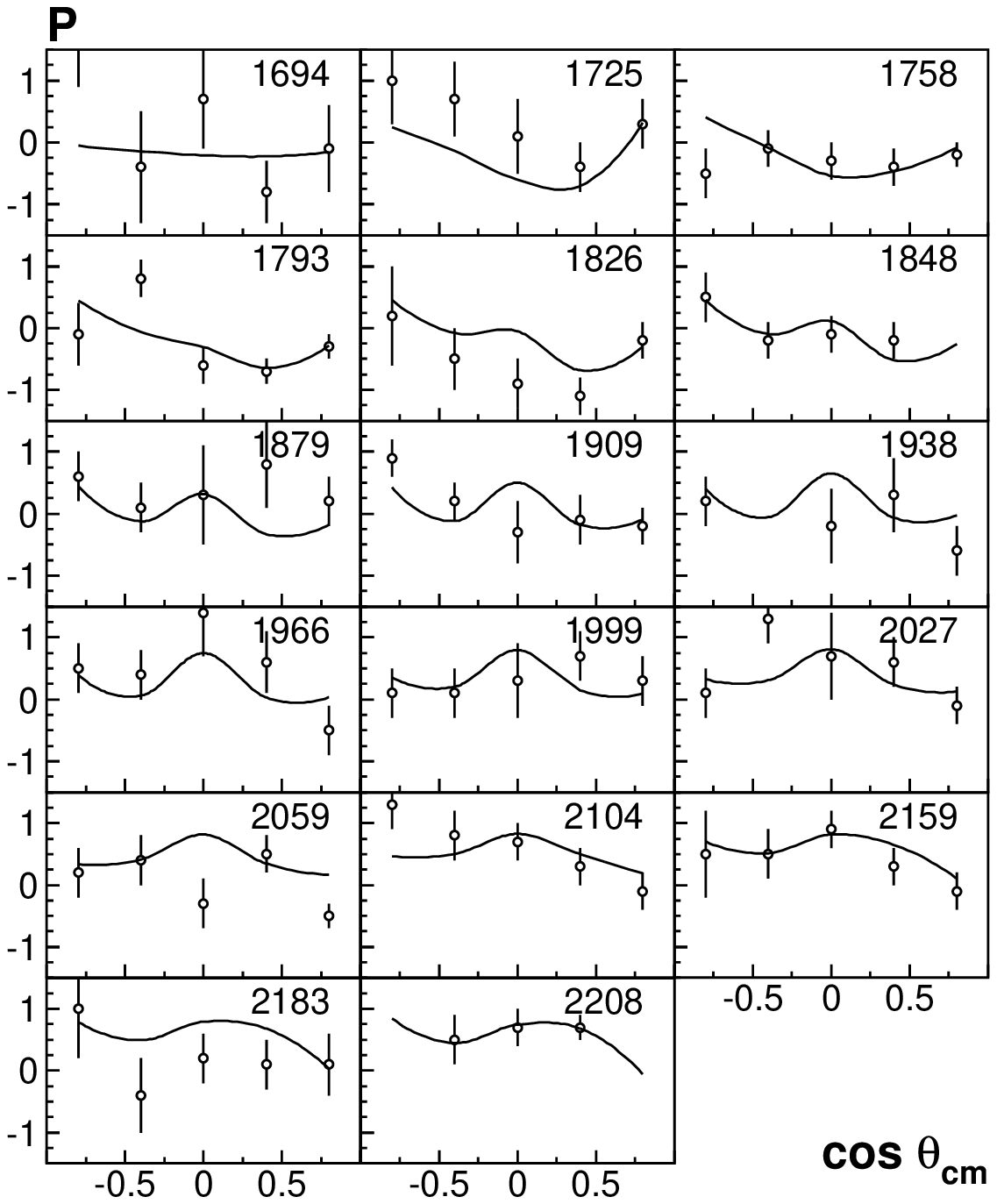,width=0.48\textwidth,height=0.38\textheight,clip=on}
\ec \caption{\label{ksig_p_00}The recoil asymmetry $P$
for the reaction $\pi^- p\,\to\,K^0\Sigma^0$. The data are taken
from \cite{Hart:1979jx}; the curves represent our fit BG2010-02. }
\end{figure}
This reaction receives contributions from the nucleon and $\Delta$
resonances with masses above 1700 MeV. The contributions of $\Delta$
resonances to $\pi^+ p\to K^+\Sigma^+$ and $\pi^- p\to K^0\Sigma^0$
are constrained by simple Clebsch-Gordan coefficients. Thus these
data supply valuable information about nucleon states decaying into
$K\Sigma$ channel and provide a consistency check of the partial
wave analysis.

\begin{figure}[pt]
\bc
\epsfig{file=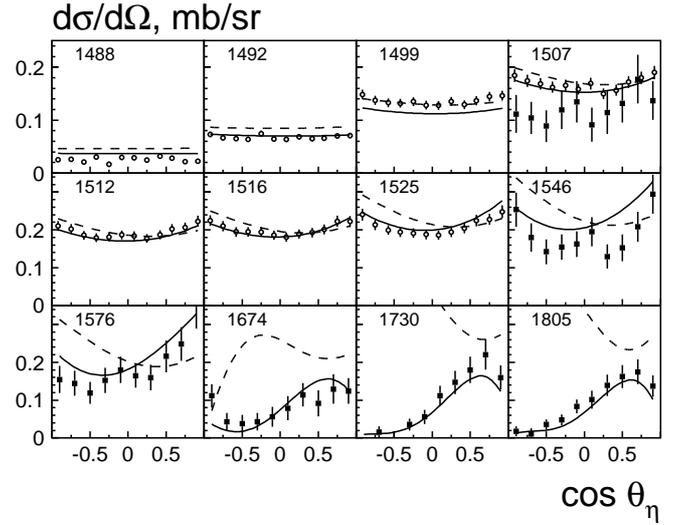,width=0.48\textwidth,clip=on}
\ec
\caption{\label{etan_dcs}The differential cross section for the
reaction $\pi^- p\,\to\,\eta n$. The open circles denote the data
from \cite{Prakhov:2005qb} and full squares the data from
\cite{Richards:1970cy}. The solid curves represent our fit
BG2010-02, the dashed curve an earlier fit \cite{Anisovich:2005tf}
which was not constrained by data on $\pi^- p\,\to\,\eta n$.}
\end{figure}
The differential cross section for the reaction $\pi^- p\to
K^0\Sigma^0$ is shown in Fig.~\ref{ksig_dcs_00} and the recoil
polarization $P$ in Fig.~\ref{ksig_p_00}. Data on the cross section
are available starting from 1879 MeV only. In the region
1879-1940\,MeV the differential cross section is very similar to
that for the $\pi^+p\to K^+\Sigma^+$ reaction, thus $\Delta$
resonances seem to play the dominant role. Above 1940\,MeV, the
structure of the differential cross section is rather complicated
due to presence of high partial waves, in particular of the
$F_{37}$-wave. Unfortunately there are no good data for the recoil
asymmetry for this reaction: the errors are too large to provide
stringent constraints for the fit.

\boldmath\subsubsection{The reaction $\pi^- p\to \eta n$}\unboldmath
We include a discussion of the reaction $\pi^- p\to \eta n$, mainly
because of negative parity nucleon resonances, even though these
will be discussed elsewhere. Differential cross section were
reported in \cite{Richards:1970cy,Prakhov:2005qb,%
Deinet:1969cd,Brown:1979ii,Debenham:1975bj,Crouch:1980vw}; however
there are large discrepancies between different data sets at least
above 1800\,MeV.

Here, we fit the precise data from the Crystal Ball collaboration
\cite{Prakhov:2005qb}; for the higher energy region, we use the data
from \cite{Richards:1970cy} since they are mostly compatible with
the Crystal Ball data and can probably be trusted up to 1800\,MeV.
Fig. \ref{etan_dcs} shows the angular distributions. They are rather
flat in the region below 1500\,MeV due to dominance of the $S_{11}$
wave. In the region 1600-1800\,MeV the angular distribution is very
asymmetric, which can come from interference of the $S_{11}$-wave
with either the $P_{11}$- or $P_{13}$-wave.

\subsection{Leading partial wave contributions}

These observations are quantitatively confirmed by the partial wave
analysis. In this paper, we focuss on $P$-wave resonances. Of
course, negative-parity resonances and higher partial waves are
required as well to fit the data. These will be discussed in detail
in a forthcoming publication. In Table \ref{res-list} we provide a
list of states which are included in this analysis but not discussed
here. Resonances above 2\,GeV are mostly introduced to improve the
description in the high-mass region even though we do not claim that
they really exist. We also summarize, in Table~\ref{background}, the
background contributions used to describe the pion- and
photo-induced data.

\begin{figure*}[pt]
\bc
\begin{tabular}{ccc}
\hspace{-2mm}\epsfig{file=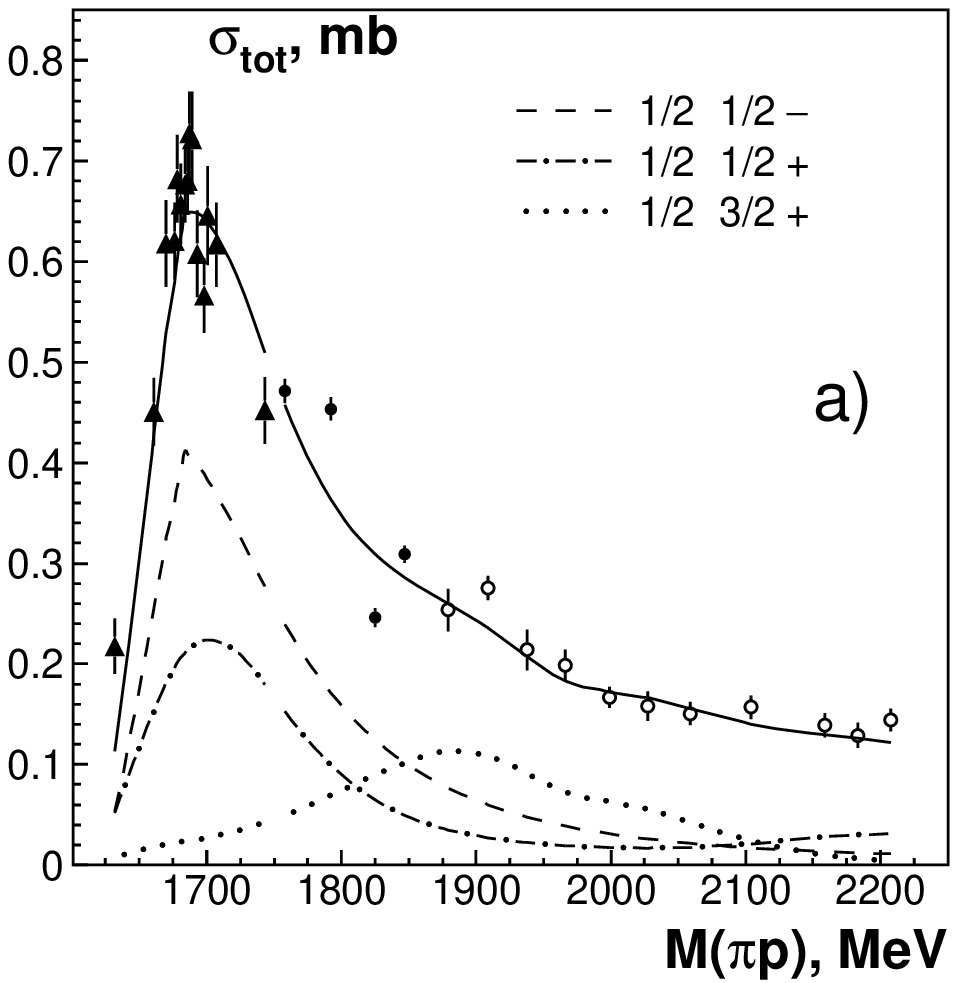,width=0.25\textwidth,clip=on}&
\hspace{-4mm}\epsfig{file=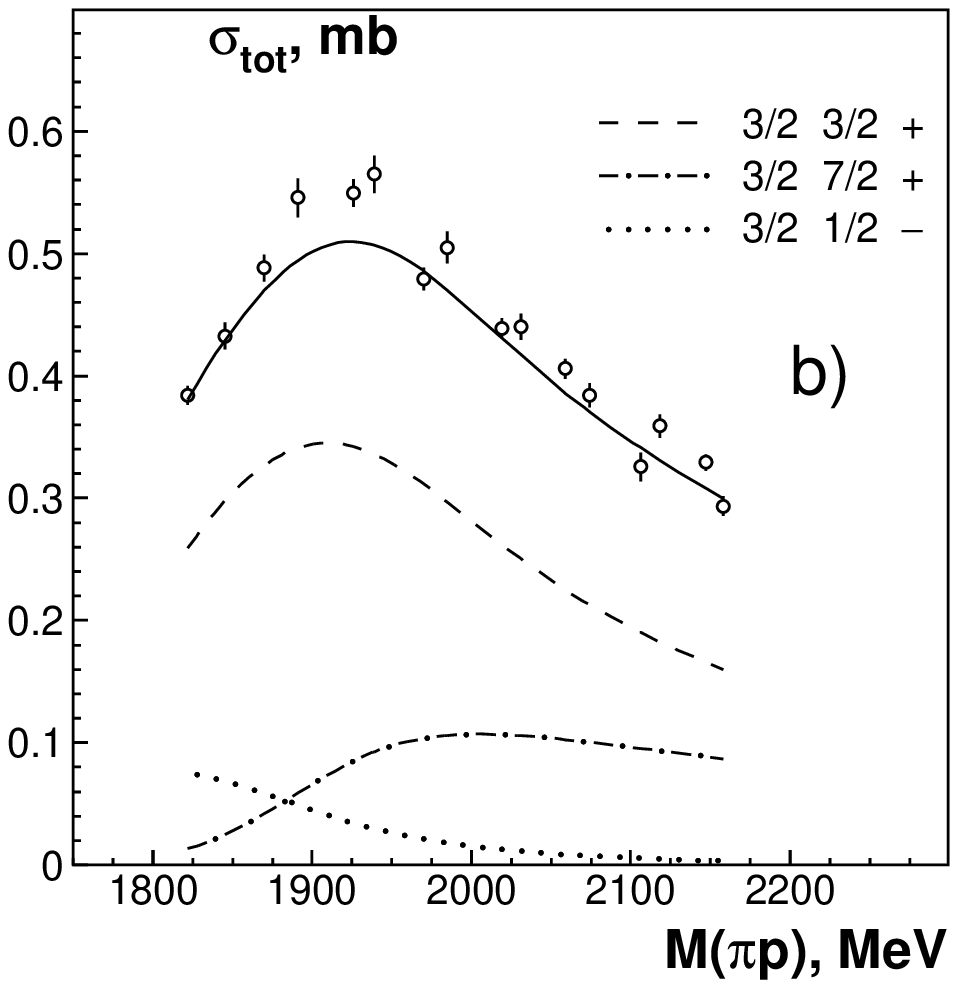,width=0.25\textwidth,clip=on}
\hspace{-1mm}\epsfig{file=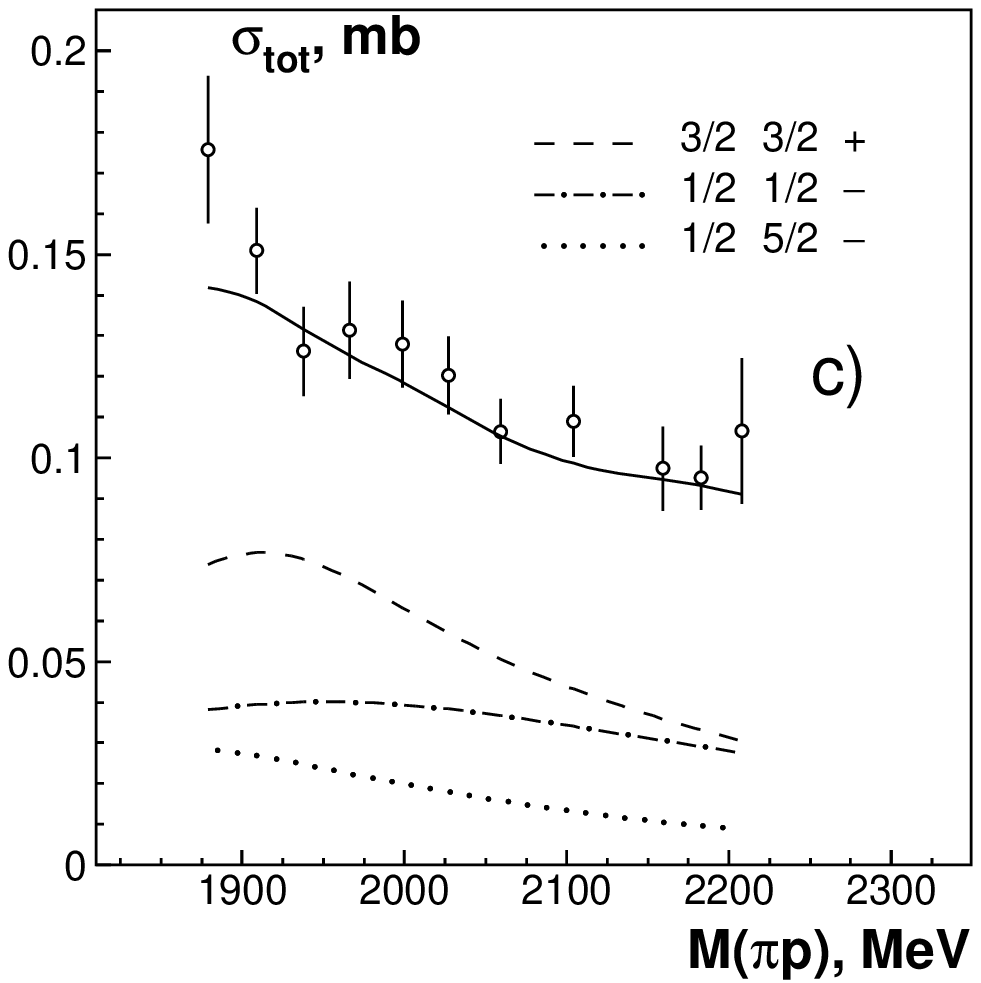,width=0.25\textwidth,clip=on}
\hspace{-2mm}\epsfig{file=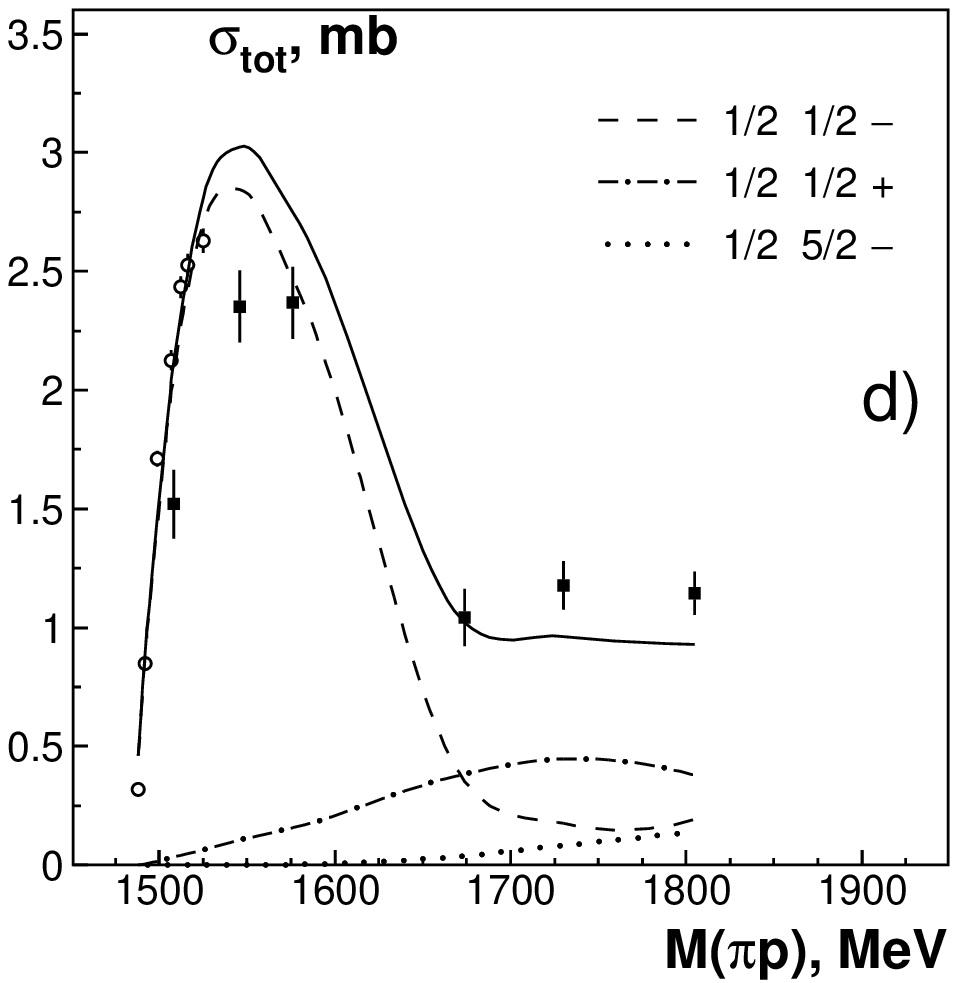,width=0.25\textwidth,clip=on}
\end{tabular}
\ec
\caption{\label{k_tot}a) The total cross section for the reaction
$\pi^- p\,\to\,K^0\Lambda$ and contributions from leading partial
waves. The full curve shows the integrated fit of the differential
cross section from the solution BG2010-02. The full triangles denote
the total cross section calculated from \cite{Knasel:1975rr} and the
open circles from \cite{Baker:1978qm},\cite{Saxon:1979xu}. b) The
total cross section for the reaction $\pi^+ p\,\to\,K^+\Sigma^+$ and
contributions from leading partial waves. The full curve shows the
integrated fit of the differential cross section  from the solution
BG2010-02. The data points show the total cross section calculated
from \cite{Candlin:1982yv}. c) The total cross section for the
reaction $\pi^- p\,\to\,K^0\Sigma^0$ and contributions from leading
partial waves. d) The total cross section for the reaction $\pi^-
p\,\to\,\eta n$ and contributions from leading partial waves. }
\end{figure*}

The contributions of the dominant waves to the reaction $\pi^- p\to
K^0\Lambda$ are shown in Fig.~\ref{k_tot}a. At threshold, the
dominant wave is $S_{11}$ which decreases fast with increasing
energy. Below 1800\,MeV, the $P_{11}$ wave is the second strongest
wave. It shows a clear resonance behavior peaking at 1720 MeV. The
$P_{13}$ wave increases steadily from threshold reaching its maximum
at 1900 MeV.

The two largest contributions to the reaction $\pi^+p\to
K^+\Sigma^+$ are assigned to the $P_{33}$ and $F_{37}$ partial waves
(see Fig.~\ref{k_tot}b). When one of the waves is assumed to be
non-resonant, angular and recoil distributions cannot be reproduced
satisfactorily. The energy dependence of the $F_{37}$ wave in this
reaction is compatible with the well known $F_{37}(1950)$ state. We
found only a rather small (below 1\%) contribution from
$\Delta(1905)F_{35}$. The $P_{33}$ wave has a maximum in the region
1900\,MeV which is expected if the $\Delta(1920)P_{33}$ state
provides a large contribution.

The contribution from $\Delta$-states to $\pi^-p\to K^0\Sigma^0$ can
be fixed, using the isotopic relations, from a fit of the $\pi^+p\to
K^+\Sigma^+$ reaction. These partial waves provide about 70\% to the
total cross section and contributions from nucleon partial waves are
needed to describe the remaining part of the data. In the
1850-1950\,MeV region one of the largest contributions is due to the
$D_{13}$ wave, which decreases fast with energy. This indicates the
presence of a resonance with a mass just below the energy region
covered by the data. The total cross section and contributions from
leading partial waves are shown in Fig.~\ref{k_tot}c.\\[-2ex]

\begin{table}[pb]
\caption{\label{res-list}List of the $S,D,F$, and $G$-wave
resonances used in the combined analysis of the data. Known states
are listed by the PDG names. Resonances with a $J^P$ subscript are
not listed in \cite{Amsler:2008zzb}.}
\bc
\begin{tabular}{cccc} \hline\hline
 $N(1535)S_{11}$ &  $N(1650)S_{11}$ & $N_{1/2^-}(1890)$ & $N(1520)D_{13}$ \\
 $N(1700)D_{13}$ & $N_{3/2^-}(1875)$ & $N_{3/2^-}(2130)$ &  $N(1675)D_{15}$ \\
 $N_{5/2^-}(2070)$ & $N(1680)F_{15}$ & $N(2000)F_{15}$ &  $N(1990)F_{17}$ \\
 $N(2190)G_{17}$ & $N(1620)S_{31}$ &  $\Delta(1900)S_{31}$ & $\Delta(1700)D_{33}$ \\
 $\Delta(1940)D_{33}$ & $\Delta_{3/2^-}(2200)$ & $\Delta(1905)F_{35}$ &  $\Delta(1950)F_{37}$  \\
\hline\hline
\end{tabular}\ec
\caption{\label{background}Background amplitudes used in the
combined analysis of the data.}
\bc
\begin{tabular}{lccc} \hline\hline
                     & pion-induced & photo-induced\\[0.5ex]
\hline\\[-2ex]
 $t$-channel exchange &
 &  $\pi^{\pm}$, $\rho^0
 /\omega$, $\rho^{\pm}$              \\[0.5ex]
 &  $K^{*0}$,  $K^{*\pm}$&  $K^{\pm}$, $K^{*0}$, $K^{*\pm}$\\[0.5ex]
 $u$-channel exchange &  & $N(938)$ ,  $\Lambda$, $\Sigma^0$,   $\Sigma^{\pm}$\\[0.5ex]
\multicolumn{3}{c}{Direct production  of final-state particles} \\[0.5ex]
 \hline\hline
\end{tabular}\ec
\end{table}

The total cross section for $\pi^-p\to \eta n$ calculated from the
fitted differential cross section is shown in Fig.~\ref{k_tot}d. At
low energies the cross section is dominated by the $N(1535)S_{11}$
contribution. The $N(1650)S_{11}$ resonance does not provide any
visible structure to the $S_{11}$ partial wave. In the
1700-1900\,MeV region, the $P$-wave amplitudes describe about 50\%
of the total cross section. However the lack of polarization
information does not allow us to determine them uniquely. There are
two sets of solutions (with BG2010-01 and BG2010-02 as best
solutions) which lead to rather different amplitudes for the
$P_{11}$ and $P_{13}$ partial waves (Fig.~\ref{pwa_etan_inel}) and
to different $P_{11}$ pole positions (see Table \ref{residues1}
below). In any case, these data impose important limits on the $\eta
N$ couplings of the $P$-wave states.

\begin{figure}[ph]
\bc
\epsfig{file=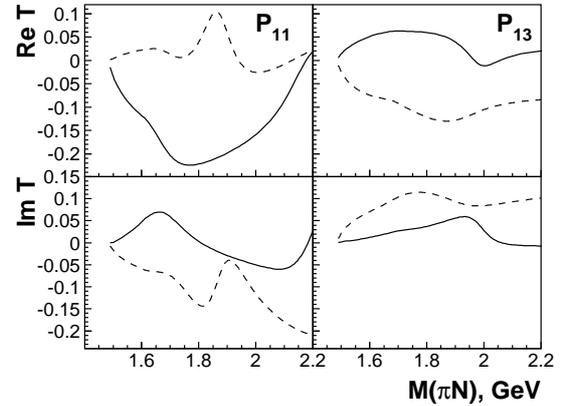,width=0.4\textwidth,clip=on}
\ec
\caption{\label{pwa_etan_inel}The $\pi^- p\to n\eta$ transition
amplitude in the $P_{11}$ and $P_{13}$ wave. The full curve
corresponds to the solution BG2010-02, the dashed curve to
BG2010-01. }
\end{figure}

 \subsection{The new data on the $\gamma p\to K\Lambda$ reaction}
The new CLAS data on the $\gamma p\to K\Lambda$ reaction
\cite{McCracken:2009ra} cover a wide mass range and exceed in
statistics all previously reported measurements of this reaction.
Also the angular range has been extended in comparison with earlier
CLAS data. We have therefore decided to use this data set only for
the  new fits.

\begin{figure}[pt]
\bc
\epsfig{file=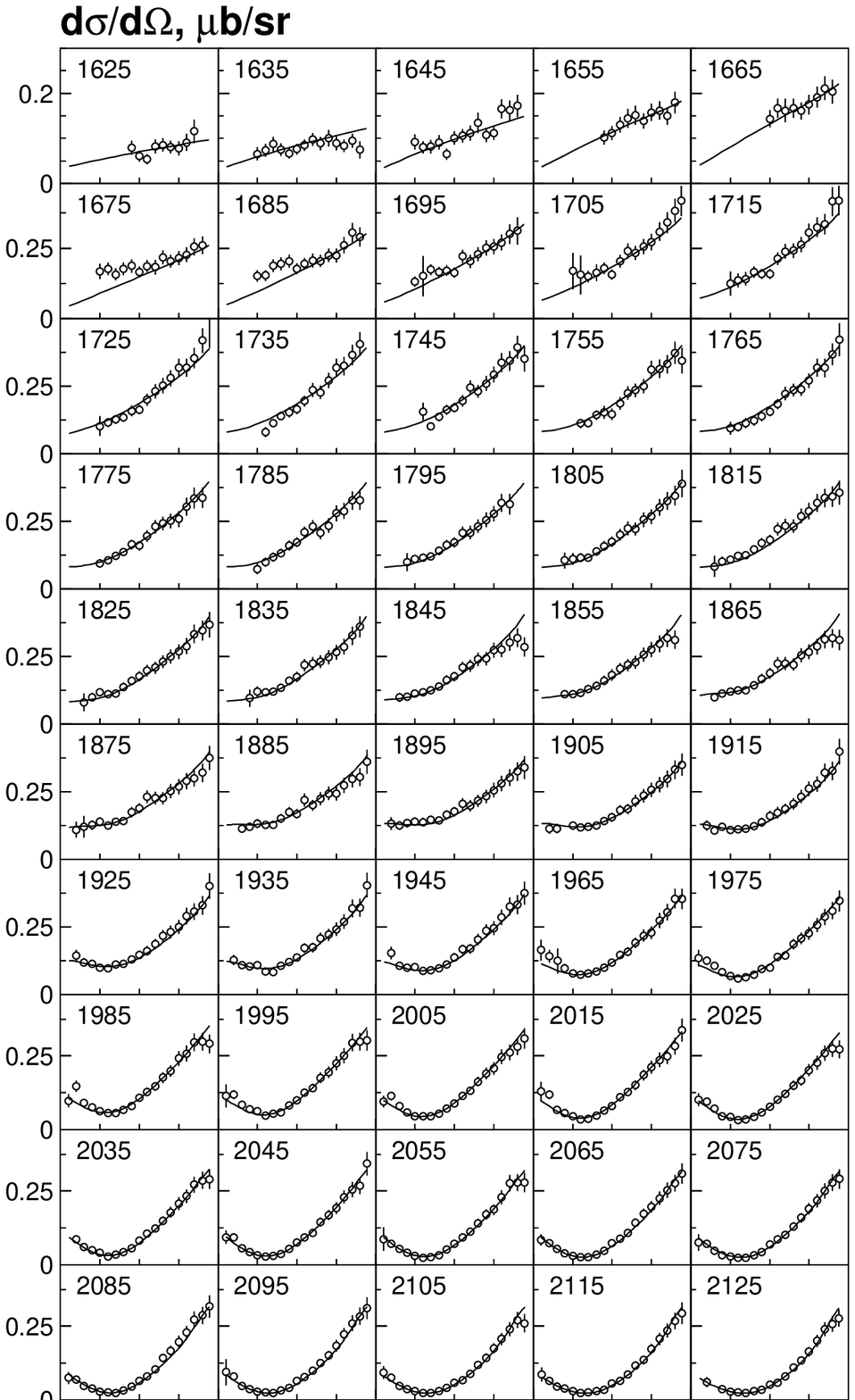,width=0.5\textwidth,clip=on}
\epsfig{file=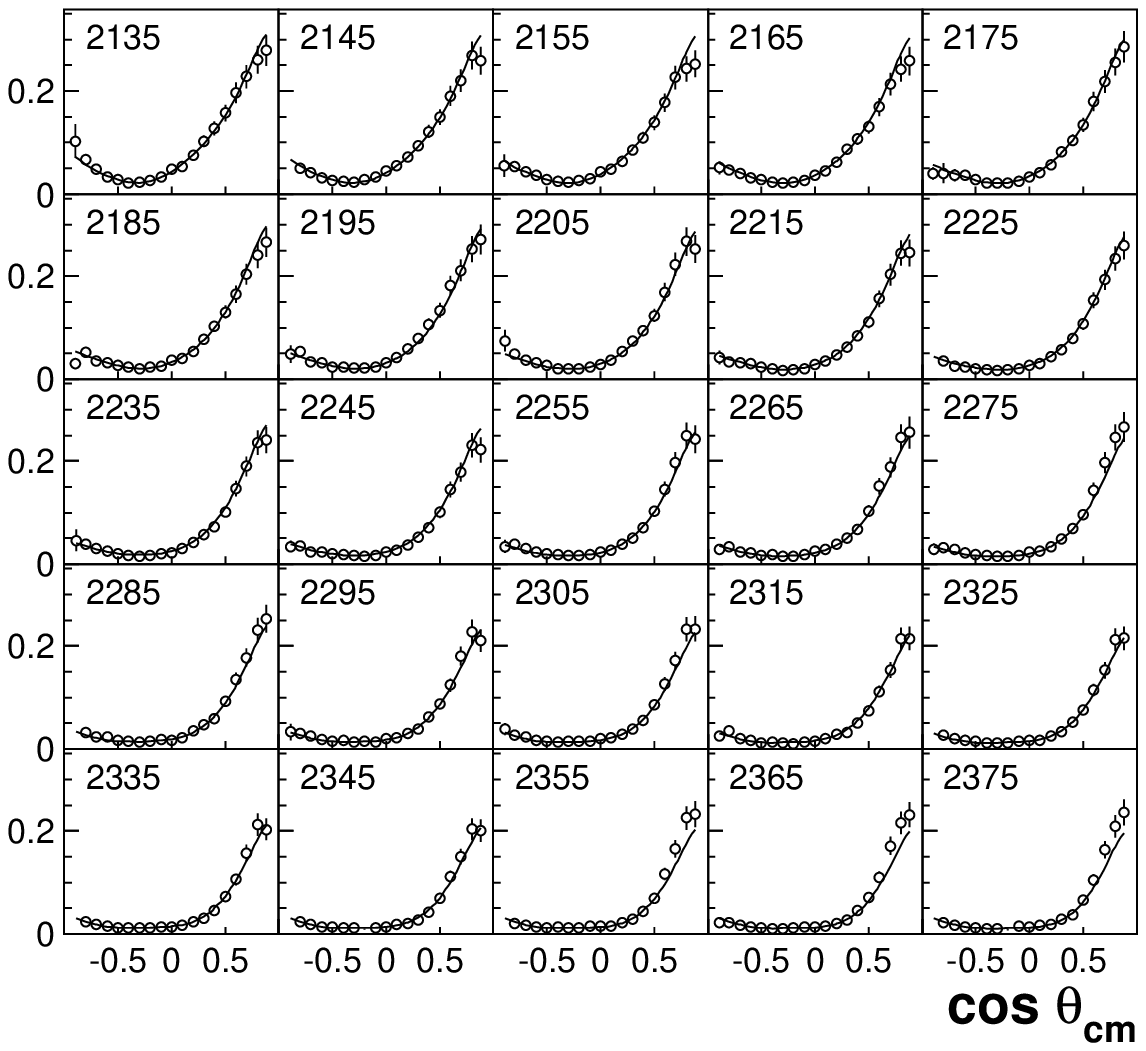,width=0.5\textwidth,clip=on}
 \ec
\caption{\label{gp_klam_dcs}Differential cross section for $\gamma
p\,\to\,K\Lambda$. The data are from \cite{McCracken:2009ra}. The
full curve corresponds to the solution BG2010-02. }
\end{figure}

\begin{figure}[pt]
\bc
\epsfig{file=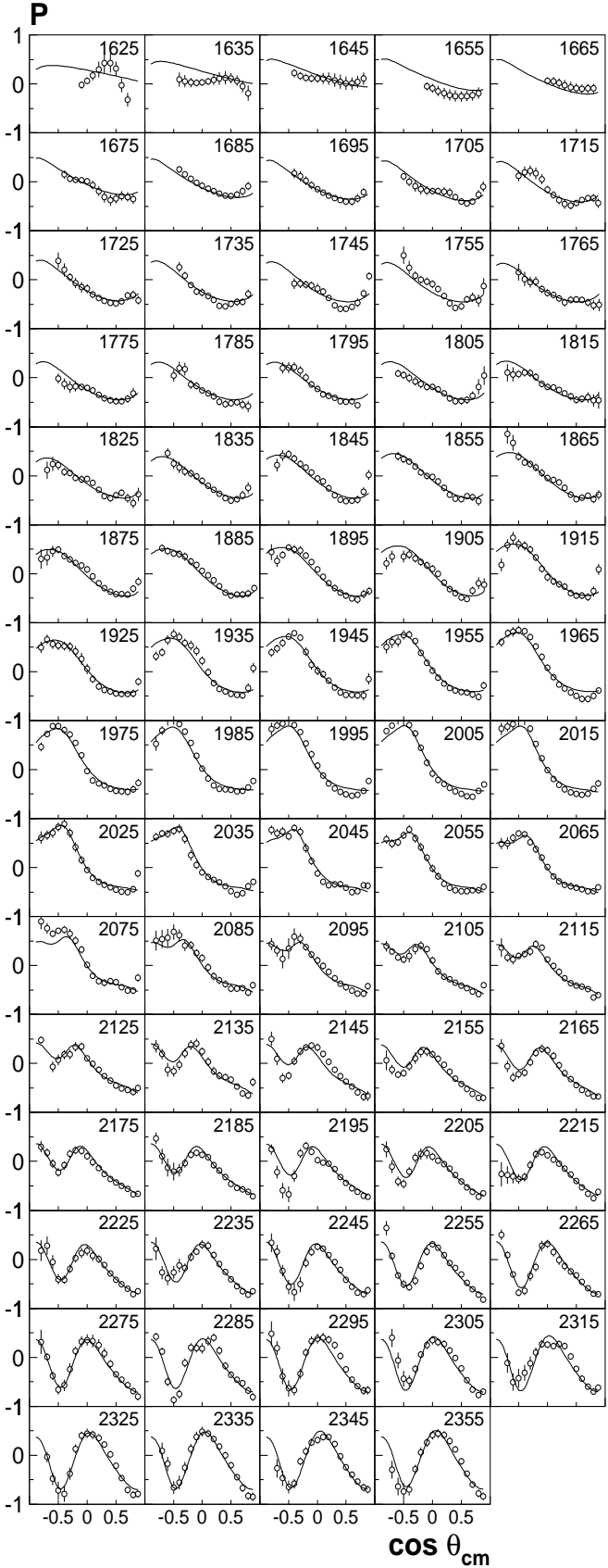,width=0.5\textwidth,height=0.952\textheight,clip=on}
\ec
\caption{\label{gp_klam_p} The $\gamma p\,\to\,K\Lambda$ recoil
asymmetry. The data are from \cite{McCracken:2009ra}. The full curve
corresponds to the solution BG2010-02. }
\end{figure}

The data cover the invariant mass region from the $K\Lambda$
threshold up to 2.8\,GeV. In the present analysis, we use these data
up to 2.4\,GeV; the higher energy region will be subject of future
investigations. At low energies the differential cross sections (see
Fig.~\ref{gp_klam_dcs}) show an almost linear dependence on
$\cos\Theta$ which can be explained by the interference of $S_{11}$
and $P$-wave amplitudes. An already satisfactory fit of a smaller
data set was obtained with the set of resonances reported in
\cite{Anisovich:2009zy}. Then we added in the fit, one by one,
different resonances parameterized as relativistic multi-channel
Breit-Wigner amplitudes. We found a significant improvement from a
$S_{11}$ state, which optimized at a mass of $1886\pm 10$\,MeV and
$80\pm15$\,MeV width. It is remarkably close to the result obtained
by H\"ohler \cite{Hohler:1979yr}. New double polarization data in
this region are urgently needed to provide a final proof for the
existence of this state.

We found a notable improvement from two other states: from
$N(1990)F_{17}$  and $N(2190)G_{17}$, both with masses compatible
with PDG values \cite{Amsler:2008zzb}. Both states also help to
achieve a good description of the new CLAS data. A full report about
properties of the $S_{11}$ partial wave and the investigation of the
$F_{17}$ and $G_{17}$ states will be given elsewhere.

The final description of the recoil asymmetry is shown in
Fig.~\ref{gp_klam_p}. There are significant problems in the
low-energy region. These are rather local and would require
introduction of new and narrow resonances. We refrain to do so
before more convincing experimental evidence exists. Systematic
deviations show up in the backward region in the 2120-2250\,MeV
invariant mass interval. However, we did not find a single resonance
which would improve notably the description. It is quite possible
that the data indicate the presence of a few new states with
different quantum numbers in this region but from the analysis of
the present data set, no firm conclusions can be drawn.

\begin{figure}[pt]
\bc
\epsfig{file=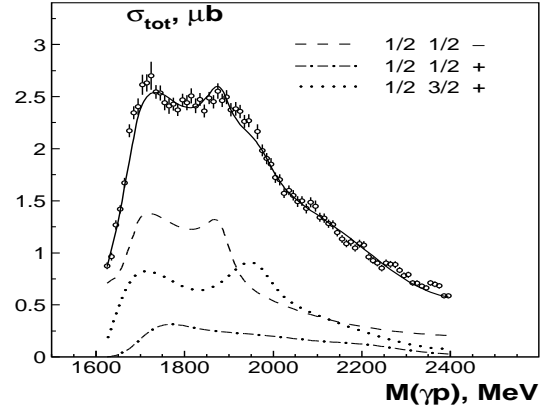,width=0.40\textwidth,height=0.3\textwidth,clip=on}
\ec
\caption{\label{gp_klam_tot} Contributions from the $S$ and $P$-wave
to $\gamma p\,\to\,K\Lambda$. The full curve shows the integrated
fit of the differential cross section from the solution BG2010-02.
The open circles denote the total cross section calculated from
\cite{McCracken:2009ra}.}
\end{figure}

The contribution of $S$ and $P$-waves to the total cross section is
shown in Fig.~\ref{gp_klam_tot}. As before the full curve is
calculated by integration of our fitted differential cross sections
over the full angular range; the data points are determined from a
summation of the measured differential cross sections over the
angular range where data exist, the uncovered angular range is taken
into account by adding the fit values for the differential cross
sections.

The $S_{11}$ and $P_{13}$ partial waves have a similar strength and
both show evidence for two peaks; both partial waves exhibit a
threshold effect and rise steeply at threshold. The threshold peaks
indicate the presence of resonances close to 1700\,MeV. The $S_{11}$
partial wave shows a second peak just below 1900\,MeV, the $P_{13}$
partial wave at about 1940\,MeV.  The $P_{11}$ partial wave is
smoother and shows a peak in the 1780\,MeV region and possibly a
broad shoulder with a possibility of a second structure at or above
1900\,MeV.

\begin{table}[pt]
\caption{\label{residues} Pole positions and residues of the
transition amplitudes (given in MeV as $|r|/\Theta$ with $Res
A=|r|\,e^{i\Theta}$). The phases are given in degrees. The helicity
couplings are given in $GeV^{-1/2}$. The errors are defined from the
spread of results found in the respective classes of solutions. The
PDG values are given in parentheses.}
\begin{footnotesize}
\renewcommand{\arraystretch}{1.00}
\bc
\begin{tabular}{lcc}
\hline\hline
\vspace{-0.3cm}\\
State                  &\hspace{-2mm}$\Delta(1232)P_{33}$ & $N(1440)P_{11}$  \\
\hline
\vspace{-0.3cm}\\
Re(pole)               &\hspace{-2mm}$1210\!\pm\!1$ ($1210\!\pm\!1$)    & $1377\!\pm\!6$ ($1365\!\pm\!15$)  \\
-2Im(pole)             &\hspace{-2mm}$100\!\pm\!1$ ($100\!\pm\!2$)       & $190\!\pm\!12$ ($190\!\pm\!30$)  \\
$A(\pi N \to \pi N)$   &\hspace{-2mm}$51.4\!\pm\!0.5$\,/-$47\!\pm\!1^o$   & $47\!\pm\!3$\,/-$83\!\pm\!10^o$  \\
$A(\pi N \to \eta N)$  &                                    & $5\!\pm\!3$\,/$160\!\pm\!60^o$     \\
$A^{1/2}(\gamma p)$    &\hspace{-2mm}$132.5\!\pm\!2$\,/$165\!\pm\!2^o$   & -$40\!\pm\!10$\,/-$37\!\pm\!10^o$   \\
$A^{3/2}(\gamma p)$    &\hspace{-2mm}$261\!\pm\!2$\,/$178\!\pm\!2^o$ & \\
\hline
\vspace{-0.3cm}\\
State &\hspace{-2mm}$N(1900)P_{13}$& $N_{3/2^+}(1975)$
\\\hline
\vspace{-0.3cm}\\
Re(pole)                  &\hspace{-2mm}$1890\!\pm\!50$ ($\sim$1900)               &$1970\!\pm\!25$ (-)      \\
-2Im(pole)                &\hspace{-2mm}$270^{+180}_{-100}$ (-)           &$250\!\pm\!60$ (-)                \\
$A(\pi N \to \pi N)$      &\hspace{-2mm} ~$3\!\pm\!2$\,/-$20\!\pm\!40^o$  & ~$2\!\pm\!1$\,/$80\!\pm\!40^o$   \\
$A(\pi N \to \eta N)$     &\hspace{-2mm} ~$6\!\pm\!3$\,/$80\!\pm\!40^o$  & ~$4\!\pm\!2.5$\,/$10\!\pm\!60^o$\\
$A(\pi N \to K\Lambda)$   &\hspace{-2mm} ~$14\!\pm\!6$\,/$145\!\pm\!25^o$ & ~$5\!\pm\!3$\,/-$10\!\pm\!40^o$   \\
$A(\pi N \to K\Sigma)$    &\hspace{-2mm} ~$6\!\pm\!4$\,/$80\!\pm\!50^o$   & ~$1.5\!\pm\!0.5$\,/-$30\!\pm\!50^o$  \\
$A^{1/2}(\gamma p)$       &\hspace{-2mm} $70\!\pm\!25$\,/~$130\!\pm\!40^o$ & -$8\!\pm\!8$\,/-$20\!\pm\!50^o$   \\
$A^{3/2}(\gamma p)$       &\hspace{-2mm} $110\!\pm\!60$\,/~$170\!\pm\!35^o$& $50\!\pm\!30$\,/~$25\!\pm\!45^o$ \\
\hline \vspace{-0.3cm}\\
State                   &\hspace{-2mm}$\Delta(1910)P_{31}$ & $\Delta(1920)P_{33}$\\
\hline
\vspace{-0.3cm}\\
Re(pole)                &\hspace{-2mm}$1900\!\pm\!40$ ($1900\!\pm\!50$)    &$1930\!\pm\!40$ ($1900\!\pm\!50$)  \\
-2Im(pole)              &\hspace{-2mm}$500^{+50}_{-100}$ ($300\!\pm\!100$) &$300\!\pm\!50$ ($300\!\pm\!100$)  \\
$A(\pi N \to \pi N)$    &\hspace{-2mm}$38\!\pm\!8$\,/$-125\!\pm\!15^o$     &$10\!\pm\!6$\,/ $20\!\pm\!60^o$     \\
$A(\pi N \to K\Sigma)$  &\hspace{-2mm}$19\!\pm\!4$\,/$-80\!\pm\!20^o$      &$10\!\pm\!5$\,/$150\!\pm\!30^o$     \\
$A^{1/2}(\gamma p)$     &\hspace{-2mm}$55\!\pm\!20$\,/~$80\!\pm\!70^o$     &$120\!\pm\!35$\,/-$120\!\pm\!35^o$ \\
$A^{3/2}(\gamma p)$     &                                     &$110\!\pm\!30$\,/-$170\!\pm\!30^o$  \\
\hline\hline
\end{tabular}
\ec
\end{footnotesize}
\renewcommand{\arraystretch}{1.0}
\end{table}

\begin{table}[pt]
\caption{\label{residues1} Pole positions and residues of the
transition amplitudes (given in MeV as $|r|/\Theta$ with $Res
A=|r|\,e^{i\Theta}$). The phases are given in degrees. The helicity
couplings are given in $GeV^{-1/2}$. The errors are defined from the
spread of results found in both classes of solutions. The PDG values
are given in parentheses. }
\begin{footnotesize}
\renewcommand{\arraystretch}{1.00}
\bc
\begin{tabular}{lcc}
\hline\hline
\vspace{-0.3cm}\\
Solution                   &\hspace{-2mm} 01     &\hspace{-2mm}02 \\
\hline\hline
\vspace{-0.3cm}\\
State                   &\hspace{-2mm} $N(1710)P_{11}$
&\hspace{-2mm}$N(1710)P_{11}$
\\
\hline
\vspace{-0.3cm}\\
Re(pole)                &\hspace{-2mm}$1690^{+25}_{-10}$ ($1720\!\pm\!50$)  &\hspace{-2mm}$1695\!\pm\!15$ ($1720\!\pm\!50$)   \\
-2Im(pole)              &\hspace{-2mm}~$210\!\pm\!25$ ($~230\!\pm\!150$) &\hspace{-2mm}~$220\!\pm\!30$ ($~230\!\pm\!150$)   \\
$A(\pi N \to \pi N)$    &\hspace{-2mm} $5\!\pm\!4$\,/-$80\!\pm\!40^o$      &\hspace{-2mm} $7\!\pm\!3$\,/$170\!\pm\!25^o$       \\
$A(\pi N \to \eta N)$   &\hspace{-2mm} $8\!\pm\!4$\,/$20\!\pm\!20^o$       &\hspace{-2mm} $6\!\pm\!4$\,/$40\!\pm\!15^o$        \\
$A(\pi N \to K\Lambda)$ &\hspace{-2mm} $17\!\pm\!4$\,/-$110\!\pm\!20^o$     &\hspace{-2mm} $16\!\pm\!5$\,/-$110\!\pm\!15^o$      \\
$A(\pi N \to K\Sigma)$  &\hspace{-2mm} $3\!\pm\!2$\,/$20\!\pm\!40^o$       &\hspace{-2mm} $5\!\pm\!3$\,/ $20\!\pm\!20^o$        \\
$A^{1/2}(\gamma p)$     &\hspace{-2mm} $80\!\pm\!25$\,/$100\!\pm\!30^o$  &\hspace{-2mm} $35\!\pm\!20$\,/-$35\!\pm\!20^o$   \\
\hline
\vspace{-0.3cm}\\
State                   &\hspace{-2mm}  $N_{1/2^+}(1875)$
&\hspace{-2mm}$N_{1/2^+}(1875)$
\\\hline
\vspace{-0.3cm}\\
Re(pole)                &\hspace{-2mm}$1860\!\pm\!20$ (~)                 &\hspace{-2mm}$1850^{+20}_{-50}$ (~)             \\
-2Im(pole)              &\hspace{-2mm}$110^{+30}_{-10}$ (~)               &\hspace{-2mm}$360\!\pm\!40$ (~)                  \\
$A(\pi N \to \pi N)$    &\hspace{-2mm} $2\!\pm\!2$\,/$30\!\pm\!30^o$      &\hspace{-2mm} $15\!\pm\!7$\,/$75\!\pm\!30^o$         \\
$A(\pi N \to \eta N)$   &\hspace{-2mm} $8\!\pm\!4$\,/~$-40\!\pm\!30^o$    &\hspace{-2mm} $11\!\pm\!4$\,/~$0\!\pm\!20^o$       \\
$A(\pi N \to K\Lambda)$ &\hspace{-2mm} $0.5\!\pm\!0.5$\,/$10\!\pm\!30^o$  &\hspace{-2mm} $14\!\pm\!5$\,/$90\!\pm\!30^o$       \\
$A(\pi N \to K\Sigma)$  &\hspace{-2mm} $3\!\pm\!2$\,/~$160\!\pm\!25^o$    &\hspace{-2mm} $10\!\pm\!3$\,/-$120\!\pm\!20^o$         \\
$A^{1/2}(\gamma p)$     &\hspace{-2mm} $12\!\pm\!6$\,/-$80\!\pm\!40^0$   &\hspace{-2mm} $55\!\pm\!15$\,/$70\!\pm\!30^0$      \\
\hline
\vspace{-0.3cm}\\
State              &\hspace{-2mm}$N(1720)P_{13}$ &\hspace{-2mm} $N(1720)P_{13}$   \\
\hline
\vspace{-0.3cm}\\
Re(pole)                    &\hspace{-2mm}$1695\!\pm\!30$ ($1675\!\pm\!15$)  &\hspace{-2mm}$1670\!\pm\!30$ ($1675\!\pm\!15$)    \\
-2Im(pole)                  &\hspace{-2mm}$400\!\pm\!60$ ($190\!\pm\!85$)    &\hspace{-2mm}$420\!\pm\!60$ ($190\!\pm\!85$)      \\
$A(\pi N \to \pi N)$        &\hspace{-2mm}~$22\!\pm\!5$\,/-$85\!\pm\!20^o$   &\hspace{-2mm}$28\!\pm\!6$\,/-$95\!\pm\!20^o$     \\
$A(\pi N \to \eta N)$       &\hspace{-2mm}~$9\!\pm\!4$\,/-$70\!\pm\!25^o$    &\hspace{-2mm}$8\!\pm\!3$\,/-$65\!\pm\!25^o$     \\
$A(\pi N \to K\Lambda)$     &\hspace{-2mm}~$15\!\pm\!8$\,/-$125\!\pm\!30^o$   &\hspace{-2mm}$12\!\pm\!4$\,/-$80\!\pm\!30^o$     \\
$A(\pi N \to K\Sigma)$      &\hspace{-2mm}~$10\!\pm\!6$\,/-$80\!\pm\!25^o$    &\hspace{-2mm}$10\!\pm\!4$\,/-$30\!\pm\!40^o$     \\
$A^{1/2}(\gamma p)$         &\hspace{-2mm}$110\!\pm\!40$\,/$20\!\pm\!40^o$    &\hspace{-2mm}$95\!\pm\!30$\,/$0\!\pm\!30^o$  \\
$A^{3/2}(\gamma p)$         &\hspace{-2mm}$130\!\pm\!50$\,/$70\!\pm\!40^o$    &\hspace{-2mm}$115\!\pm\!30$\,/$70\!\pm\!25^o$   \\
\hline
\vspace{-0.3cm}\\
State                   &\hspace{-2mm}$\Delta(1600)P_{33}$ &\hspace{-2mm}$\Delta(1600)P_{33}$\\
\hline
\vspace{-0.3cm}\\
Re(pole)                &\hspace{-2mm}$1480\!\pm\!30$ ($1600\!\pm\!100$)  &\hspace{-2mm}$1480\!\pm\!30$ ($1600\!\pm\!100$) \\
-2Im(pole)              &\hspace{-2mm}$240\!\pm\!40$ ($300\!\pm\!100$)    &\hspace{-2mm}$230\!\pm\!40$ ($300\!\pm\!100$)  \\
$A(\pi N \to \pi N)$    &\hspace{-2mm} $14\!\pm\!3$\,/-$170\!\pm\!15^o$   &\hspace{-2mm} $14\!\pm\!3$\,/-$170\!\pm\!15^o$   \\
$A(\pi N \to K\Sigma)$  &\hspace{-2mm} $2\!\pm\!1$\,/-$80\!\pm\!20^o$     &\hspace{-2mm} $3\!\pm\!1$\,/-$110\!\pm\!20^o$    \\
$A^{1/2}(\gamma p)$     &\hspace{-2mm}$26\!\pm\!6 $\,/-$80\!\pm\!30^o$    &\hspace{-2mm}$16\!\pm\!5 $\,/ $110\!\pm\!30^o$  \\
$A^{3/2}(\gamma p)$     &\hspace{-2mm}$15\!\pm\!6 $\,/ $0\!\pm\!20^o$     &\hspace{-2mm}$10\!\pm\!5 $\,/-$150\!\pm\!30^o$    \\
\hline\hline
\end{tabular}
\ec
\end{footnotesize}
\renewcommand{\arraystretch}{1.0}
\vspace{5mm}\end{table}

\section{Discussion of partial waves}
In this section, we discuss the individual partial waves, $P_{33}$,
$P_{31}$, $P_{13}$, and $P_{11}$, and their resonant contributions.
The pole positions of the transition amplitudes between channels
$i,j=\pi N,\;\eta N,\; K\Lambda,\; K \Sigma\ldots$
 \be
T^\pm_{L}(i\to j)=\sqrt{\rho_i} \sum\limits_a K^{\pm L}_{ia}(I-i\rho
K^{\pm L})^{-1}_{aj}\sqrt{\rho_j}
\label{res_amp}
\ee
are calculated as residues in the energy complex plane and given in
Tables~\ref{residues} and \ref{residues1}. Note that our helicity
couplings $A_{1/2}$ and $A_{3/2}$ are defined at the pole position
and are complex numbers. They become real only in the limit of an
isolated Breit-Wigner amplitude representation of a resonance.

In spite of the large data base, the partial wave analysis does not
converge to a unique solution. Instead, we found two classes of
solutions which have rather different imaginary parts for the third
$P_{11}$ pole: the first set of solutions has a pole at $1860\pm
20-i55^{+15}_{-5}$\,MeV (the best solution from this class is called
BG2010-01) and second set has a pole at $1850^{+20}_{-50}-i180\pm
20$ (best solution: BG2010-02). The second solution is well
compatible with the result reported in \cite{Sarantsev:2005tg}.
Within the two classes of solutions, there is a large number of
sub-solutions which depend on the model details: the number of poles
in the $P_{13}$ wave has an impact on all fit values. For most
resonances, the two classes of solutions give compatible results.
Masses, widths, residues and their respective errors for states
which are similar in BG2010-01 and BG2010-02 are listed in
Table~\ref{residues}, the states which have different properties are
listed in Table~\ref{residues1}. The spread of results in both sets
of solutions is used to estimate the errors.

\boldmath\subsection{$(I)J^P=(3/2)3/2^+$ }\unboldmath

This wave is, of course, dominated by the well-known $\Delta(1232)$
isobar. This leading resonance is well below the threshold for
strangeness production but since the elastic $\pi N$ scattering and
single pion photoproduction data are included in this analysis, the
fit returns pole position and Breit-Wigner parameters for this
resonance as well. Hence the $\Delta(1232)$ parameters are included
in Table \ref{residues}.

As discussed in the Introduction, the mass region above the
$\Delta(1232)$ isobar is controversial. Two further resonances,
$\Delta(1600)P_{33}$ and $\Delta(1920)P_{33}$, are seen in the
analyses \cite{Hohler:1979yr,Cutkosky:1980rh,Manley:1992yb} and in
inelastic reactions \cite{Manley:1992yb,Penner:2002md,Horn:2007pp}
while in~\cite{Arndt:2006bf} -- which relies on the most extensive
set of data on $\pi N$ elastic scattering -- $\Delta(1600)P_{33}$ is
observed as broad structure and $\Delta(1920)P_{33}$ is missing.

\begin{figure}[pt]
\bc
\begin{tabular}{ccc}
\hspace{-2mm}\epsfig{file=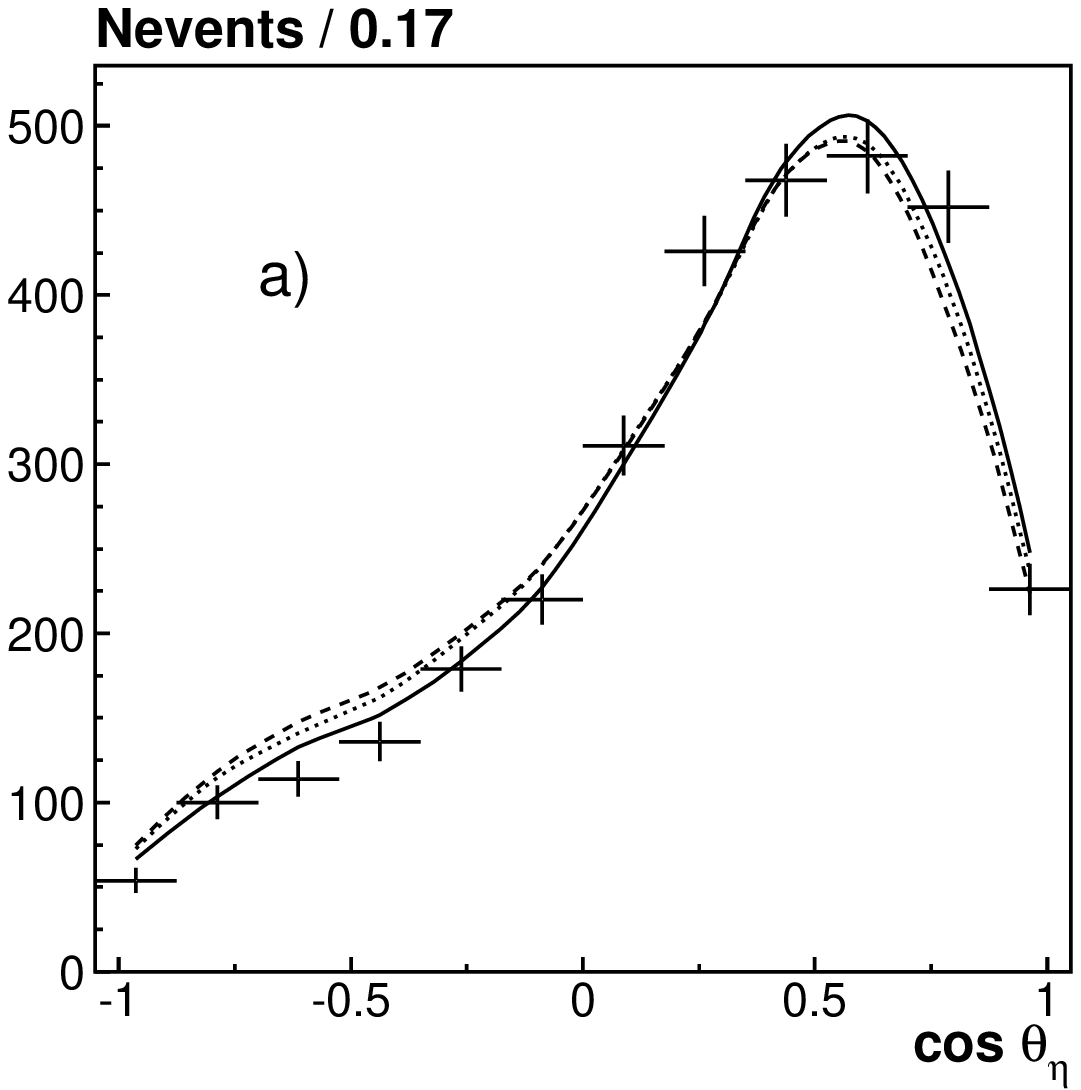,width=0.24\textwidth,clip=on}&
\hspace{-4mm}\epsfig{file=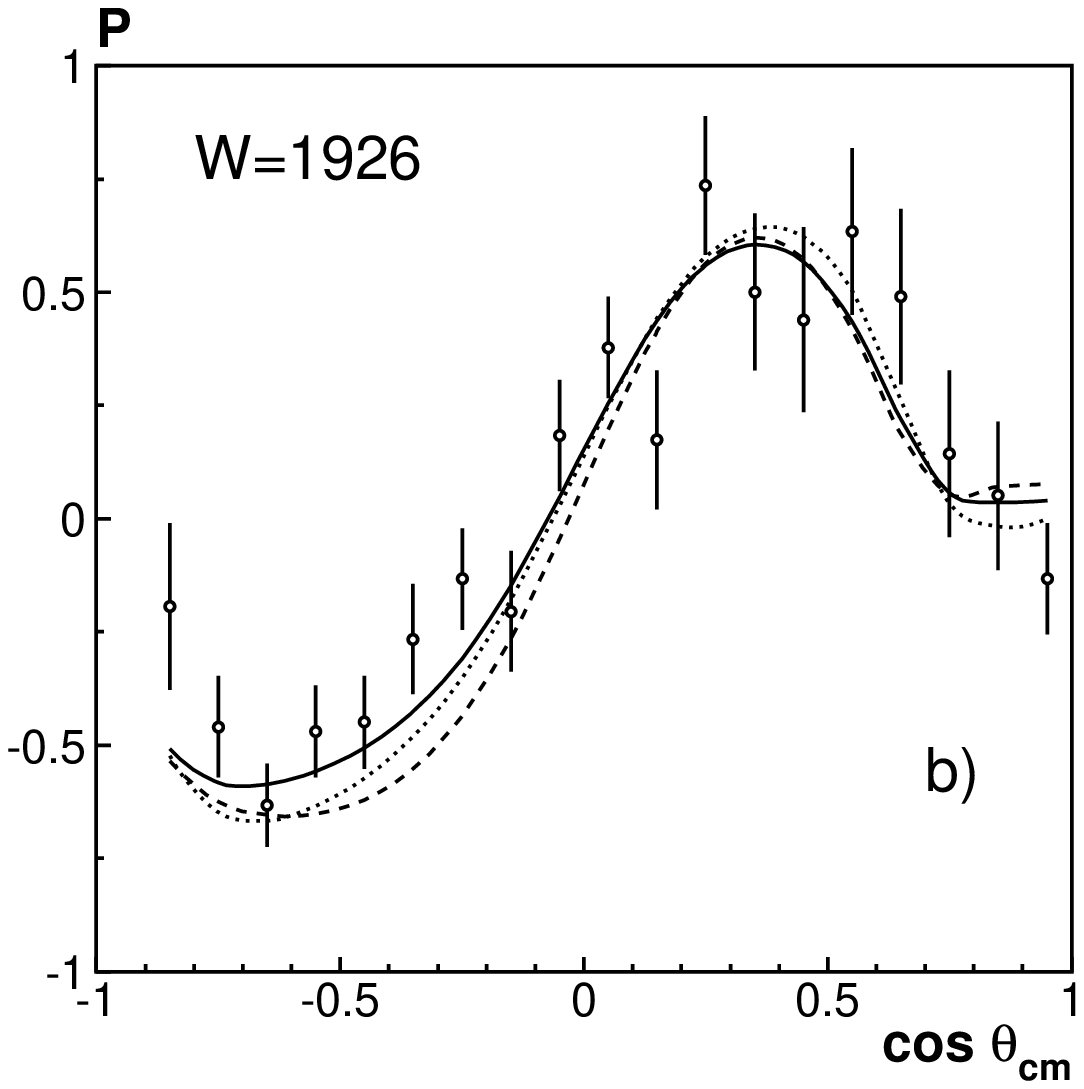,width=0.24\textwidth,clip=on}\\
\hspace{-2mm}\epsfig{file=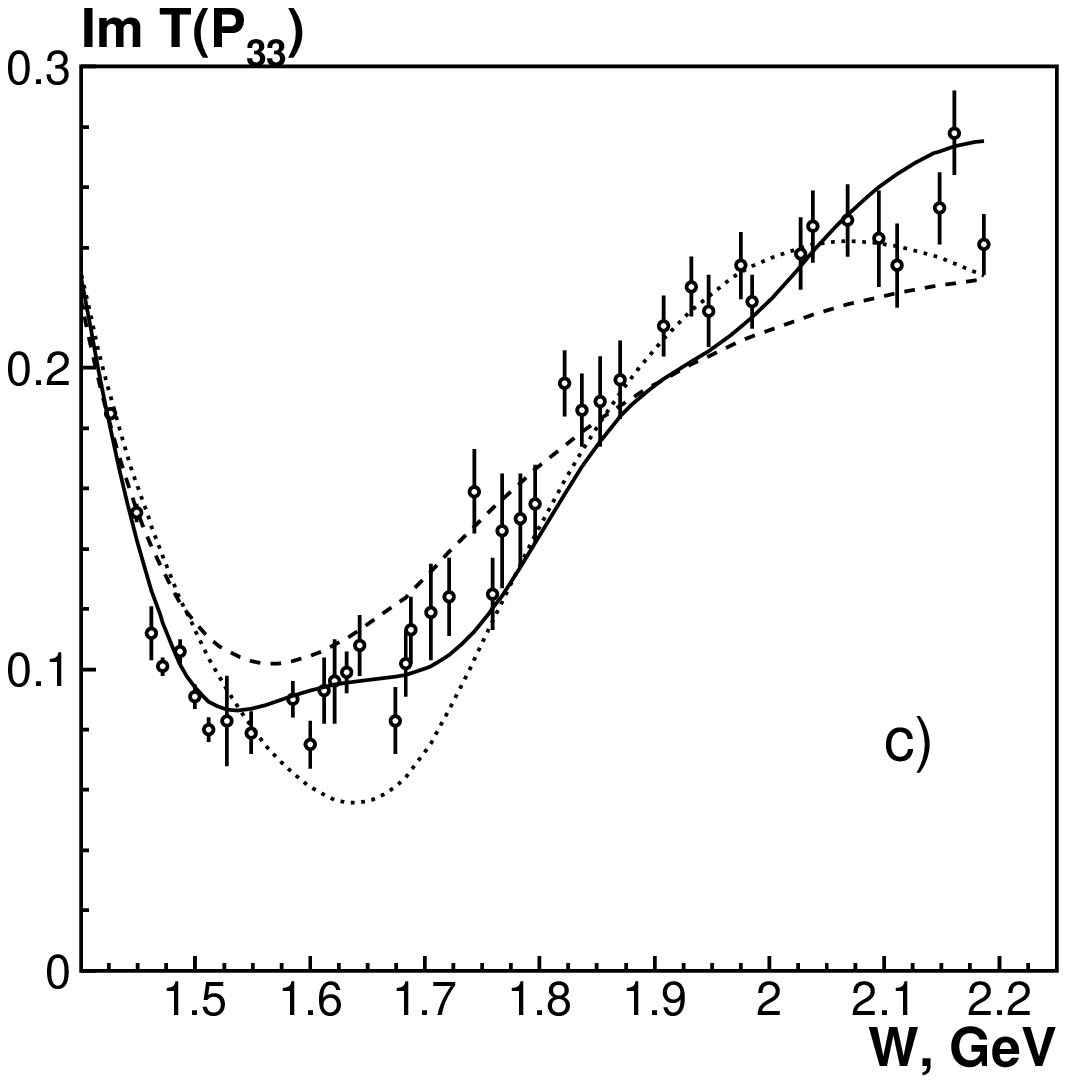,width=0.24\textwidth,clip=on}&
\hspace{-4mm}\epsfig{file=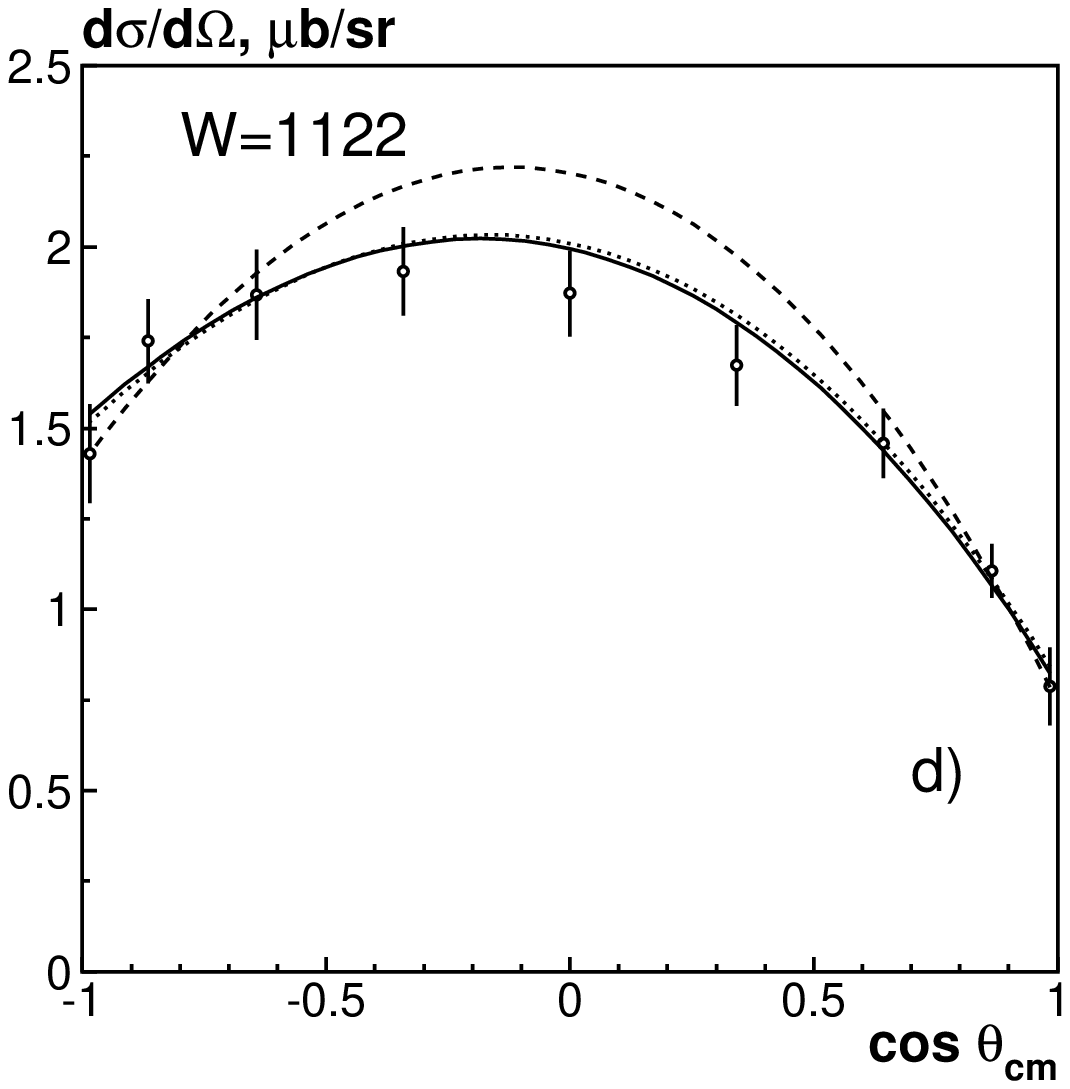,width=0.24\textwidth,clip=on}\\
\end{tabular}
\ec
\caption{\label{p33-structure} a) Angular distribution of the $\eta$ meson
for the reaction $\gamma p\to \pi\eta p$,
b) the recoil asymmetry for the reaction $\pi^+ p\to K^+\Sigma^+$, c)
imaginary part of the $\pi N$ elastic amplitude and d) $\gamma p\to
\pi^0 p$ differential cross section at low energies. The full curve
corresponds to the solution BG2010-02 with two $P_{33}$ states above
$\Delta(1232)$, the dotted curve to a solution one state, the dashed
curve to a solution with both these states removed from the fit. The
data shown contribute only a small fraction of the total $\chi^2$
improvement (Table~\ref{ln}). For the three hypotheses, solution
BG2010-02 yields, respectively: a) $\chi^2$ = 19, 34, 46 for 12 data
points, b) $\chi^2$ = 22.5, 29, 42 for 19 data points, c) 113, 358,
448 for 40 data points, and d) 4.0, 4.6, 21.2 for 9 data points.}
\end{figure}

We have already reported evidence for the $\Delta(1920)P_{33}$
resonance by a study of the reaction $\gamma p\to \pi^0\eta p$
\cite{Horn:2007pp}. In a coupled-channel analysis of a large set of
data \cite{Anisovich:2009zy}, the mass of this state was found to be
$1950\pm 40$\,MeV. The present analysis defines the pole position
around 1925 MeV, a compromise between the fit to $\gamma p\to
\pi^0\eta p$ and the fit to $\pi^+ p\to K^+\Sigma^+$ data. In the
3-pole 6-channel K-matrix fit we find the pole at
$1925\!\pm\!40-i150\!\pm\!25$. The pole position and residues for
the elastic and transition amplitudes are given in
Table~\ref{residues}. If the two $P_{33}$ resonances are excluded
from the fit, the differential cross section for this reaction shows
systematic deviations. A selection of data in comparison to fit
results is shown in Fig. \ref{p33-structure}. When one resonance is
removed, the fit decides in favor of $\Delta(1920)P_{33}$ (replacing
$\Delta(1600)P_{33}$ by background terms). The total $\chi^2$
deteriorates significantly even though the effect in Fig.
\ref{p33-structure} is less visible.

Nevertheless, the $\Delta(1600)P_{33}$ state plays an important role
for the description of all fitted data, too. The pole position was
found to be $1480\!\pm\!30-i120\!\pm\!20$\,MeV. If this state is
excluded, systematic discrepancies are observed in the description
of the $P_{33}$ elastic amplitude (see Fig. \ref{p33-structure}c).
Imposing a very large weight on the elastic data we can reproduce
the elastic amplitude also without $\Delta(1600)P_{33}$. Without
$\Delta(1600)P_{33}$, the pole representing $\Delta(1920)P_{33}$
moves to a higher mass and becomes broader. However, this fit fails
to reproduce accurately the $\pi^+p\to K^+\Sigma^+$ data. This is an
example for the main point of this analysis: the elastic amplitude
may yield a satisfactory solution (in terms of $\chi^2$) without
introduction of a resonance but the inclusion of inelastic reactions
may reveal that resonant contributions are required.

To check the stability of the solution we also introduced a further
higher-mass $P_{33}$ state, first as a Breit-Wigner resonance and
then as forth K-matrix pole.  Both parameterizations optimized with
a pole at about 2150\,MeV mass and 500\,MeV width. However, only a
marginal improvement was observed in the description of the data.

\begin{figure}[pt]
\bc
\epsfig{file=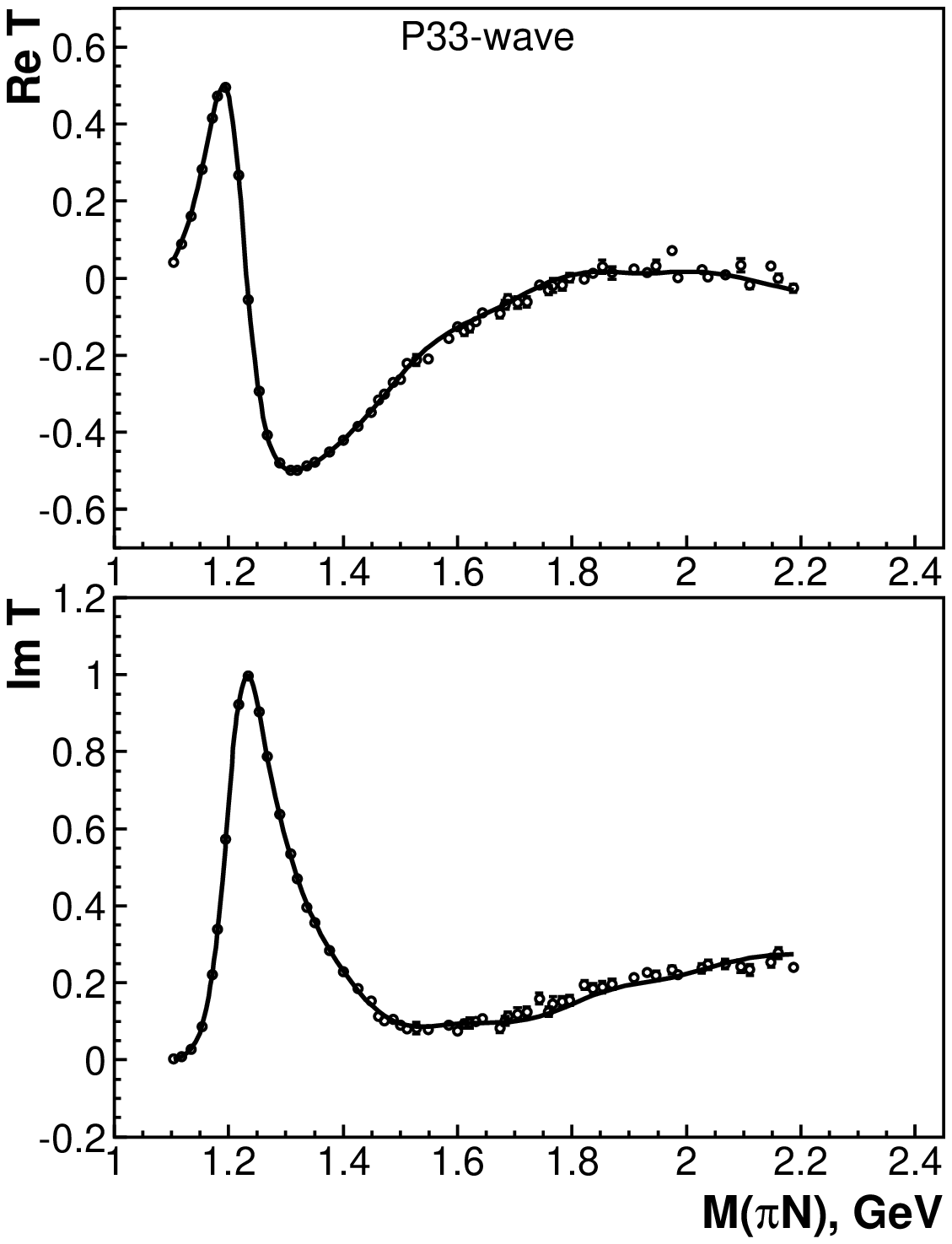,width=0.35\textwidth,height=0.40\textwidth,clip=on}
\ec
\caption{\label{p33} Description of the $P_{33}$ elastic amplitude
extracted by SAID from the energy-independent partial wave analysis.
The full curve shows our solution BG2010-02.}
\bc
\epsfig{file=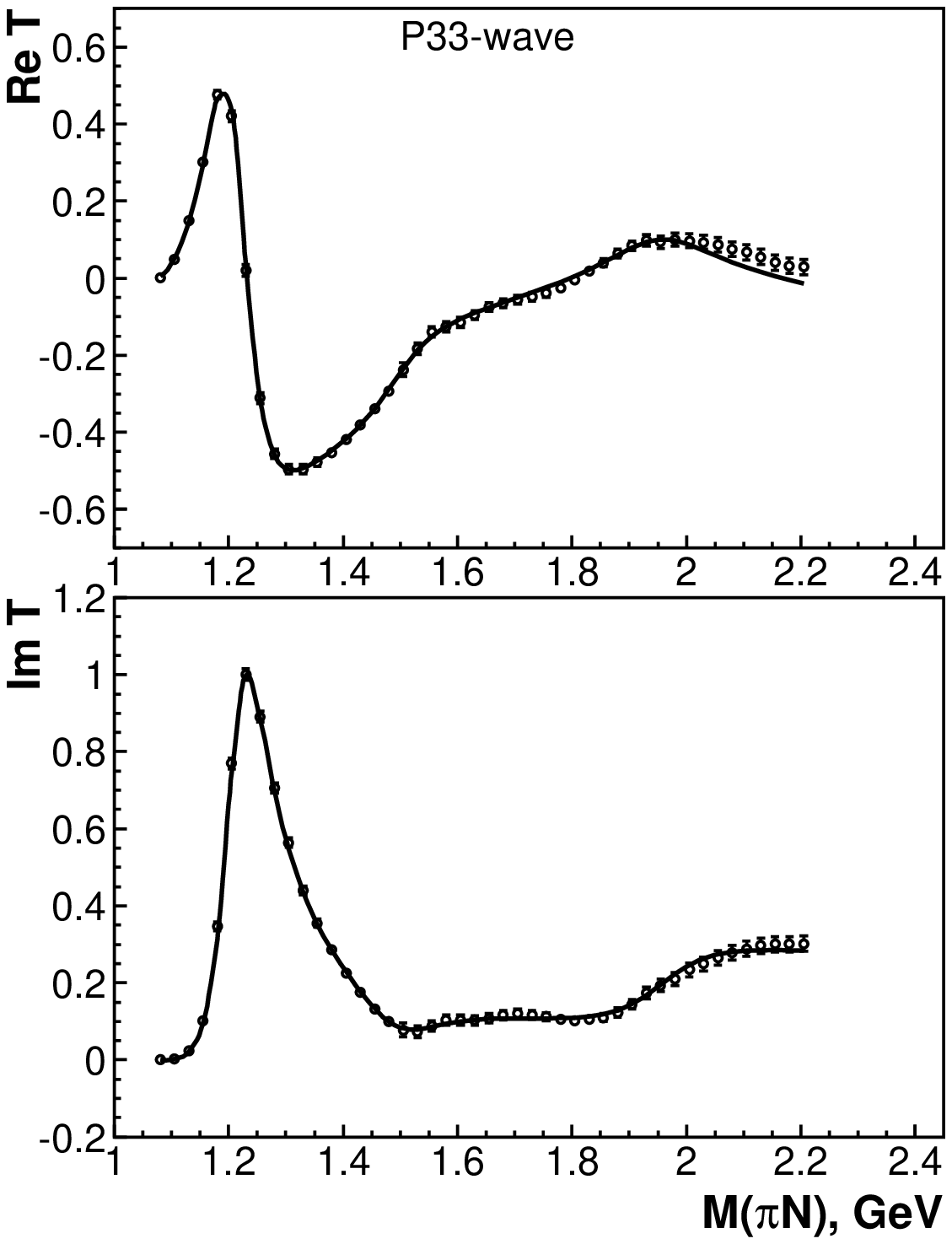,width=0.35\textwidth,height=0.40\textwidth,clip=on}
\ec
\caption{\label{p33_h} Fit of H\"ohler's solution for the $P_{33}$
elastic wave. For the fit we introduced 2\% errors for H\"ohler's
values below 1500 MeV and 5\% errors above. The full curve shows our
modified solution BG2010-02.}
\end{figure}

The three pole structure, observed in the $P_{33}$ wave below 2\,GeV
is not in conflict with the SAID energy-independent partial wave
amplitude. The comparison of our curve and the result of the SAID
energy-independent phase shift analysis is shown in Fig.~\ref{p33}.

The H\"ohler solution for the $P_{33}$ wave shows a clear structure
in the mass region 1600-2000 MeV which is compatible with our
finding. We have added for the fits to real- and imaginary part of
the amplitude 2\% errors below 1500\,MeV and 5\% errors above. The
result of the fit is shown in Fig.~\ref{p33_h}. Note that curves in
Figs.~\ref{p33} and \ref{p33_h} represent different fits. However,
when the SAID amplitude for the $P_{33}$-wave is substituted by the
H\"ohler amplitude, only the elastic $P_{33}$ amplitude changes
significantly but other results change very little.

The likelihood is not a direct criterium for the fit quality. More
interesting are the changes in likelihood when resonances are
removed from the fit and the data refitted. In Table \ref{ln} we
give, for convenience, changes $\Delta\chi^2=-2\Delta(L_{\rm tot})$.
Solution BG2010-01 is slightly worse than solution BG2010-02,
$\Delta\chi^2=152$.

\begin{table}[ph]
\caption{\label{ln}Changes in $\chi^2$ for solution BG2010-02 when
resonances are removed and the data refitted.}
\bc
\begin{tabular}{cccc}
\hline\hline\\[-2.2ex]
removed & $\Delta\chi^2$ & removed & $\Delta\chi^2$\\[0.2ex]
\hline\\[-2.2ex]
\hspace{-2mm}$\Delta(1600)P_{33}$\hspace{-3mm}&\hspace{-3mm}2843\hspace{-2mm}&\hspace{-2mm}$\Delta(1600)P_{33}$
and
$\Delta(1920)P_{33}$\hspace{-2mm}&\hspace{-2mm}6420\hspace{-3mm}\\[0.2ex]
\hspace{-2mm}$N_{1/2^+}(1875)$\hspace{-3mm}&\hspace{-3mm} 2154\hspace{-2mm}&\hspace{-2mm}$N_{1/2^+}(1875)$  and $N(1710)P_{11}$\hspace{-2mm}&\hspace{-3mm} 4756\\
\hspace{-2mm}$N_{3/2^+}(1975)$\hspace{-3mm}&\hspace{-3mm} 1332\hspace{-2mm}&\hspace{-2mm}$N(1900)P_{13}$ and $N_{3/2^+}(1975)$\hspace{-2mm}&\hspace{-3mm} 3187\\
\hline\hline\vspace{-10mm}
\end{tabular}
\ec
\end{table}
\boldmath\subsection{$(I)J^P=(3/2)1/2^+$}\unboldmath

The $P_{31}$ partial wave provides only a rather broad contribution
to the $\pi^+ p\to K^+\Sigma^+$ reaction with a maximum around 2
GeV. Moreover, we found a very small photoproduction contribution
from this wave. First we tried a fit with a two-pole K-matrix
parameterization: one pole in the 1600\,MeV region and second one in
the 2000\,MeV region. In the best fit the poles are optimized at
$1480-i90$ and $1890-i220$\,MeV. The second pole is needed to
reproduce the data, and we have no doubts that it exists. The
position of the first pole is not well defined by the data, the
corresponding K-matrix pole has very large couplings to the two-pion
decay channels. If the pole is taken out of the fit and and replaced
by non-resonant terms for direct two-pion production, we do not
observe a notable change in the quality of the data description.
Thus we conclude that the present data base does not provide a proof
for the existence of a state with $(I)J^P=(3/2)1/2^+$ quantum
numbers in the 1500-1800\,MeV mass region. The final solution was
obtained with one K-matrix pole around 1900\,MeV.

\boldmath\subsection{$(I)J^P=(1/2)3/2^+$ }\unboldmath

The existence of the four-star $N(1720)P_{13}$ resonance is, of
course, beyond doubt and absolutely required in our fits. Its
properties, in particular its $N\eta$ coupling, are controversial.
At about the same mass, $N(1710)P_{11}$ can decay into the same
final states, and the assignment of decays to one resonance or to
the other one can be ambiguous.

This is, in particular, true for their $N\eta$ decays. In
\cite{Anisovich:2005tf,Bartholomy:2007zz}, we obtained a better fit
to the reaction $\gamma p\to p\eta$ when $N(1720)P_{13}$ was
introduced as intermediate resonance than by using $N(1710)$
$P_{11}$. This was in contrast to the results reported in
\cite{Vrana:1999nt,Batinic:1995kr,Penner:2002ma} who assigned most
of the $N\eta$ intensity to $N(1720)P_{13}$ but consistent with
\cite{Aznauryan:2003zg}.

In the present analysis, data on $\pi^-p\to \eta n$ are included.
The increase in data base now allows us to have contributions from
both, $N(1710)P_{11}$ and $N(1720)P_{13}$, in the fit without
running into fit instabilities. The $\pi^-p\to \eta n$ data exclude
the large $N(1720)P_{13}\to\eta N$ coupling found in
\cite{Anisovich:2005tf} (see Fig.~\ref{etan_dcs}). We confirmed,
however, when this data are excluded and only one resonance is
admitted, a preference for $N(1720)P_{13}$. The present data sets
still do not yield an unambiguous ans\-wer for the $N\eta$ branching
ratios. The ambiguity will be discussed below. In any case, the
example shows the merits of using a large body of data in a
coupled-channel analysis but also that ambiguities can still
survive. In particular for such sensitive questions how much
intensity one has to assign to specific states, the forthcoming
double polarization data on $\pi$ and $\eta$ photoproduction are
indispensable.

The $N(1900)P_{13}$ resonance was first introduced in our analysis
\cite{Nikonov:2007br}. It was required to achieve a good description
of the double polarization observables $C_x$, $C_z$
\cite{Bradford:2006ba} in photoproduction of kaon-hyperon final
states. The solution predicted very well the data on further double
polarization observables, $O_x$, $O_z$, reported one year later
\cite{Lleres:2008em}.

\begin{figure}[pt]
\bc
\begin{tabular}{ccc}
\hspace{-2mm}\epsfig{file=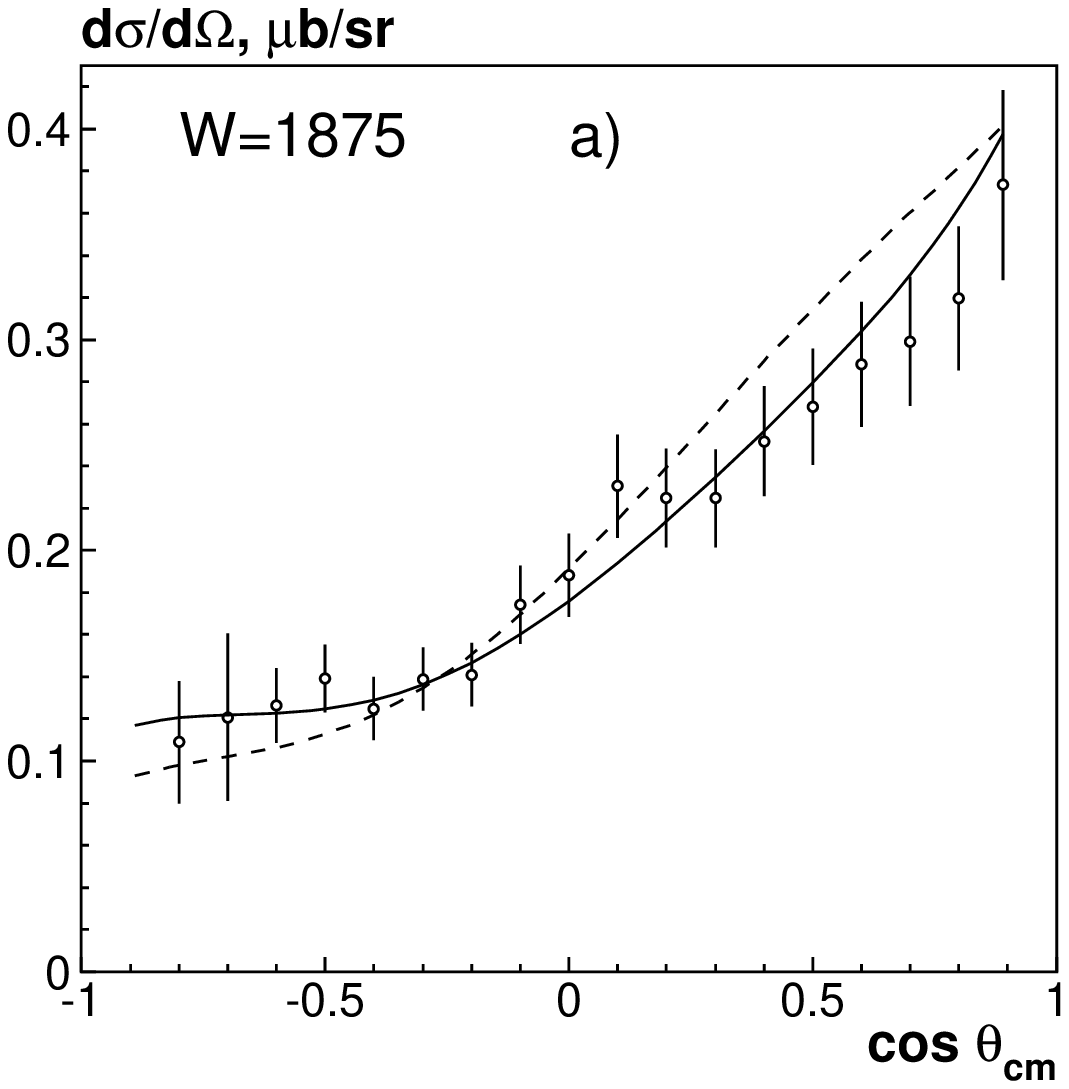,width=0.24\textwidth,clip=on}&
\hspace{-4mm}\epsfig{file=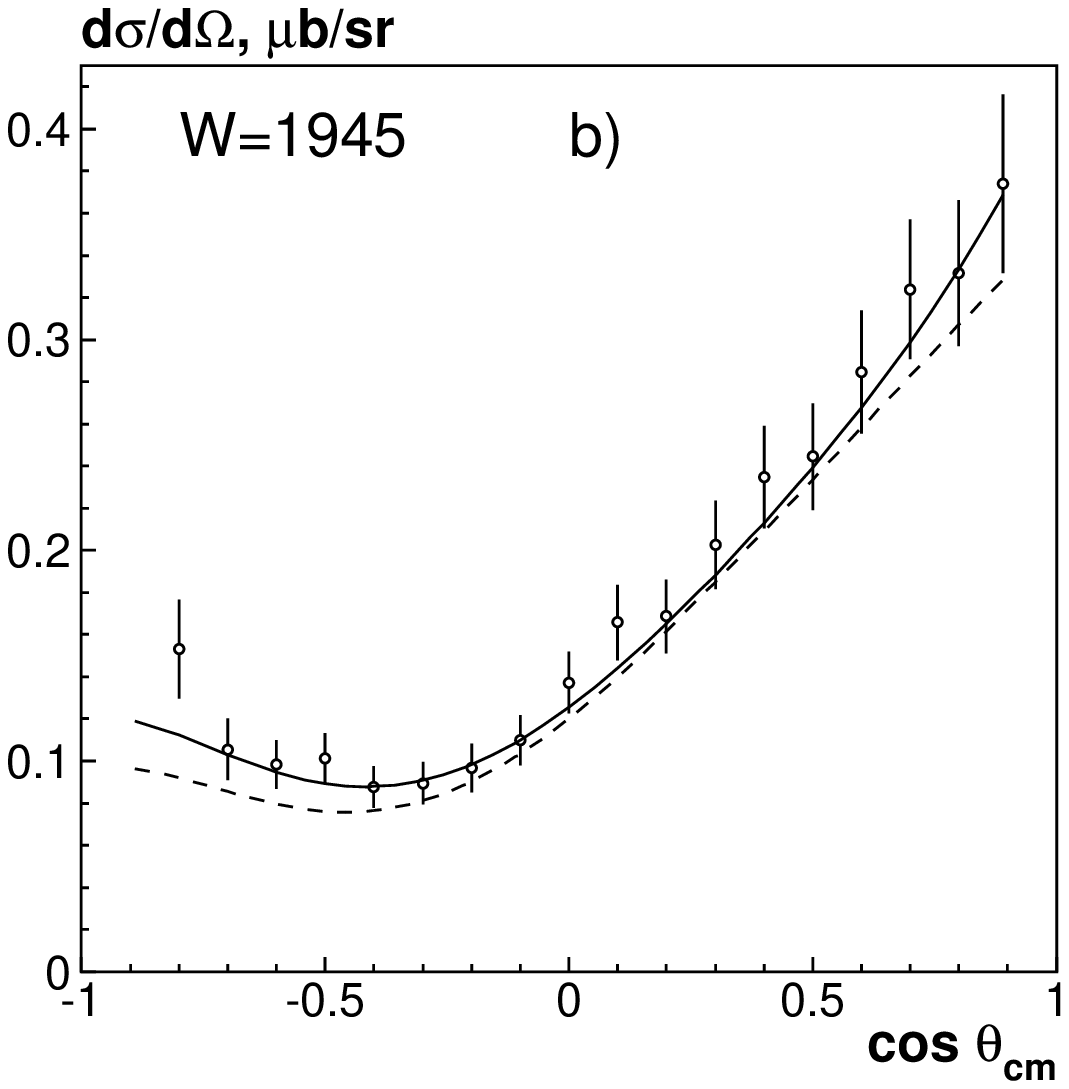,width=0.24\textwidth,clip=on}\\
\hspace{-2mm}\epsfig{file=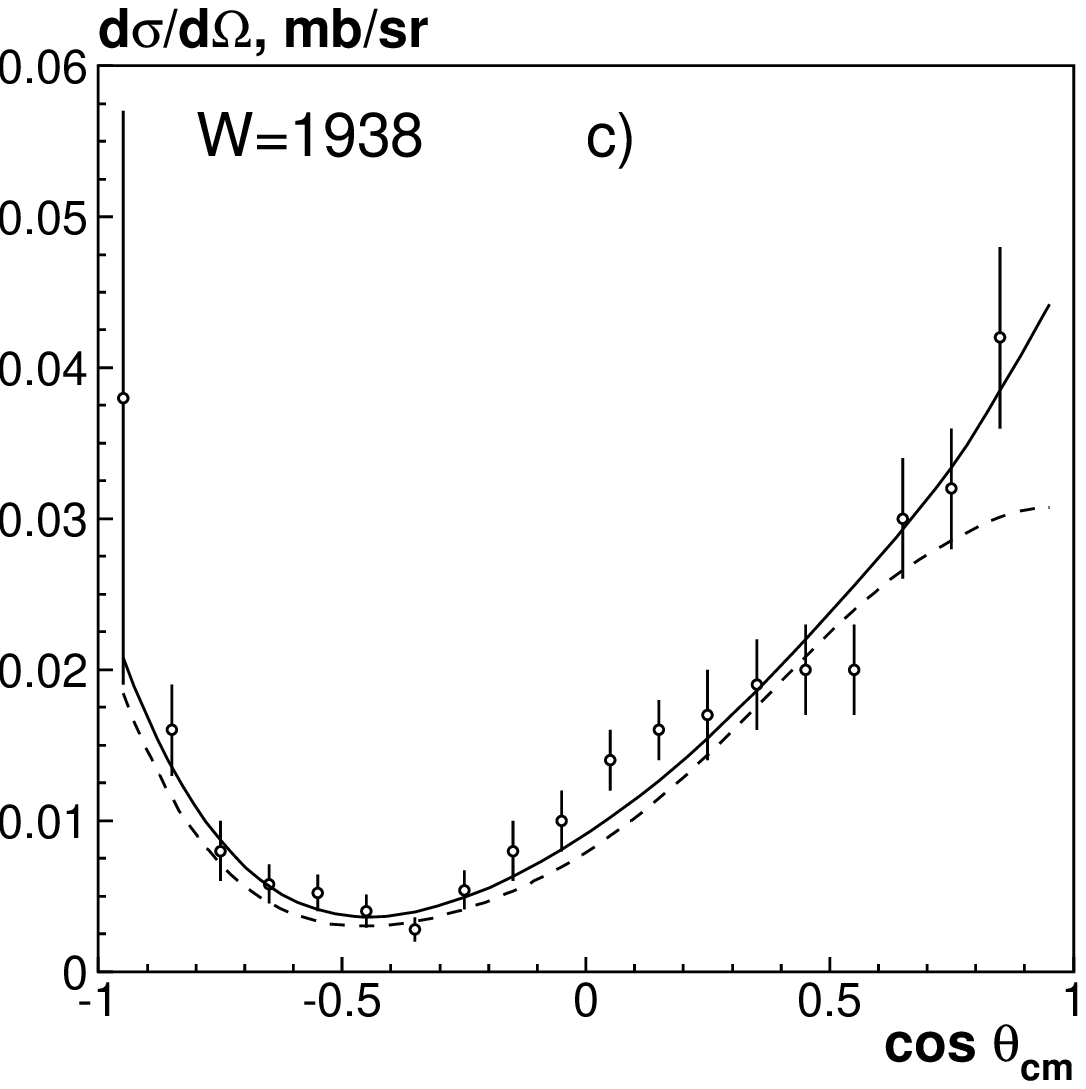,width=0.24\textwidth,clip=on}&
\hspace{-4mm}\epsfig{file=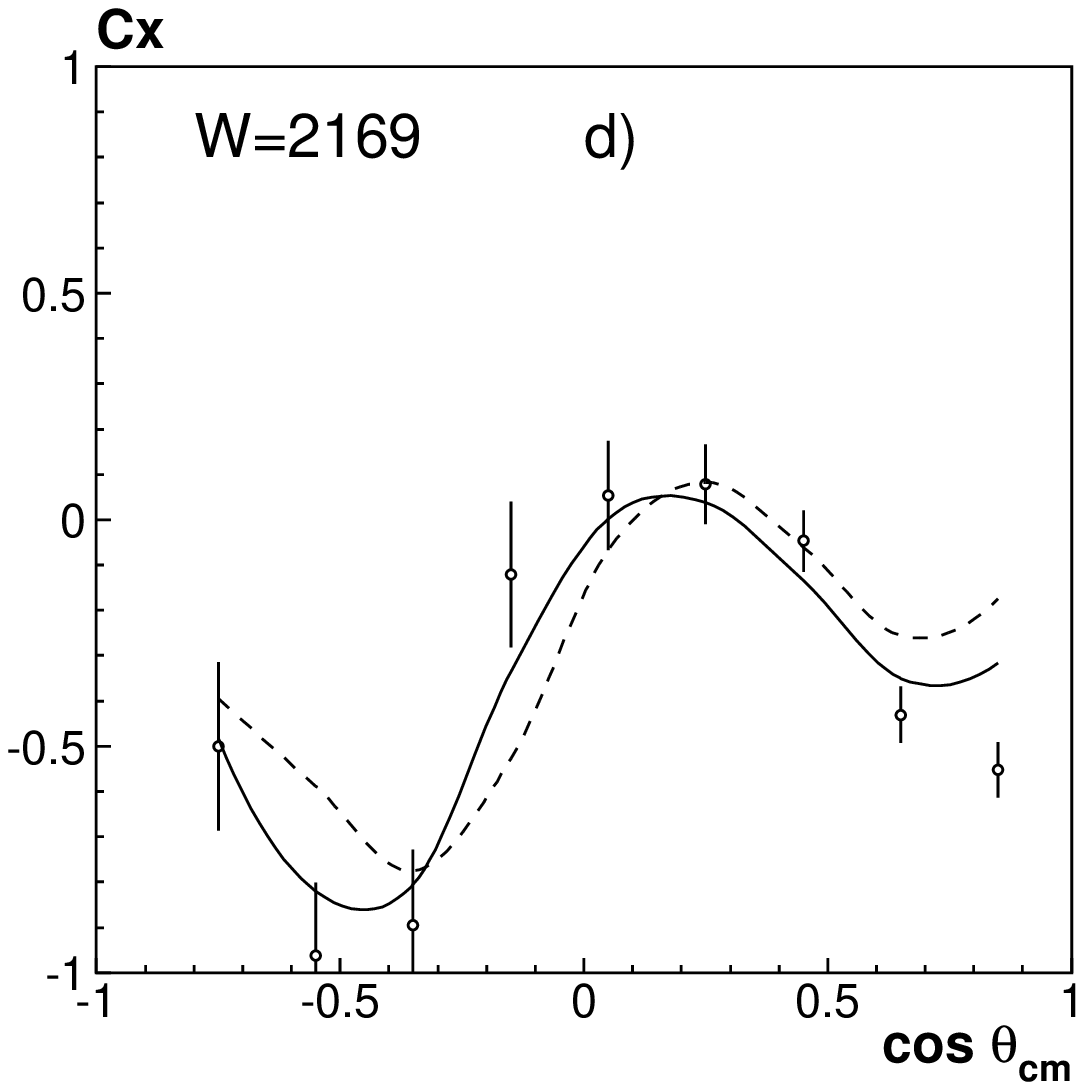,width=0.24\textwidth,clip=on}\\
\end{tabular}
\ec
\caption{\label{p13-structure}Differential cross sections for
$\gamma p\to K^+\Lambda$ (a,b) and $\pi^-p\to K^0\Lambda$ (c). $C_x$
for $\gamma p\to K^+\Lambda$. The full curve corresponds to the
solution BG2010-02 with two close-by $P_{13}$ states above
$N(1720)P_{13}$, the dashed curve to a solution with the high-mass
$P_{13}$ resonances removed. The data shown contribute only a small
fraction of the total $\chi^2$ improvement (Table~\ref{ln}). For the
two hypotheses, solution BG2010-02 yields, respectively: a) $\chi^2$
= 8.3 / 24 for 18 data points, b) $\chi^2$ = 8.6/ 28 for 18 data
points, c) 17 / 33 for 19 data points, and d) 22 / 60 for 9 data
points.\vspace{-4mm}}
\bc
\epsfig{file=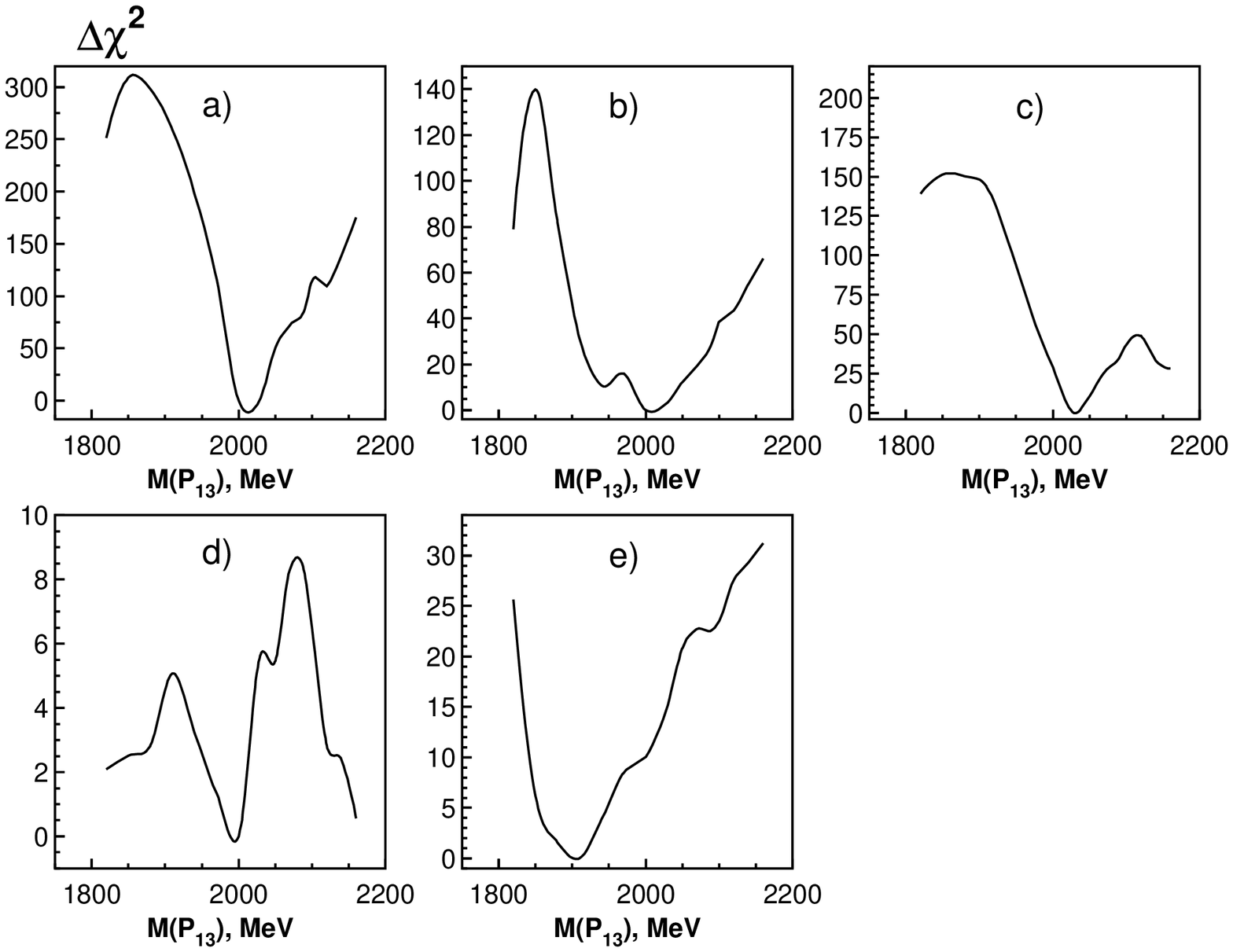,width=0.5\textwidth,clip=on}
\ec
\caption{\label{p13_scan} Mass scan of the $P_{13}$ Breit-Wigner
amplitude. Change of $\chi^2$ for the fit of a) differential cross
sections (DCS) for $\gamma p\to K^+\Lambda$ \cite{McCracken:2009ra},
b) DCS for $\gamma p\to p\eta$ \cite{Crede:2009zzb}, c) $C_x$, $C_z$
for $\gamma p\to K\Lambda$ \cite{Bradford:2006ba}, d) $C_z$ for
$\gamma p\to K\Sigma$ \cite{Bradford:2006ba}, e) $P$ for $\pi^- p\to
K^0\Lambda$ \cite{Baker:1978qm,Saxon:1979xu}. }
\end{figure}

$N(1900)P_{13}$ provides one of the dominant contributions to the
$\gamma p\to K\Lambda$ and a notable contribution to the $\gamma
p\to K^+\Sigma^0$ total cross sections. In the $\pi N\to K\Lambda$
reaction the $P_{13}$ partial wave has a clear peak around
1900\,MeV. If the $P_{13}(1900)$ state is excluded from the fit, the
$\chi^2$ deteriorates significantly. Fig. \ref{p13-structure} shows
a few examples. The  $\chi^2$ change from the four figures is 110,
the total $\chi^2$ change 6179. We consider the existence of
$N(1900)P_{13}$ to be solidly confirmed by the present analysis;
splitting into two resonances is a likely possibility.

A mass scan of the $P_{13}$ wave in the 1.75-2.1\,GeV mass region
for selected reactions is shown in Fig.~\ref{p13_scan}. For the mass
scan, the $P_{13}$ amplitude was given by one K-matrix pole
representing $N(1720)P_{13}$; above 1.75\,GeV the amplitude was
parameterized as Breit-Wigner amplitude. The mass was then increased
in steps, all other variables were refitted. All mass scans have a
clear minimum in the region 1900-2050\,MeV, but the behavior of the
curve is more complicated than that expected for a single pole. Some
minima suggest a resonance mass of about 1900\,MeV, others seem to
prefer 2000\,MeV. Hence we tried fits with a 3-pole 8-channel
K-matrix. The fit produces the rather complicated structure in the
region 1900\,MeV and produced two close-by poles. The first pole is
very stable and has a position close to the one reported in
\cite{Nikonov:2007br}. It has an appreciable coupling to the $\gamma
p$ channel (see Table~\ref{residues}). The second pole is broader
and higher in mass. In most solutions it has very small helicity
couplings and thus decouples from the photoproduction data. The
overall $\chi^2$ change is significant:  $\Delta\chi^2=6179$ when
both resonances are removed and $\Delta\chi^2=2451$ when only one
$P_{13}$ resonance is removed. There is a possiblity that a more
flexible parameterization of the $P_{13}$ partial wave can resolve
this problem and describe all data with a single pole. However, at
present, this double pole structure is needed to achieve a good
description of all data. Thus we claim that there are indications
for the existence of two close-by states even though the evidence
for a double-pole structure is certainly not yet conclusive. The
introduction of a fourth K-matrix pole in the region 2100-2300\,MeV
does not change the picture of the singularities in the 1900\,MeV
region.

\begin{figure}[pt]
\bc
\epsfig{file=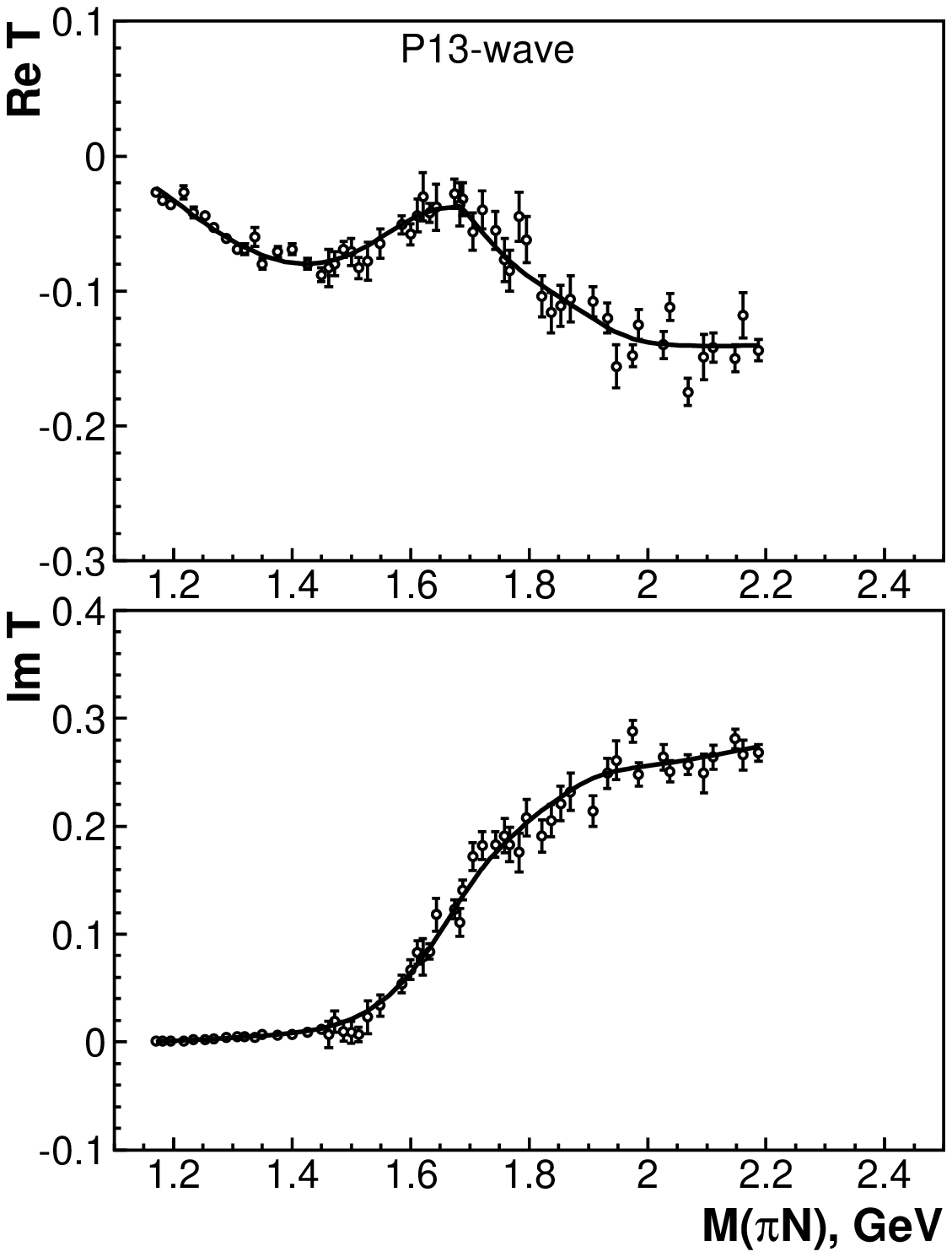,width=0.35\textwidth,height=0.40\textwidth,clip=on}
\ec
\caption{\label{p13} Description of the $P_{13}$ elastic amplitude
(solution BG2010-02) extracted by SAID from the energy-independent
partial wave analysis.}
\bc
\epsfig{file=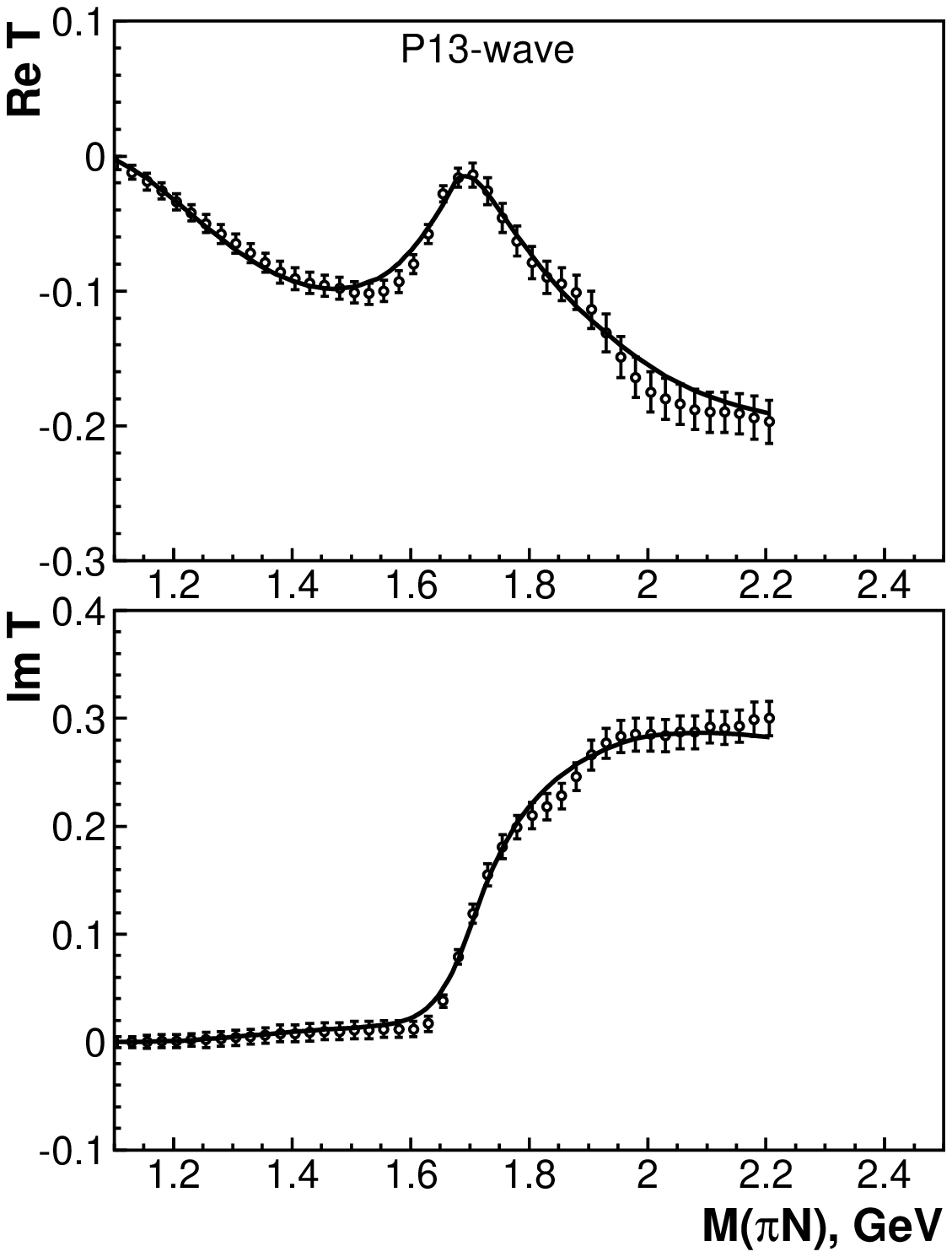,width=0.35\textwidth,height=0.40\textwidth,clip=on}
\ec
\caption{\label{p13_h} Fit of H\"ohler's solution for the $P_{13}$
elastic wave. For the fit of we introduced 2\% errors for H\"ohler's
values below 1750\,MeV and 5\% errors above. The curve shows the
modified solution BG2010-02.}
\end{figure}

The presence of pole singularities above 1700\,MeV is not in
conflict with the SAID energy-independent partial wave analysis of
the elastic data: our description of the $P_{13}$ partial wave is
shown in Fig.~\ref{p13}. H\"ohler's result for the $P_{13}$ wave,
shown in Fig.~\ref{p13_h}, has no clear structure above 1800\,MeV
neither. We would like to mention that H\"ohler's result is
appreciably different from the SAID result in the region 1700\,MeV:
it shows a much narrower structure. The substitution of the SAID
amplitude by H\"ohler's amplitude thus leads to a narrower
$N(1720)P_{13}$ state and a better definition of the singularities
around 1900\,MeV. For the fit using H\"ohler's solution we found
helicity couplings and amplitudes residues for $N(1900)P_{13}$ to be
larger than in the fit to the SAID amplitudes. Hence we increased
the final errors in these quantities. Other partial waves were
changed very little and the corresponding values are included in the
errors.

The position in the complex plane and the amplitude residues for the
two highest $P_{13}$ poles are given in Table~\ref{residues}. In
some of our solutions the $N(1720)P_{13}$ state has a large coupling
to the $S$-wave $N(1520)D_{13}\pi$ channel which opens fast at the
resonance position. As the result, we observe in different Riemann
sheets a two-pole Flatte-like structure. For both these poles the
closest physical region is at the $N(1520)D_{13}\pi$ threshold and
the information extracted from residues of the pole on one sheet can
be misleading. However, in most solutions the coupling to
$N(1520)D_{13}\pi$ is smaller and residues from the pole closest to
the physical region provide the characteristics of this state. These
values are given in Table~\ref{residues1}.

\boldmath\subsection{$(I)J^P=(1/2)1/2^+$ }\unboldmath

The $P_{11}$ wave has a maximum in the $\pi^-p\to K^0\Lambda$
reaction at 1720\,MeV which suggests a contribution from the
$N(1710)P_{11}$ state, the long tail may indicate the need for a
further state at even higher mass. We therefore used a 4-pole
K-matrix amplitude to fit the data: a first pole at the nucleon mass
as Born term, the well known Roper resonance at 1440\,MeV, and two
further poles.

\begin{figure}[pt]
\bc
\begin{tabular}{ccc}
\hspace{-2mm}\epsfig{file=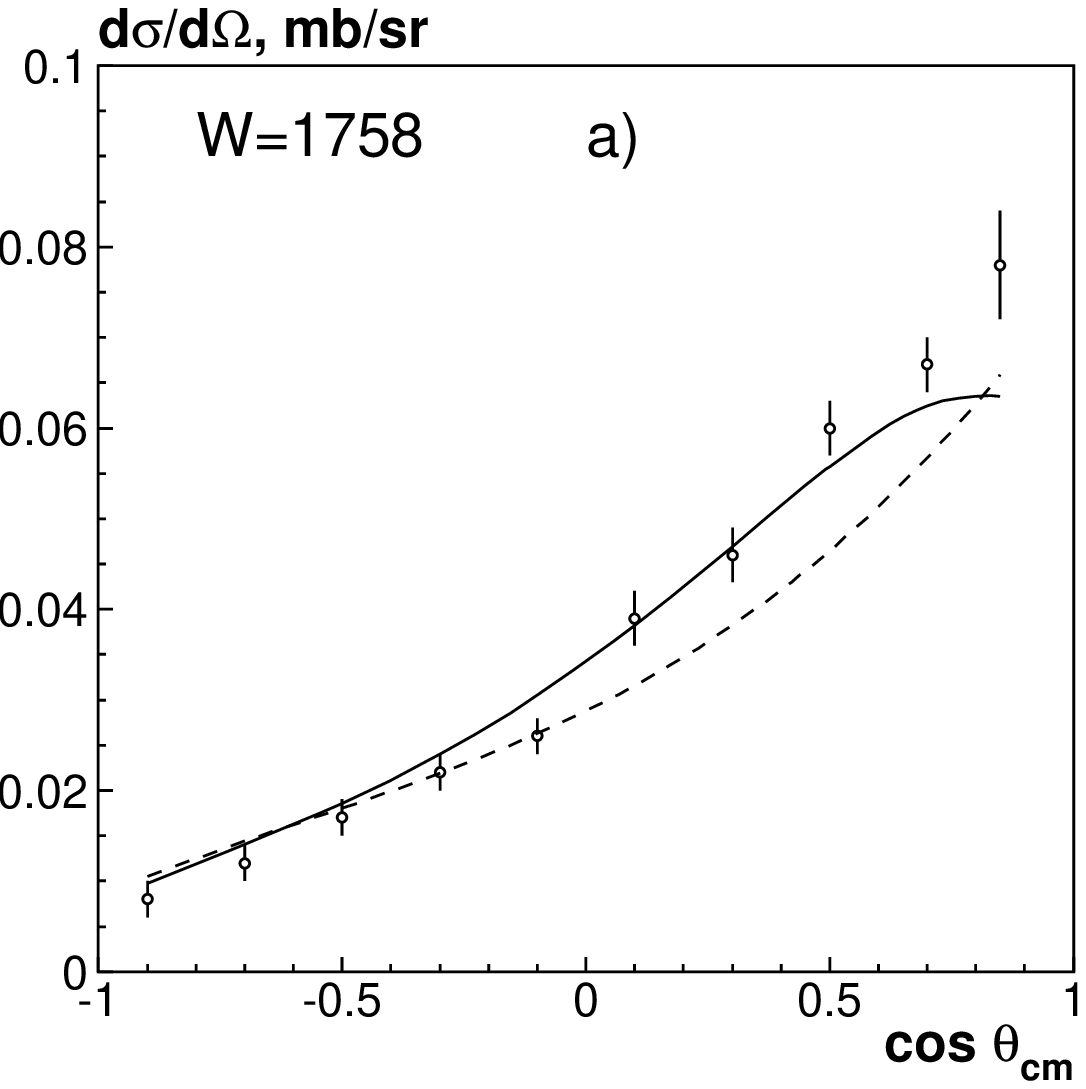,width=0.24\textwidth,height=0.18\textwidth,clip=on}&
\hspace{-4mm}\epsfig{file=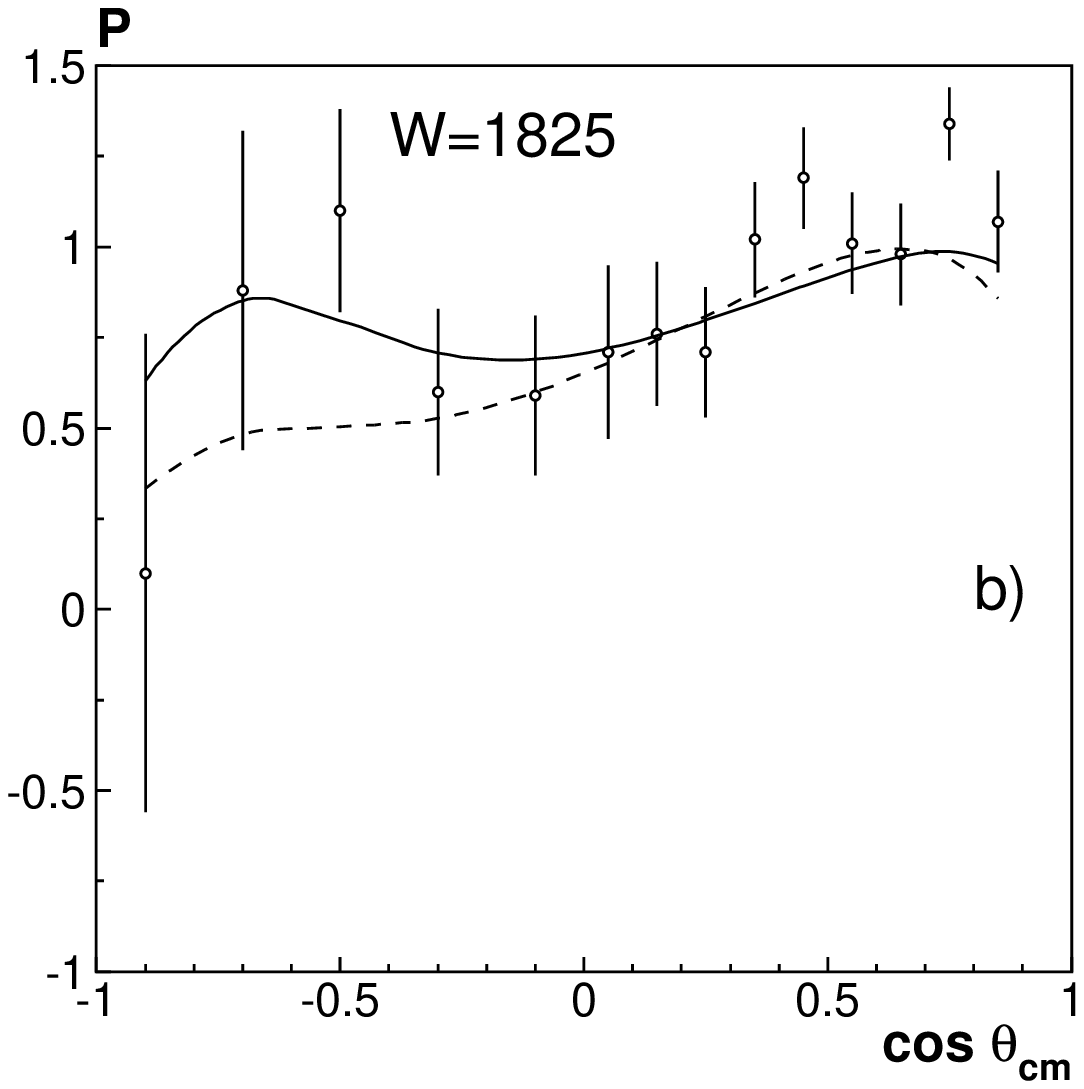,width=0.24\textwidth,height=0.18\textwidth,clip=on}\\
\hspace{-2mm}\epsfig{file=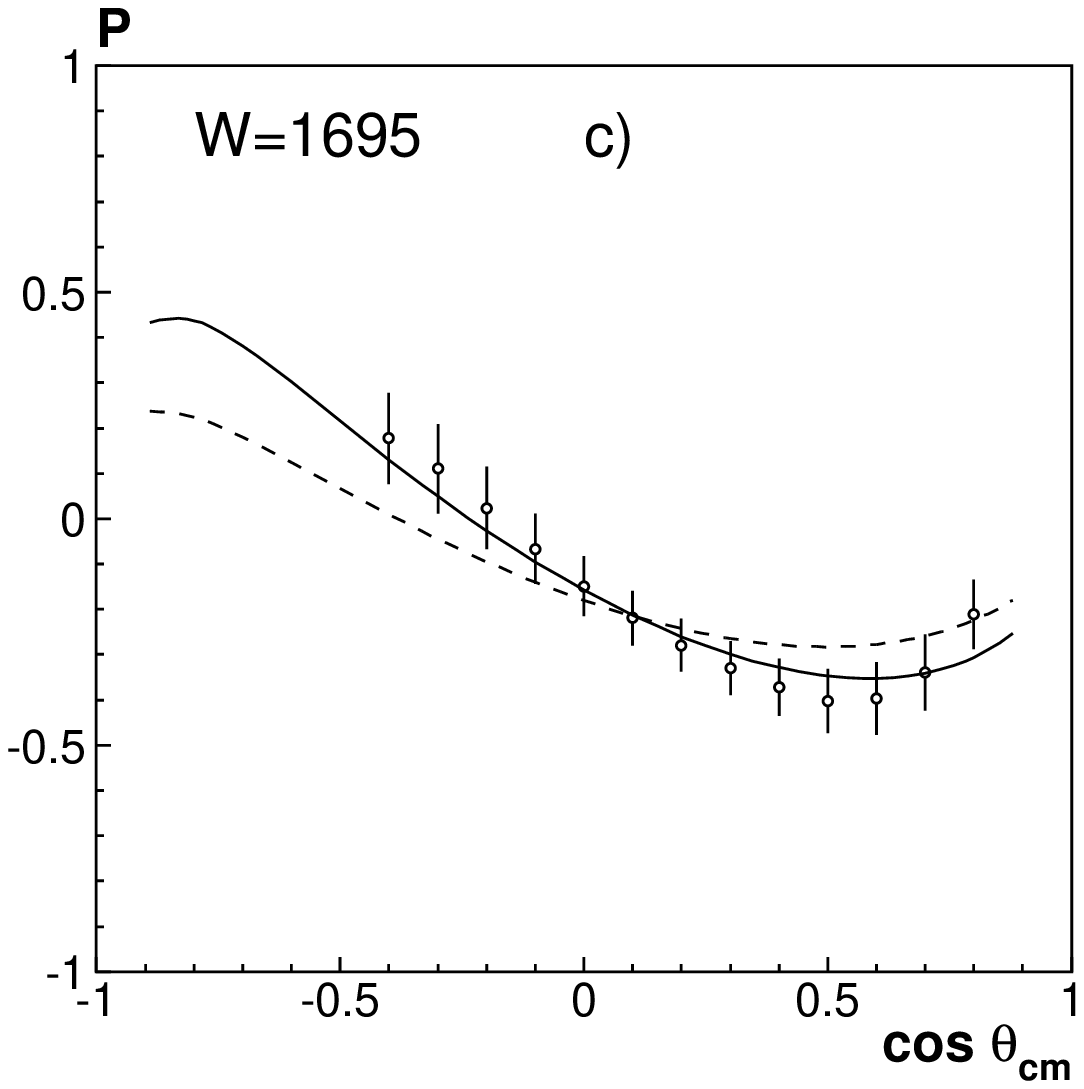,width=0.24\textwidth,height=0.18\textwidth,clip=on}&
\hspace{-4mm}\epsfig{file=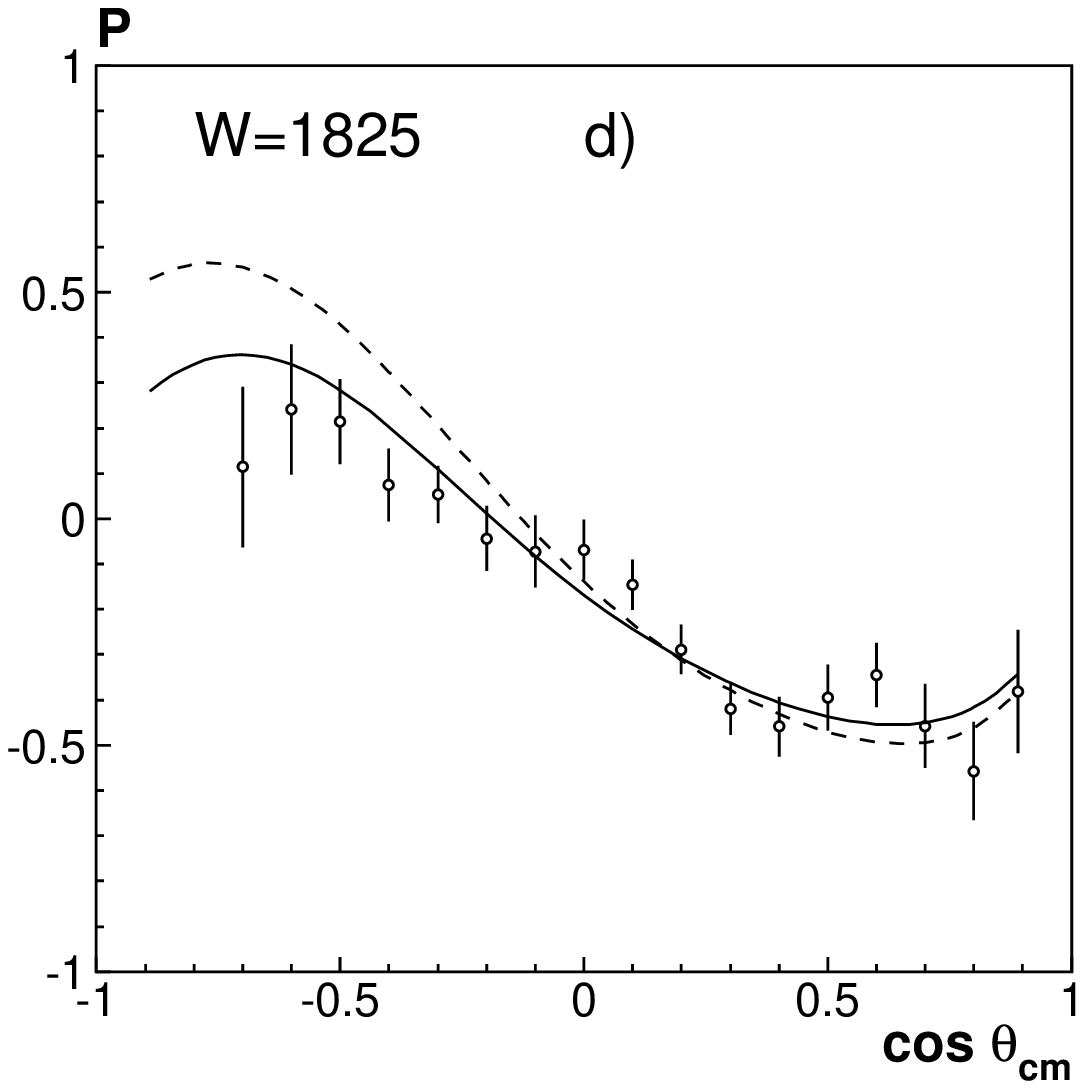,width=0.24\textwidth,height=0.18\textwidth,clip=on}\\
\hspace{-2mm}\epsfig{file=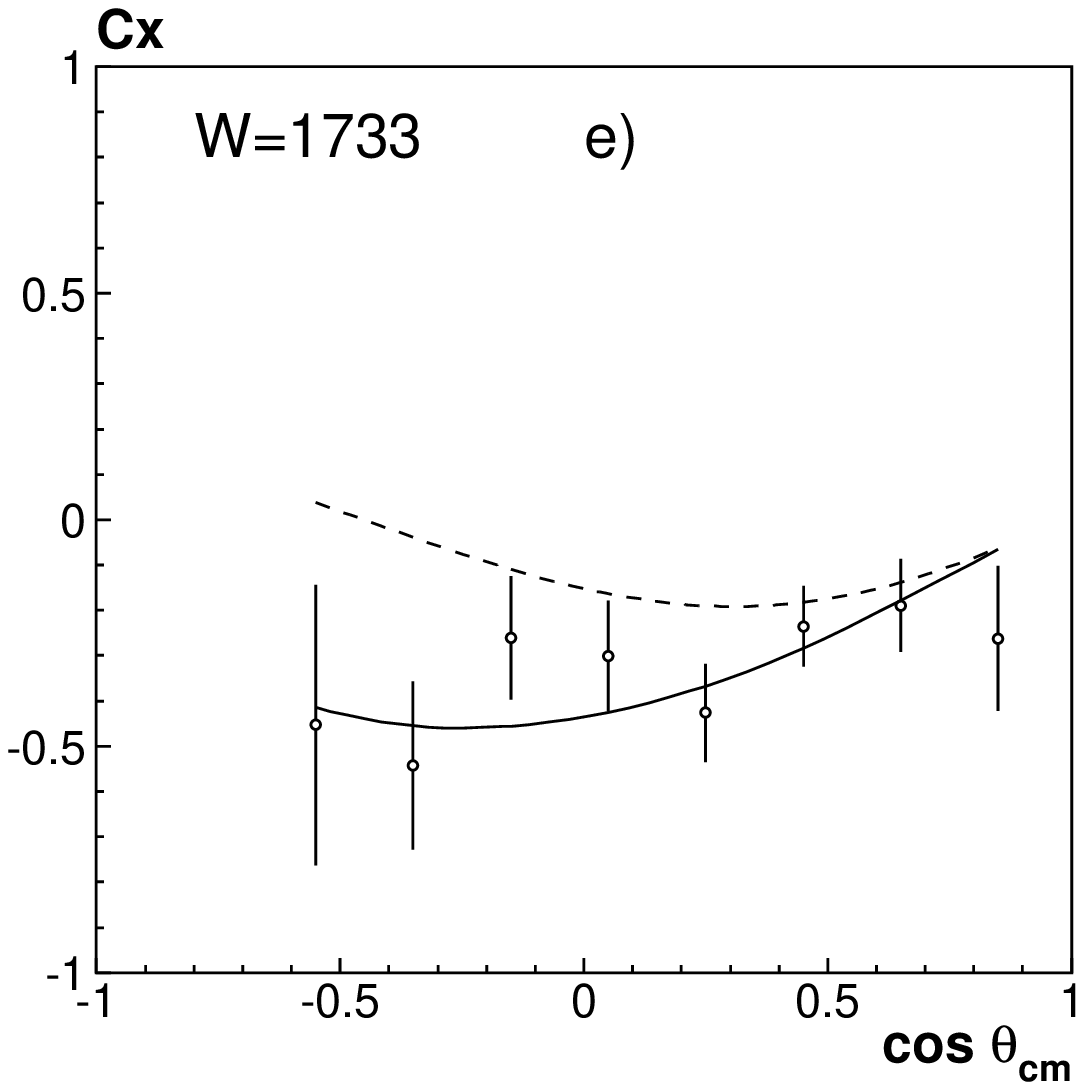,width=0.24\textwidth,height=0.18\textwidth,clip=on}&
\hspace{-4mm}\epsfig{file=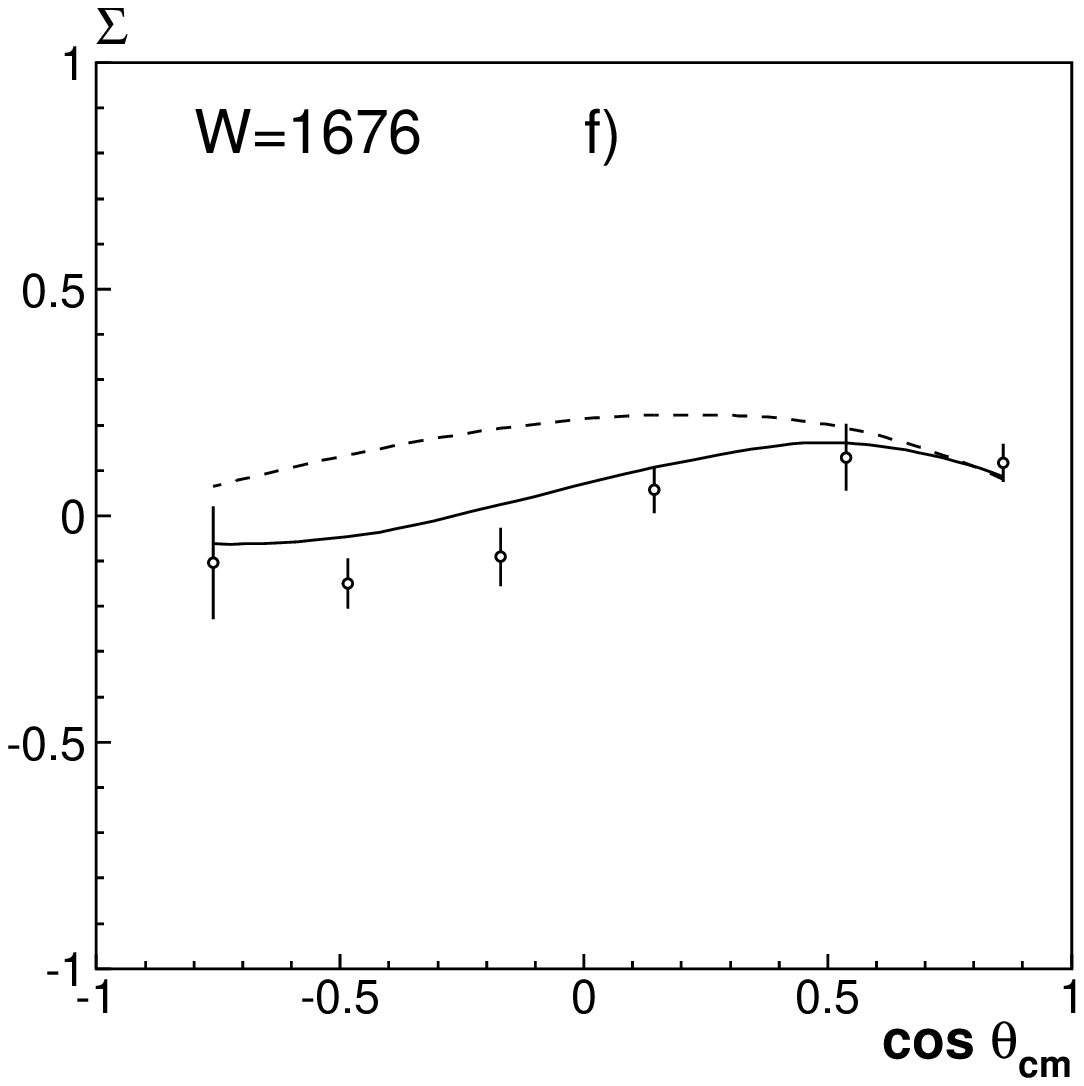,width=0.24\textwidth,height=0.18\textwidth,clip=on}
\end{tabular}
\ec
\caption{\label{pip_klam_3x3}a) Differential cross section and b)
the $\Lambda$ recoil polarization \cite{Baker:1978qm,Saxon:1979xu}
for $\pi^- p\,\to\,K^0\Lambda$ \cite{McCracken:2009ra} for four
selected energy bins. c,d) Recoil polarization
\cite{McCracken:2009ra}, e) $C_x$ \cite{McCracken:2009ra} and e)
beam asymmetry $\Sigma$ \cite{Lleres:2007tx} for $\gamma
p\,\to\,K^+\Lambda$. The full curve corresponds to the solution
BG2010-02 with two $P_{11}$ states above the Roper resonance, the
dashed curve to a solution with both $P_{11}$ resonances removed. }
\end{figure}

One pole was found at $1690\!\pm\!25-i110\!\pm\!15$\,MeV which we
identify with $N(1710)P_{11}$ and a second pole representing
$N_{1/2^+}(1875)$, a resonance which was first suggested in
\cite{Castelijns:2007qt}. If both resonances are omitted from the
fit, the contributions from the $P_{11}$ wave to $\pi^-p\to
K^0\Lambda$ and $\pi^-p\to K^0\Sigma$ become smaller and
featureless, and the fit does not reproduce well the differential
cross section (Fig.~\ref{pip_klam_3x3}a,b) and exhibits problems
with the description of the recoil asymmetry
(Fig.~\ref{pip_klam_3x3}c-f).

The description improves when $N(1710)P_{11}$ is introduced but the
fit also demands the presence of a further pole in the 1870\,MeV
region: the interference of this state with the $N(1710)$ $P_{11}$
resonance provides a correct energy dependence for this partial
wave. The visible effect of this additional resonance in the
individual plots are small but the overall improvement in $\chi^2$
is significant (Table~\ref{ln}). As mentioned above, there are two
equivalent solutions here. In the first solution, $N_{1/2^+}(1875)$
is a rather narrow state with a small coupling to the $\pi N$
channel and a rather small helicity coupling. In the second solution
this state has appreciable $\pi N$ and helicity couplings. In both
solutions the hadron couplings for the $N(1710)P_{11}$ state are
found to be very similar, however the helicity coupling differs in
sign.

\begin{figure}[pt]
\bc
\begin{tabular}{ccc}
\hspace{-2mm}\epsfig{file=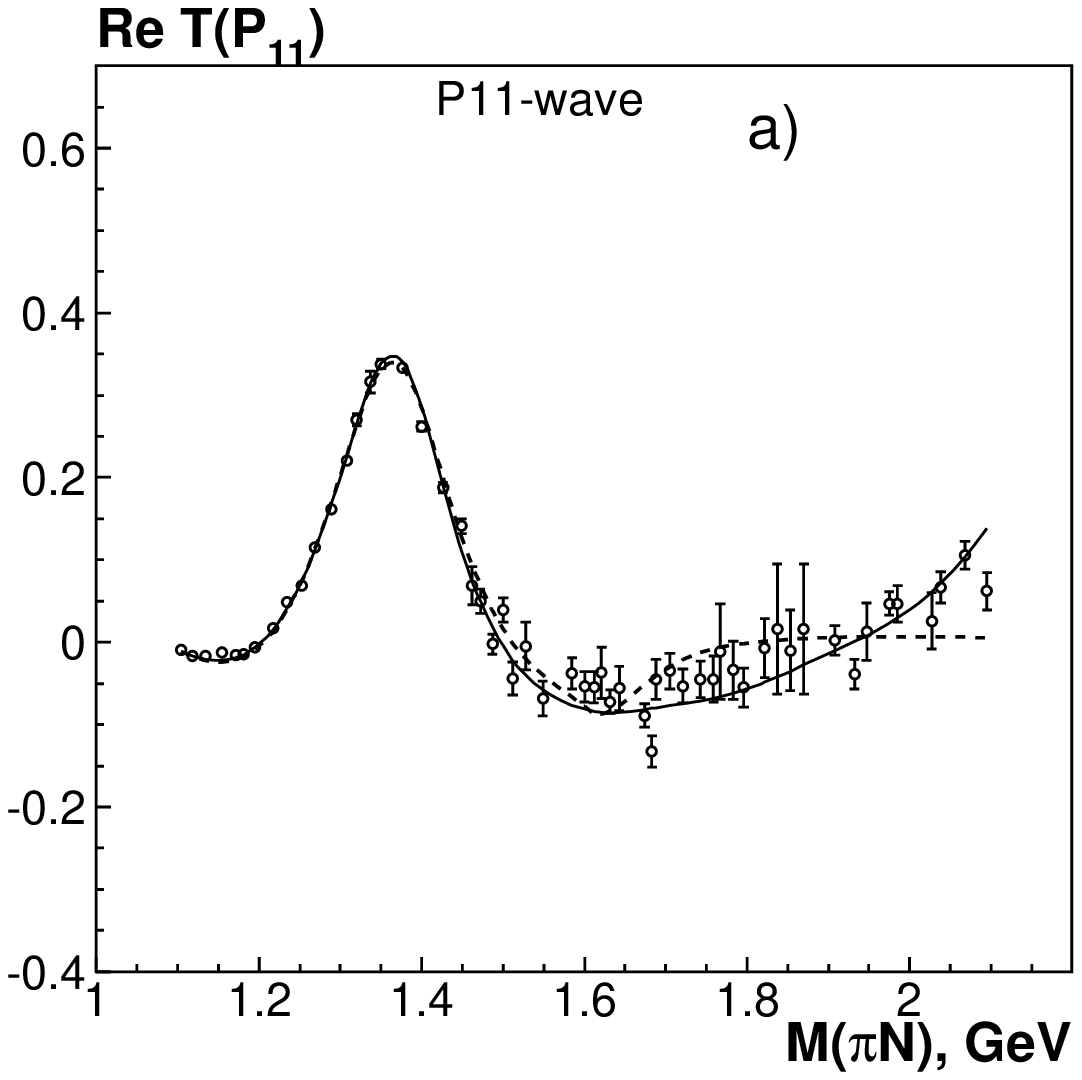,width=0.24\textwidth,clip=on}&
\hspace{-4mm}\epsfig{file=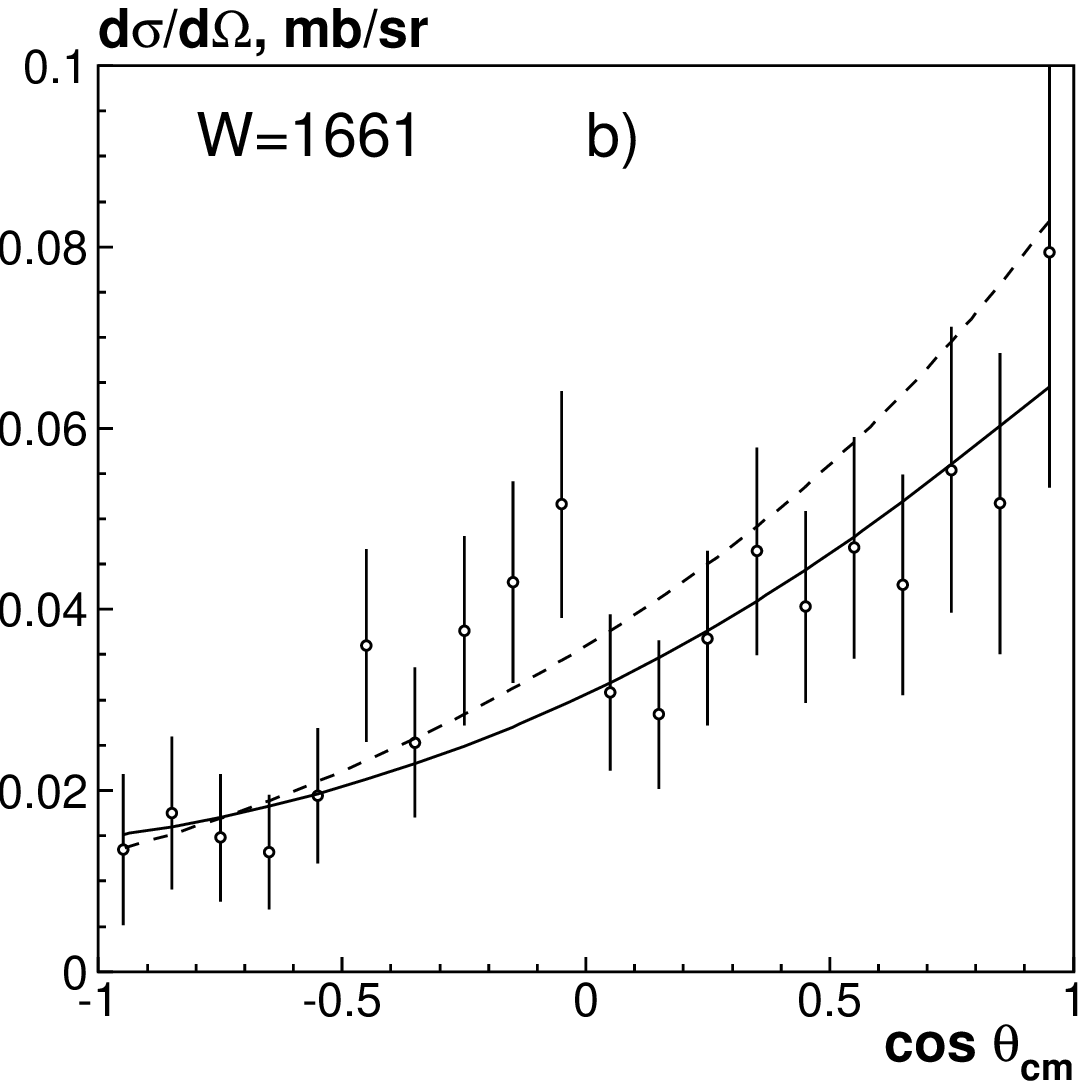,width=0.24\textwidth,clip=on}
\end{tabular}
\ec
\caption{\label{nop11_2}a) Real part of the elastic $P_{11}$
amplitude and b) the differential cross section \cite{Knasel:1975rr}
for $\pi^- p\,\to\,K^0\Lambda$. The full curve corresponds to the
solution BG2010-02 with two $P_{11}$ states above the Roper
resonance, the dashed curve to a solution with $N(1710)P_{11}$
removed.}
\end{figure}

Fig.~\ref{pip_klam_3x3} shows the differential cross section, the
observable $Pd\sigma/(d\Omega\sin\Theta)$, and the $\Lambda$ recoil
polarization for $\gamma p\,\to\,K\Lambda$
for selected bins in the region 1720-1900 MeV. The solid curve
represents our full fit with two $P_{11}$ states above the Roper
resonance, the dashed curve a fit with both these $P_{11}$ states
removed from the fit and when only non-resonant terms were admitted.
The fit with only one $P_{11}$ state in the region 1600-2000\,MeV
produced a pole close to the $P_{11}(1710)$ state. Such a fit has a
systematically worse description in a large number of fitted
reactions and in some places, there are visible systematical
deviations (see Fig.~\ref{nop11_2}). The inclusion of a third pole
at 2100\,MeV does not yield a visible effect.

To check stability of the poles we have introduced an additional
$P_{11}$ state in the higher mass region, first as a Breit-Wigner
state and then as fifth K-matrix pole. Both Breit-Wigner and
K-matrix amplitudes produced a compatible result with a pole at a
mass around 2100\,MeV and a rather large width of about 500\,MeV.
Although this state provides some improvement for the overall
description of the data it also introduces a convergency problem to
the fit. So, this state, if it exists, should be confirmed by other
data, e.g. by double polarization data and/or data on multi-meson
final states. Nevertheless´, the state could be the resonance
suggested by H\"ohler \cite{Hohler:1979yr} and by Cutkosky
\cite{Cutkosky:1980rh}.

Our conclusions concerning the existence of two $P_{11}$ states
above the Roper resonance are fully compatible with the data on
elastic scattering, with the data from the analysis of H\"ohler and
collaborators \cite{Hohler:1979yr} as well as with the data from the
analysis from Arndt and collaborators \cite{Arndt:2006bf}.
\begin{figure}[pb]
\bc
\epsfig{file=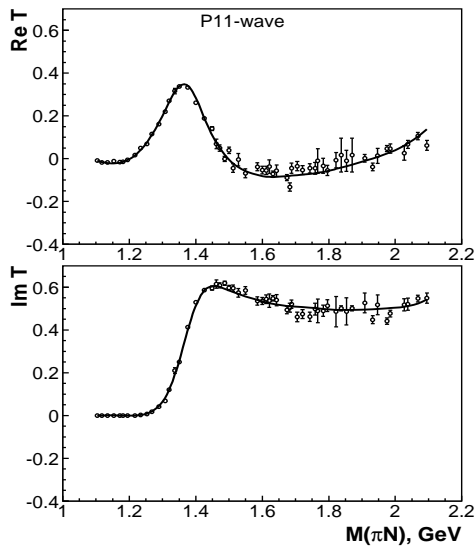,width=0.35\textwidth,height=0.40\textwidth,clip=on}
\ec
\caption{\label{p11}The $P_{11}$ elastic amplitude extracted by SAID
in an energy independent partial wave analysis \cite{Arndt:2006bf}
and our energy-dependent fit (BG2010-02 solution).}
\end{figure}
\begin{figure}[pt]
\bc
\epsfig{file=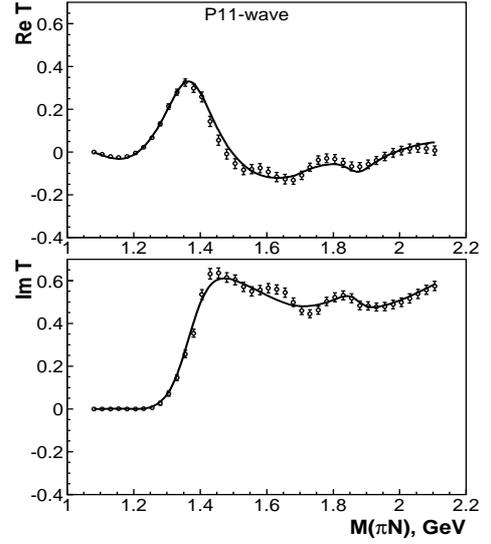,width=0.35\textwidth,height=0.40\textwidth,clip=on}
\ec
\caption{\label{p11_h}The $P_{11}$ elastic amplitude extracted by
H\"ohler in an energy independent partial wave analysis
\cite{Hohler:1979yr} and our energy-dependent fit (modified
BG2010-02). The H\"ohler result is shown as
 points with 2\% errors below 1500\,MeV and 5\% above.\vspace{-4mm} }
\end{figure}
Figure \ref{p11} shows the comparison of the real and imaginary
parts of our $P_{11}$ elastic amplitude with the energy-independent
analysis from \cite{Arndt:2006bf}. In some solutions the fit shows a
slight structure above the Roper resonance which arises from the
combined analysis of the elastic and inelastic data. However a
smooth behavior of elastic amplitude is also compatible with
existence of resonances above Roper. The result of our analysis is
hardly affected if we substitute the SAID amplitudes by those of
H\"ohler \cite{Hohler:1979yr}. The only change in this solution are
larger $\pi N$ coupling constants of the two high-mass $P_{11}$
resonances. We use the $\pi N$ coupling constants from the two
analyses to define the final errors. The data of H\"ohler and our
fit are shown in Fig.~\ref{p11_h}. H\"ohler's data are given without
errors. We use 2\% errors below 1600\,MeV and 5\% errors above
1600\,MeV
\begin{figure}[pt]
\bc
\epsfig{file=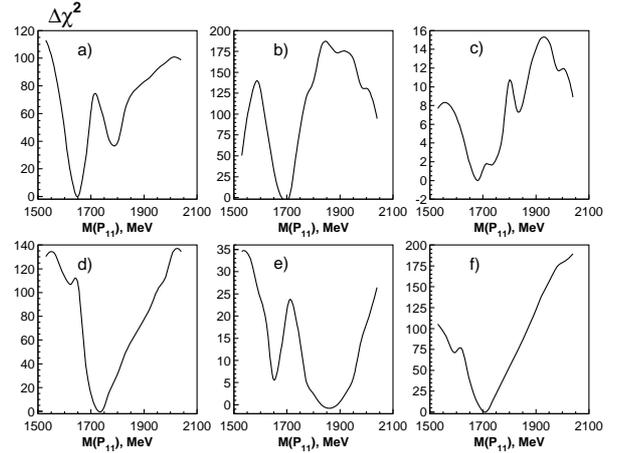,width=0.45\textwidth,clip=on}
\ec
\caption{\label{p11_scan} Mass scan of the $P_{11}$ Breit-Wigner
amplitude. The $P_{11}$ wave was parameterized as two pole K-matrix
(nucleon pole and Roper resonance) plus one Breit-Wigner resonance.
Change of $\chi^2$ for the fit of a) differential cross section and
b) the recoil asymmetry for $\gamma p\to K\Lambda$
\cite{McCracken:2009ra}, c) of the differential cross section for
the $\pi^- p\to \eta n$ \cite{Richards:1970cy}, of the differential
cross section for the $\pi^-p\to K\Lambda$, separated into
low-energy d) \cite{Knasel:1975rr} and high-energy e) region
\cite{Baker:1978qm,Saxon:1979xu}, f) recoil asymmetry for $\pi^-p\to
K^0\Lambda$ \cite{Baker:1978qm,Saxon:1979xu}.\vspace{-5mm} }
\end{figure}

We performed a mass scan of the $P_{11}$ wave in the 1.55-2.10\,GeV
mass region. In this case the $P_{11}$ wave was parameterized as two
pole K-matrix (nucleon pole and Roper resonance) and one
relativistic multi-channel Breit-Wigner amplitude. This amplitude
gave the best $\chi^2$ for a Breit-Wigner mass at 1690\,MeV and for
300\,MeV width. Then, the mass of the Breit-Wigner amplitude was
changed in 30\,MeV steps, fixed, and all other parameters refitted.
The $\chi^2$ changes for 6 reactions with significant contributions
from $P_{11}$ partial wave are shown in Fig.~\ref{p11_scan}. All
curves have a clear minimum in the region 1670-1700\,MeV, including
the fit to $\gamma p\to K\Lambda$ \cite{McCracken:2009ra}. To
emphasize the need for a forth $P_{11}$ resonance we have split the
differential cross section on $\pi^-p\to K^0\Lambda$ into a low
energy region (below 1750\,MeV) from \cite{Knasel:1975rr}, and the
high-energy part (above 1750\,MeV) from
\cite{Baker:1978qm,Saxon:1979xu}. The low-energy region exhibits a
clear minimum at about 1730\,MeV; the high-energy  part still
``sees'' the low-mass peak (now at 1650\,MeV) but a second minimum
in the region 1850-1880\,MeV is clearly present.

\begin{table}[pt]
\caption{\label{residuespdg} Elastic pole residues and comparison
with PDG values. }
\begin{footnotesize}
\renewcommand{\arraystretch}{1.00}
\bc
\begin{tabular}{lccc}
\hline\hline
\vspace{-0.3cm}\\
$N(1440)P_{11}$ & $47\pm3$ & -$(83\pm10)^\circ$ & This work\\
                & $38$     & -$98^\circ$       & \cite{Arndt:2006bf}\\
                & $40$     &                    & \cite{Hohler:1993xq}\\
                & $52\pm5$ & -$(100\pm35)^\circ$& \cite{Cutkosky:1980rh}\\ \hline\vspace{-0.3cm}\\
$N(1710)P_{11}$ & $5\pm4$ & -$(80\pm40)^\circ$ & BG2010-01\\
                & $7\pm3$ & -$(190\pm25)^\circ$& BG2010-02\\
                & $15$     &                    & \cite{Hohler:1993xq}\\
                & $8$      & -$167^\circ$       & \cite{Cutkosky:1980rh}\\ \hline\vspace{-0.3cm}\\
$N(1720)P_{13}$ & $22\pm5$ & -$(85\pm20)^\circ$ & BG2010-01\\
                & $28\pm6$ & -$(85\pm20)^\circ$  & BG2010-02\\
                & $25$     & -$94^\circ$        & \cite{Arndt:2006bf}\\
                & $15$     &                    & \cite{Hohler:1993xq}\\
                & $8\pm2$ & -$(160\pm30)^\circ$ & \cite{Cutkosky:1980rh}\\ \hline\vspace{-0.3cm}\\
$\Delta(1232)P_{33}$& $51.4\pm0.5$ & -$(47\pm1)^\circ$ & This work\\\vspace{-0.3cm}\\
                & $52$     & -$47^\circ$        & \cite{Arndt:2006bf}\\
                & $50$     & -$48^\circ$        & \cite{Hohler:1993xq}\\
                & $52\pm5$ & -$47^\circ$        & \cite{Cutkosky:1980rh}\\ \hline\vspace{-0.3cm}\\
$\Delta(1600)P_{33}$& $14\pm3$ & -$(170\pm15)^\circ$ & This work\\
                & $44$     & -$147^\circ$        & \cite{Arndt:2006bf}\\
                & $17\pm4$ & -$(150\pm30)^\circ$& \cite{Cutkosky:1980rh}\\\hline\vspace{-0.3cm}\\
$\Delta(1910)P_{31}$& $38\pm8$ & -$(120\pm15)^\circ$ & This work\\\vspace{-0.3cm}\\
                & $38$     &                    & \cite{Hohler:1993xq}\\
                & $20\pm4$ & -$(90\pm30)^\circ$ & \cite{Cutkosky:1980rh}\\\hline\vspace{-0.3cm}\\
$\Delta(1920)P_{33}$& $10\pm6$ & $(20\pm60)^\circ$ & This work\\
                & $24\pm4$ & -$(150\pm30)^\circ$& \cite{Cutkosky:1980rh}\\\vspace{-0.3cm}\\
\hline\hline
\end{tabular}
\ec
\end{footnotesize}
\renewcommand{\arraystretch}{1.0}
\end{table}

In Table~\ref{residuespdg} we collect elastic pole residues of those
resonances where our results can be compared with PDG values. In
most cases, the agreement is excellent. Only for
$\Delta(1920)P_{33}$ we find an opposite sign, and there are
significant differences for $N(1710)P_{11}$ in the two solutions
BG2010-01 and BG2010-02.

\section{Extraction of the Breit-Wigner parameters}

The dynamical Breit-Wigner parameterization of the $\pi N$ elastic
amplitude can be written as:
\be
A_{BW}=\frac{(g^{BW}_{\pi N})^2\rho_{\pi
N}(s)}{M^2_{BW}-s-i\sum\limits_i (g^{BW}_i)^2\rho_i(s)}.
\label{dyn_bw}
\ee
Here $s$ is the invariant energy squared, $g^{BW}_i$ is the
Breit-Wigner coupling and $\rho_i(s)$ is the phase volume for the
channel $i$.

The expressions for the phase volumes $\rho_i(s)$ are given in
\cite{Anisovich:2006bc}. To extract the Breit-Wigner parameters from
the multi-channel multi-pole K-matrix parameterization we introduced
the following procedure:

\begin{table}[pt]
\caption{\label{residues2} Effective masses, widths (in\,MeV) and
$|(\Gamma_i\Gamma_f)^{\frac 12}/\Gamma_{tot}|$ calculated at the
effective mass and given in percents.}
\begin{footnotesize}
\renewcommand{\arraystretch}{1.00}
\bc
\begin{tabular}{lccc}
\hline\hline
\vspace{-0.3cm}\\
                  \multicolumn{4}{c}{$P_{11}(1440)$}\\
\hline
\vspace{-0.3cm}\\
 Channel                  & $M_{eff}$                 &  $\Gamma_{eff}$ & $|(\Gamma_i\Gamma_f)^{\frac 12}/\Gamma_{tot}|$
\\
\hline
\vspace{-0.3cm}\\
$\pi N \to \pi N$             & $1389\!\pm\!6$        &$197\!\pm\!10$  & $70\pm 4 \%$  \\
$\pi N \to \eta N$            & $1612\!\pm\!5$        &$420\!\pm\!25$  & $5\pm 3 \%$ \\
\hline \hline
\vspace{-0.3cm}\\
                   \multicolumn{4}{c}{$N_{3/2^+}(1975)$} \\
\hline
\vspace{-0.3cm}\\
 Channel                  & $M_{eff}$                 &  $\Gamma_{eff}$ & $|(\Gamma_i\Gamma_f)^{\frac 12}/\Gamma_{tot}|$  \\
 \hline
\vspace{-0.3cm}\\
$\pi N \to \pi N$             & $1960\!\pm\!20$       & $220\!\pm\!50$ & $1\pm 1 \%$  \\
$\pi N \to \eta N$            & $1970\!\pm\!20$       & $220\!\pm\!50$ & $3\pm 2 \%$ \\
$\pi N \to K\Lambda$          & $1970\!\pm\!20$       & $220\!\pm\!50$ & $4\pm 2 \%$   \\
$\pi N \to K\Sigma$           & $1975\!\pm\!20$       & $210\!\pm\!50$ & $1.5\pm 1 \%$ \\
\hline \hline
\vspace{-0.3cm}\\
                   \multicolumn{4}{c}{$N(1900)P_{13}$}\\
\hline
 Channel                  & $M_{eff}$                 & $\Gamma_{eff}$ & $|(\Gamma_i\Gamma_f)^{\frac 12}/\Gamma_{tot}|$ \\
\hline
\vspace{-0.3cm}\\
$\pi N \to \pi N$             & $1900\!\pm\!20$       & $310\!\pm\!70$ & $1.5\pm 1 \%$  \\
$\pi N \to \eta N$            & $1910\!\pm\!15$       & $305\!\pm\!70$ & $2  \pm 1 \%$ \\
$\pi N \to K\Lambda$          & $1910\!\pm\!15$       & $280\!\pm\!70$ & $3.5\pm 1.5 \%$   \\
$\pi N \to K\Sigma$           & $1920\!\pm\!15$       & $280\!\pm\!70$ & $2.5\pm 1 \%$ \\
\hline \hline
\vspace{-0.3cm}\\
                   \multicolumn{4}{c}{$\Delta(1910)P_{31}$}\\
\hline
\vspace{-0.3cm}\\
 Channel                  & $M_{eff}$                 &  $\Gamma_{eff}$ & $|(\Gamma_i\Gamma_f)^{\frac 12}/\Gamma_{tot}|$ \\
\hline
\vspace{-0.3cm}\\
$\pi N \to \pi N$             & $1895\!\pm\!25$    & $460\pm 40$  & $17\pm 3 \%$  \\
$\pi N \to K\Sigma$           & $1950\!\pm\!25$    & $390\pm 30$  & $4\pm 1 \%$ \\
\hline \hline
\vspace{-0.3cm}\\
                   \multicolumn{4}{c}{$\Delta(1232)P_{33}$}\\
\hline
\vspace{-0.3cm}\\
 Channel                  & $M_{eff}$                 &  $\Gamma_{eff}$ &
 $|(\Gamma_i\Gamma_f)^{\frac 12}/\Gamma_{tot}|$\\
\hline
\vspace{-0.3cm}\\
$\pi N \to \pi N$             & $1229\pm 3$    & $98^{+4}_{-6}$  & $100 \%$  \\
\hline\hline
\vspace{-0.3cm}\\
                   \multicolumn{4}{c}{$\Delta(1600)P_{33}$}\\
\hline
\vspace{-0.3cm}\\
 Channel                  & $M_{eff}$                 &  $\Gamma_{eff}$ & $|(\Gamma_i\Gamma_f)^{\frac 12}/\Gamma_{tot}|$
\\
\hline
\vspace{-0.3cm}\\
$\pi N \to \pi N$             & $1500\pm 20$    & $240^{+15}_{-40}$  & $15\pm 3 \%$  \\
\hline\hline
\vspace{-0.3cm}\\
                   \multicolumn{4}{c}{$\Delta(1920)P_{33}$}\\
\hline
\vspace{-0.3cm}\\
 Channel                  & $M_{eff}$                 &  $\Gamma_{eff}$ & $|(\Gamma_i\Gamma_f)^{\frac 12}/\Gamma_{tot}|$
\\
\hline
\vspace{-0.3cm}\\
$\pi N \to \pi N$             & $1935\pm 35$          & $310\pm 40$  & $7^{+6}_{-3} \%$  \\
$\pi N \to K\Sigma$           & $1950\!\pm\!20$       & $280\pm 30$  & $4.5\pm 2 \%$ \\
\vspace{-0.3cm}\\
\hline\hline
\end{tabular}
\ec
\end{footnotesize}
\renewcommand{\arraystretch}{1.0}
\end{table}

First, Breit-Wigner couplings $g^{BW}_{i}$ were introduced which are
proportional to the couplings calculated as residues in the pole
(see Table~\ref{residues}):
\be
g^{BW}_{i}=f g_{i}
\ee
where $f$ is a global scaling factor for all decay channels. This
factor as well as the Breit-Wigner mass are chosen to match the
Breit-Wigner pole and the pole of the full amplitude resulting from
the fit. Then, helicity Breit-Wigner couplings were calculated to
reproduce the helicity residues in the pole position. Yet, in the
case of rapidly increasing phase volumes, the Breit-Wigner mass and
width are not easily compared with results of other analyses.
Therefore we also introduced an effective Breit-Wigner mass which
corresponds to the maximum of the squared $\pi N\to {final\ state}$
transition amplitude, and an effective width which corresponds to
the full width half maximum of the squared-amplitude distribution.
Further, we report the absolute value of the amplitude at the value
of the effective mass. In the case of a simple Breit-Wigner
amplitude, these numbers correspond to the mass, width and square
root of the product of initial- and final-state branching ratios.
The corresponding numbers for the $P$-wave resonances are given in
Table~\ref{residues2}-\ref{residues4}. The effective mass can be
well above the true pole position, in particular in final states
with a rapidly increasing phase volume.

\section{Discussion and conclusions}

The classical path to explore nucleon and $\Delta$ resonances is the
energy-independent partial wave analysis of elastic $\pi N$
scattering data. The stakes are high: for a full reconstruction of
the scattering amplitudes, required are high-precision data -- over
the full energy range and with complete angular coverage -- (i)~on
differential cross sections, (ii)~on the angular asymmetry of the
outgoing pion when the target nucleon polarization is reversed, and
(iii)~on the polarization of the outgoing nucleon, again over the
full energy range and with complete angular coverage.

\begin{table}[pt]
\caption{\label{residues3} Effective masses, widths (in\,MeV) and
$|(\Gamma_i\Gamma_f)^{\frac 12}/\Gamma_{tot}|$ calculated at the
effective mass and given in percents for solution 01}
\begin{footnotesize}
\renewcommand{\arraystretch}{1.00}
\bc
\begin{tabular}{lccc}
\hline\hline
\vspace{-0.3cm}\\
                   \multicolumn{4}{c}{$N(1710)P_{11}$}
\\
\hline
\vspace{-0.3cm}\\
 Channel                  & $M_{eff}$                 &  $\Gamma_{eff}$ & $|(\Gamma_i\Gamma_f)^{\frac 12}/\Gamma_{tot}|$
\\
\hline
\vspace{-0.3cm}\\
$\pi N \to \pi N$             & $1692\!\pm\!5$        &$200\!\pm\!15$  & $4.5\pm 2.5 \%$  \\
$\pi N \to \eta N$            & $1706\!\pm\!5$        &$200\!\pm\!15$  & $3^{+2}_{-1} \%$ \\
$\pi N \to K\Lambda$          & $1730\!\pm\!6$        &$205\!\pm\!20$  & $8\pm 3 \%$ \\
\hline\hline
\vspace{-0.3cm}\\
                   \multicolumn{4}{c}{$N_{1/2^+}(1875)$}
\\
\hline
\vspace{-0.3cm}\\
 Channel                  & $M_{eff}$                 &  $\Gamma_{eff}$ & $|(\Gamma_i\Gamma_f)^{\frac 12}/\Gamma_{tot}|$
\\
\hline
\vspace{-0.3cm}\\
$\pi N \to \pi N$             & $1859\!\pm\!4$       & $116\pm 15$ & $5\pm 3 \%$  \\
$\pi N \to \eta N$            & $1860\!\pm\!7$       & $115\pm 15$ & $15^{+5}_{-2} \%$ \\
$\pi N \to K\Lambda$          & $1862\!\pm\!7$       & $115\pm 15$ & $4\pm 2 \%$   \\
$\pi N \to K\Sigma$           & $1864\!\pm\!9$       & $115\pm 15$ & $7\pm 4 \%$ \\
\hline\hline
\vspace{-0.3cm}\\
                   \multicolumn{4}{c}{$N(1720)P_{13}$}
\\
\hline
\vspace{-0.3cm}\\
 Channel                  & $M_{eff}$                 &  $\Gamma_{eff}$ & $|(\Gamma_i\Gamma_f)^{\frac 12}/\Gamma_{tot}|$
\\
\hline
\vspace{-0.3cm}\\
$\pi N \to \pi N$             & $1677^{+30}_{-10}$    & $310\pm 30$ & $10\pm 2 \%$  \\
$\pi N \to \eta N$            & $1700^{+40}_{-10}$    & $300\pm 35$ & $7\pm 3 \%$ \\
$\pi N \to K\Lambda$          & $1740\!\pm\!25$       & $410\pm 40$ & $3\pm 2 \%$   \\
$\pi N \to K\Sigma$           & $1840\!\pm\!25$       & $480\pm 50$ & $8^{+5}_{-2} \%$ \\
\vspace{-0.3cm}\\
\hline\hline
\end{tabular}
\ec
\end{footnotesize}
\renewcommand{\arraystretch}{1.0}
\end{table}
These data do not exist. Hence one needs to rely on further
theoretical input. Dispersion relations provide a powerful and
flexible tool to constrain the low-energy region by our knowledge of
strong interactions at high energies. However, approximations need
to be made, and obviously, these approximations have a significant
impact on the results. At least, the approximations made in the
three classical energy-independent partial wave analyses
\cite{Hohler:1979yr,Cutkosky:1980rh,Arndt:2006bf} lead to
significantly different amplitudes, at least in several partial
waves and above the first resonance in a given partial wave.

\begin{table}[pt]
\caption{\label{residues4} Effective masses, widths (in\,MeV) and
$|(\Gamma_i\Gamma_f)^{\frac 12}/\Gamma_{tot}|$ calculated at the
effective mass and given in percents for solution 02}
\begin{footnotesize}
\renewcommand{\arraystretch}{1.00}
\bc
\begin{tabular}{lccc}
\hline\hline
\vspace{-0.3cm}\\
                   \multicolumn{4}{c}{$N(1710)P_{11}$}
\\
\hline
\vspace{-0.3cm}\\
 Channel                  & $M_{eff}$                 &  $\Gamma_{eff}$ & $|(\Gamma_i\Gamma_f)^{\frac 12}/\Gamma_{tot}|$
\\
\hline
\vspace{-0.3cm}\\
$\pi N \to \pi N$             & $1700\!\pm\!6$        &$210\!\pm\!20$  & $5\pm 2 \%$  \\
$\pi N \to \eta N$            & $1714\!\pm\!6$        &$215\!\pm\!20$  & $7\pm 3 \%$ \\
$\pi N \to K\Lambda$          & $1740\!\pm\!10$       &$210\!\pm\!20$  & $16\pm 4 \%$ \\
\hline\hline
\vspace{-0.3cm}\\
                   \multicolumn{4}{c}{$N_{1/2^+}(1875)$}
\\
\hline
\vspace{-0.3cm}\\
 Channel                  & $M_{eff}$                 &  $\Gamma_{eff}$ & $|(\Gamma_i\Gamma_f)^{\frac 12}/\Gamma_{tot}|$
\\
\hline
\vspace{-0.3cm}\\
$\pi N \to \pi N$             & $1845\!\pm\!25$      & $340\pm 30$ & $8\pm 4 \%$  \\
$\pi N \to \eta N$            & $1855\!\pm\!30$      & $320\pm 35$ & $9\pm 3 \%$ \\
$\pi N \to K\Lambda$          & $1865\!\pm\!30$      & $330\pm 30$ & $10\pm 4\%$   \\
$\pi N \to K\Sigma$           & $1880\!\pm\!30$      & $300\pm 30$ & $6\pm 2 \%$ \\
\hline\hline
\vspace{-0.3cm}\\
                   \multicolumn{4}{c}{$N(1720)P_{13}$}
\\
\hline
\vspace{-0.3cm}\\
 Channel                  & $M_{eff}$                 &  $\Gamma_{eff}$ & $|(\Gamma_i\Gamma_f)^{\frac 12}/\Gamma_{tot}|$
\\
\hline
\vspace{-0.3cm}\\
$\pi N \to \pi N$             & $1660\pm 25$          & $310\pm 30$ & $12\pm 2 \%$  \\
$\pi N \to \eta N$            & $1690\pm 25$          & $300\pm 30$ & $5\pm 2 \%$ \\
$\pi N \to K\Lambda$          & $1730\!\pm\!20$       & $400\pm 35$ & $4\pm 1 \%$   \\
$\pi N \to K\Sigma$           & $1840\!\pm\!20$       & $480\pm 40$ & $9\pm 3 \%$ \\

\vspace{-0.3cm}\\
\hline\hline
\end{tabular}
\ec
\end{footnotesize}
\renewcommand{\arraystretch}{1.0}
\end{table}

Of course, energy-independent partial wave analyses are not the
final step to determine the number and the properties of resonances
in a particular partial wave. Real and imaginary part of the
amplitude have to be fitted within an energy-dependent model which
parameterizes resonant and non-resonant terms. The model tries to
describe the amplitudes with a minimum number of resonances; further
resonances coupling weakly to $N\pi$ can always exist. But then,
they have to show up in other reactions.

In the energy dependent analyses, the total number of nucleon and
$\Delta$ resonances suggested by
\cite{Hohler:1979yr,Cutkosky:1980rh} is larger by more than a factor
two than in \cite{Arndt:2006bf}. In \cite{Arndt:2006bf}, more recent
precision data were used and their predictions of the spin rotation
parameters agree much better with recent data than those based on
the amplitudes from \cite{Hohler:1979yr,Cutkosky:1980rh}. Thus the
existence of many states reported by PDG is seriously challenged.

The difference is very important for our understanding of the baryon
spectrum. Quark models predict a large number of resonances in each
partial wave. The high-mass resonances are predicted to have small
couplings to $N\pi$, hence they may be difficult to identify in $\pi
N$ elastic scattering, but it is surprising when only one resonance
is observed per partial wave. A modern view questions the usefulness
of quarks to describe the nucleon excitation spectrum. Instead,
resonances are generated dynamically from the interaction of baryons
and mesons. In this approach, excitations above the lowest-mass
state in a given partial wave are not necessarily expected. The
results of \cite{Arndt:2006bf} are thus a strong support for the
conjecture that quark degrees of freedom may play only a minor role
in the spectroscopy of light baryons. We just mention that a modern
quark-model variant, AdS/QCD, provides a link between perturbative
high-energy scattering and spectroscopy (see, e.g.
\cite{Brodsky:2010ur} and references therein). In a special variant,
AdS/QCD reproduces with surprising precision the full spectrum of
nucleon and $\Delta$ resonances \cite{Forkel:2008un}, including all
those resonances which are challenged in \cite{Arndt:2006bf}. If
these all do not exist, the coincidence between AdS/QCD and data
would be extremely fortuitous.

$P$-wave nucleon and $\Delta$ resonances provide a fruitful area in
which the alternatives \cite{Hohler:1979yr,Cutkosky:1980rh} and
\cite{Arndt:2006bf} can be tested. Above the famous $\Delta(1232)$,
two further states, $\Delta(1600)P_{33}$ and $\Delta(1920)$
$P_{33}$, were reported in \cite{Hohler:1979yr,Cutkosky:1980rh};
$\Delta(1920)P_{33}$ is absent in \cite{Arndt:2006bf}. There is no
$N(1710)P_{11}$ above the Roper resonance in \cite{Arndt:2006bf} and
no $N(1900)$ $P_{13}$. The four-star $\Delta(1910)P_{31}$ is
required in \cite{Hohler:1979yr,Cutkosky:1980rh}, and is
questionable in \cite{Arndt:2006bf}.

In this paper, we reported evidence for these debated states from a
coupled-channel fit which includes inelastic reactions. We have
studied the mass spectrum of $P$-wave nucleon and $\Delta$
resonances in a coupled-channel analysis of a large body of data on
pion- and photo-induced reactions. Individual partial waves are
typically described by 6-channel K-matrices with up to five poles.
The coupled-channel fit minimizes the number of parameters needed to
represent resonant and background contributions. The $\pi N$ elastic
amplitudes from energy-independent partial wave analyses are
included in the data base. Hence we fit the same data as
\cite{Hohler:1979yr,Arndt:2006bf} but in our fits, the
inelasticities are no black box but constrained by data from pion-
and photo-induced inelastic reactions in which the final states are
fully reconstructed. Particularly useful are here data on hyperon
production. $\Lambda$ and $\Sigma$ baryons produced in a reaction
reveal their polarization state in their decay. In recent
experiments on photo-production of hyperons, polarized photons were
used which allow the experiments to study the polarization transfer
from the initial to the final state. These experiments are phase
sensitive and constrain drastically the final solution. This use of
inelastic reactions is, in our view, a major step forward compared
to \cite{Hohler:1979yr,Arndt:2006bf} since in the latter analyses,
the inelasticities are not really under control.

The surprising result of our analysis is that the $P$-wave
resonances reported by \cite{Hohler:1979yr,Cutkosky:1980rh} and
challenged by \cite{Arndt:2006bf} are found with properties which
are not very different from those reported by
\cite{Hohler:1979yr,Cutkosky:1980rh}. In particular we confirm
existence and properties of $N(1710)P_{11}$ and $N(1900)P_{13}$ and
of  $\Delta(1600)P_{33}$ and $\Delta(1920)$ $P_{33}$. We do not find
evidence for $\Delta(1750)P_{31}$, a resonance not seen by
\cite{Hohler:1979yr,Cutkosky:1980rh} (but in some other analyses)
and, last not least, we confirm with good evidence $N_{1/2^+}(1875)$
\cite{Castelijns:2007qt,Anisovich:2007bq} and find evidence for a
tentative $N_{3/2^+}(1975)$.

\subsection*{Acknowledgements}
We would like to thank the members of SFB/TR16 for continuous
encouragement. We acknowledge financial support from the Deutsche
Forschungsgemeinschaft (DFG) within the SFB/ TR16 and from the
Forschungszentrum J\"ulich within the FFE program.

\end{document}